\documentclass[twocolumn]{aastex62}

\usepackage{hyperref}
\usepackage{amsmath}

\newcommand{\angstrom}{\textup{\AA}}

\accepted{May 14, 2019}
\submitjournal{ApJ}

\shorttitle{Stronger Constraints on the Evolution of the $M_{\rm{BH}}-\sigma_*$ Relation up to $z\sim0.6$}
\shortauthors{Sexton et al. 2019}

\begin{document}

\title{STRONGER CONSTRAINTS ON THE EVOLUTION OF THE $M_{\rm{BH}}-\sigma_*$ RELATION UP TO $z\sim0.6$}

\correspondingauthor{Remington O. Sexton}
\email{remington.sexton@email.ucr.edu}

\author[0000-0003-3432-2094]{Remington O. Sexton}
\affil{Department of Physics and Astronomy, University of California, Riverside, CA 92521}

\author[0000-0003-4693-6157]{Gabriela Canalizo}
\affil{Department of Physics and Astronomy, University of California, Riverside, CA 92521}

\author{Kyle D. Hiner}
\affil{Department of Physics and Astronomy, University of California, Riverside, CA 92521}
\affil{Departamento de Astronom´ıa, Universidad de Concepci´on, Chile}

\author[0000-0002-9214-4428]{Stefanie Komossa}
\affil{Max-Planck-Institut f{\"u}r Radioastronomie, Auf dem H{\"u}gel 69, D-53121 Bonn, Germany}

\author[0000-0002-8055-5465]{Jong-Hak Woo}
\affil{Seoul National University, Republic of Korea}

\author[0000-0001-7568-6412]{Ezequiel Treister}
\affil{Instituto de Astrofísica, Facultad de Física, Pontificia Universidad Católica de Chile, Casilla 306, Santiago 22, Chile}

\author{Sabrina Lyn Hiner Dimassimo}
\affil{Institute for Defense Analyses, USA}



\begin{abstract}
We revisit the possibility of redshift evolution in the $M_{\rm{BH}}-\sigma_*$ relation with a sample of 22 Seyfert 1 galaxies with black holes (BHs) in the mass range $10^{6.3}-10^{8.3}~M_\odot$ and redshift range $0.03<z<0.57$ with spectra obtained from spatially resolved Keck/Low-Resolution Imaging Spectrometer observations.  Stellar velocity dispersions were measured directly from the \ion{Mg}{1b} region, taking into consideration the effect of \ion{Fe}{2} contamination, active galactic nucleus (AGN) dilution, and host-galaxy morphology on our measurements.  BH masses are estimated using the H$\beta$ line width, and the luminosity at 5100 \angstrom~ is estimated from surface brightness decomposition of the AGN from the host galaxy using high-resolution imaging from the \textit{Hubble Space Telescope}.  Additionally, we investigate the use of the [\ion{O}{3}]$\lambda5007$ emission line width as a surrogate for stellar velocity dispersion, finding better correlation once corrected for \ion{Fe}{2} contamination and any possible blueshifted wing components.  Our selection criteria allowed us to probe lower-luminosity AGNs and lower-mass BHs in the non-local universe than those measured in previous single-epoch studies.  We find that any offset in the $M_{\rm{BH}}-\sigma_*$ relation up to $z\sim0.6$ is consistent with the scatter of local BH masses, and address the sources of biases and uncertainties that contribute to this scatter.
\end{abstract}



\section{Introduction} \label{sec:intro}
 
Since their initial discovery nearly two decades ago, black hole (BH) scaling relations have motivated extensive study in the role of central supermassive black holes (SMBHs) in the evolution of their host galaxies over cosmic time.  Measurements of gas kinematics in inactive galaxies revealed a strong correlation between the mass $M_{\rm{BH}}$ of the central SMBH and the stellar velocity dispersion $\sigma_*$ of the central spheroid of its host galaxy \citep{Ferrarese2000,Gebhardt2000a,Tremaine2002,McConnell2013}.  This fundamental relationship was soon established for local active galaxies as well by exploiting the visible broad-line regions (BLRs) in type 1 active galactic nuclei (AGNs) as a direct probe of virial BH mass \citep{Gebhardt2000b,Ferrarese2001,Onken2004,Greene2006a,Woo2010,Bennert2011a,Woo2013,Bennert2015,Woo2015}.  Today, the $M_{\rm{BH}}-\sigma_*$ relation remains the strongest and most fundamental correlation between SMBHs and their host galaxies (see \citet{Kormendy2013} for a comprehensive review).\\
\indent There exists an ongoing debate as to whether SMBHs co-evolve in tandem with their host galaxies over time or if the scaling relations we observe today are an evolutionary endpoint, such that host galaxies grow over time to ``catch up'' to their SMBHs formed at much earlier times.  Co-evolution would imply some feedback mechanism powered by the central AGN, which acts to self-regulate the growth of the SMBH and host galaxy \citep{Fabian2012,King2015}.  In the latter scenario, scaling relations as the result of an evolutionary endpoint call into question how the seeds of today's SMBHs grew so rapidly in the early universe \citep{Volonteri2010,Greene2012}.  Alternatively, the emergence of BH scaling relations could be non-causal in nature, and could be explained through the hierarchical assembly of BH and stellar mass via mergers \citep{Jahnke2011}.  To address the controversy, numerous attempts have been made to measure $M_{\rm{BH}}-\sigma_*$ in the non-local universe to determine which SMBH evolutionary track may be responsible for local observations.\\
\indent Early attempts by \citet{Woo2006} and \citet{Woo2008} to measure the $M_{\rm{BH}}-\sigma_*$ relation of broad-line Seyfert 1 galaxies at $z=0.36$ and $z=0.57$ resulted in a significant positive offset of 0.43 dex and 0.63 dex in $M_{\rm{BH}}$, respectively, implying that BHs were ``overmassive'' relative to their host galaxies at earlier times.  If we quantify the required stellar mass assembly as inferred from stellar velocity dispersion \citep{Zahid2016}, the results by Woo et al.\ imply that host bulges must grow by a factor of $\sim4$ within 4 Gyr ($z=0.36$), and a factor of $\sim6$ within 5.5 Gyr ($z=0.57$), to be consistent with the $M_{\rm{BH}}-\sigma_*$ relation at $z=0$. This is problematic since it means that bulges must undergo significant stellar mass assembly in a relatively short amount of time, and the possible mechanisms for doing so without significantly growing their BHs remain largely speculative.  Similar studies by \citet{Canalizo2012} using dust-reddened 2MASS quasi-stellar objects (QSOs) at $0.14<z<0.37$, and \citet{Hiner2012} using post-starburst QSOs at $z\sim0.3$, found a similar significant positive offset from the local relation, which further exacerbated the problem.  \\
\indent It is however possible that the observed offset in the $M_{\rm{BH}}-\sigma_*$ relation at higher redshifts is not of physical origin, but the result of selection bias.  \citet{Lauer2007} explained that AGNs selected by a luminosity threshold preferentially selects overmassive BHs relative to their hosts due to a steep drop in the luminosity function of galaxies. In addition to this, \citet{Shen2010} suggested that single-epoch (SE) samples can be biased toward high BH masses due to uncorrelated variations between continuum luminosity and line widths in reverberation mapping studies.  These two biases can act independently and in conjunction with one another to create the observed offset from the local $M_{\rm{BH}}-\sigma_*$ relation and give a false indication of host-galaxy evolution.  Selecting samples at both low and high redshift using consistent criteria can help to mitigate these biases.  In addition to this, since previous non-local studies primarily sampled BHs at the high-mass regime of the $M_{\rm{BH}}-\sigma_*$ relation, it would be ideal to sample the low-mass regime of the $M_{\rm{BH}}-\sigma_*$ relation as a function of redshift.  Since selection criteria based on AGN luminosity necessarily bias samples toward the more massive BHs of AGNs, we have historically lacked a sample of lower-mass BHs of comparable galaxy sizes as those previously studied, especially in the non-local universe. \\
\indent In this paper we attempt to address the aforementioned biases using a new set of selection criteria based on the broad H$\beta$ emission line width to select lower-mass BHs in the non-local universe.  In Section \ref{sec:data_acq} we discuss our sample selection, observations/data acquisition, and reduction procedure.  In Section \ref{sec:analysis} we describe in detail how measurements of $\sigma_*$, line widths, and AGN luminosity are performed to calculate BH mass.  We also investigate the use of the [\ion{O}{3}] width as a proxy for $\sigma_*$ in the context of BH scaling relations following the precedent of previous studies  \citep{Brotherton1996,McIntosh1999,Veron-Cetty2001,Shields2003,Greene2005,Woo2006,Komossa2007,Bennert2018}.  In Section \ref{sec:results} we present our results for our sample on the $M_{\rm{BH}}-\sigma_*$ relation and investigate the possible evolution as a function of redshift.  We discuss any systematic uncertainties and selection biases which may affect our results in Section \ref{sec:systematics}.  Finally, we discuss the implications of our results in Section \ref{sec:discussion}.\\
\indent Throughout this paper, we assume a standard cosmology of $\Omega_m=0.27$, $\Omega_\Lambda=0.73$, and $H_0=71$ km s$^{-1}$ Mpc$^{-1}$.  We refer to individual objects by their abbreviated object designations (i.e., J000338, etc.).

\section{Data Acquisition} \label{sec:data_acq}

\subsection{Sample Selection} \label{sec:sample}

To construct the sample, objects were selected from the SDSS DR7 \citep{sdss} database which satisfied the following properties: (1) a redshift within the range $0.0 < z<0.9$ to ensure that the broad H$\beta$ and \ion{Mg}{1b} complexes were within the observed spectral range of the SDSS, (2) a broad H$\beta$ FWHM within the range $500\text{~km s}^{-1}\leq\mathrm{FWHM}_{\mathrm{H}\beta}\leq2000 \text{~km s}^{-1}$ to select lower BH mass objects, and (3) visible stellar absorption features (typically \ion{Ca}{0}H+K equivalent width $\mathrm{EW}_{\mathrm{Ca H+K}}>0.5$ \angstrom~) to ensure that $\sigma_*$ could be accurately measured.  The resulting 2539 objects were then cross-referenced with \textit{HST} archival data to ensure that high-resolution images were available for detailed deconvolution of the AGN point-spread function (PSF) and its respective host galaxy.  Relatively deep (1000-2000+ s) \textit{HST} imaging was found for 32 objects, performed using a variety of instruments and filters, and spanning the redshift range $0.03<z<0.57$.  Observational time constraints allowed for spatially resolved optical spectroscopy of 29 of the 32 objects using the Keck Low-Resolution Imaging Spectrometer (LRIS; see Section \ref{sec:observations}).  Modeling of the power-law AGN continuum to determine the luminosity at 5100 \angstrom\; could not be performed on seven objects due to the high fraction of stellar light from the host galaxy and were omitted from the final sample.  \\
\indent The final sample of 22 objects are listed in Table \ref{tab:obs_table}.  Of these, eight satisfy the H$\beta$ width criteria for narrow-line Seyfert 1 (NLS1) galaxies ($500\text{~km s}^{-1}\leq\mathrm{FWHM}_{\mathrm{H}\beta}\leq2000 \text{~km s}^{-1}$; \citet{Goodrich1989}), while the remaining 14 are classified as broad-line Seyfert 1 (BLS1) galaxies ($\mathrm{FWHM}_{\mathrm{H}\beta}>2000 \text{~km s}^{-1}$).  It is possible that more objects in our sample satisfy the broad-line width criterion for NLS1s since these objects tend to exhibit Lorentzian profiles \citep{Veron-Cetty2001}; however we still require the Gaussian FWHM model to determine BH mass.\\
\indent We note that the definition of ``NLS1'' can extend beyond the H$\beta$ line width criteria given above.  Previous studies have selected NLS1s based on the flux ratio [\ion{O}{3}]/H$\beta_{\rm{broad}}<3$ \citep{Shuder1981,Osterbrock1985}, which ensures that NLS1s have larger H$\beta$ widths than forbidden lines; however this criterion does not exclude BLS1s.  All 22 objects in our sample satisfy the [\ion{O}{3}]/H$\beta_{\rm{broad}}<3$ criterion by virtue of the fact that all of the objects in our sample are Type 1 AGNs.  Another commonly cited characteristic of NLS1 galaxies include strong \ion{Fe}{2} emission in the presence of weak [\ion{O}{3}] emission.  However, more recent studies with larger samples of NLS1s have found that correlations of \ion{Fe}{2} with other emission line properties are not as unique to NLS1s as previously thought.  For instance, \citet{Veron-Cetty2001} found that any anti-correlation between \ion{Fe}{2} and [\ion{O}{3}] is weak at best, and concluded that all objects with broad H$\beta<2000$ km s$^{-1}$ are genuine NLS1s.  Similarly, \citet{Xu2012} and \citet{Valencia-S2012} found that the same correlations between \ion{Fe}{2} and other emission line properties commonly found in NLS1s are as common among BLS1s, implying that these selection criteria are not unique to NLS1s, but rather that strong \ion{Fe}{2} emission is a common property across the arbitrarily chosen line width criteria that distinguish NLS1s and BLS1s.  We therefore find that our definition of NLS1 based on solely on line width is justified.\\
\indent The small fraction of NLS1 objects obtained in the final sample can be traced back to the simplistic algorithm used to perform emission line fits in SDSS DR7, particularly when applied to SDSS-classified QSO spectra.  The large (1000 pixel) mean/median filter used for continuum subtraction does not perform well in the presence of a strong and rapidly varying stellar continuum.  Additionally, the DR7 algorithm does not simultaneously fit narrow and broad components, and can therefore produce inaccurate results if there is a strong narrow-line emission present atop a broad-line component.  The DR7 algorithm performs optimally when fitting SDSS-classified QSOs which exhibit a weaker stellar continuum relative to the AGN continuum and weaker narrow-line emission relative to broad-line emission.  This is opposite of what is seen of typical NLS1 galaxies, which have a stronger stellar continuum relative the AGN continuum, and narrow emission lines of comparable widths to the broad-line emission components.  Because the DR7 algorithm is not optimized for the peculiar spectra of NLS1 galaxies, we do not recommend using the DR7 emission-line database (\texttt{specLine} table) to query NLS1 objects.

\subsection{Observations} \label{sec:observations}

\begin{figure*}[pht!]
\centering
\includegraphics[trim={0 0 0 0},scale=0.7,clip]{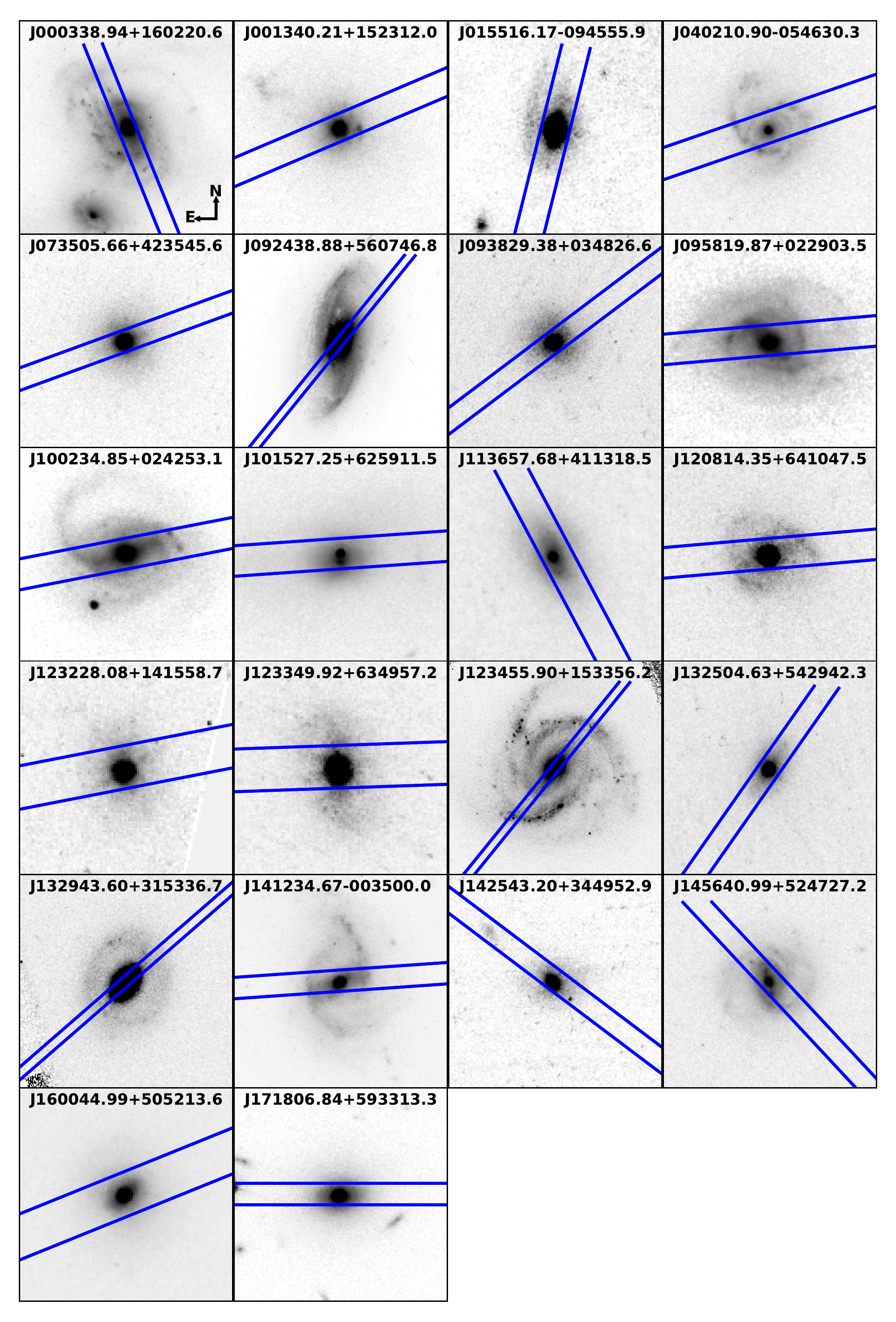}
\caption{ \textit{HST} imaging cutouts of our sample.  Each image is aligned with north pointed up.  Blue bars represent the Keck/LRIS 1$''$ slit aperture placed on each object and aligned according to the PA (in degrees N of E) listed in Table \ref{tab:obs_table}.
\label{fig:slitgrid}}
\end{figure*}

Long-slit spectroscopy was performed on 2015 March 24-25 and 2015 December 3-4 using the LRIS (\citet{lris1}) on the Keck I Telescope atop the summit of Maunakea in Hawai'i.  Weather conditions for all nights were clear, with subarcsecond seeing ranging between $0\,\farcs6$ and $0\,\farcs8$.  A 1$''$ slit was chosen to spatially resolve both the central region close to the AGN and the host galaxy bulge within the effective (half-light) radius.  Figure \ref{fig:slitgrid} shows the position angle of the slit, chosen to be aligned with the semi-major axis of the bulge component of the host galaxy.  After passing through the slit, the beam is then collimated and split by a dichroic designated by a wavelength cutoff.  Wavelengths below the dichroic cutoff are passed through a grism and into the LRIS-B camera, while wavelengths above the cutoff are passed through a grating of a specified blaze angle into the LRIS-R camera.  Both the LRIS-B and LRIS-R (\citet{lris2}) CCD detectors have a pixel scale of 0.135$''$pixel$^{-1}$.  Table \ref{tab:obs_table} lists the dichroic, grating, and central wavelength used for each object to ensure that the region around H$\beta$ was captured on the LRIS-R detector.  The 1200/7500, 900/5500, and 600/5000 lines mm$^{-1}$ gratings provide logarithmically rebinned (constant velocity) spectral resolutions of $\sim20$, $27$, and $40$ km s$^{-1}$, respectively.  All observations with the LRIS-B detector utilized the 600/4000 lines mm$^{-1}$ grism, which has a spectral resolution of $\sim32$ km s$^{-1}$.\\

\movetabledown=1.25in
\begin{rotatetable*}
\begin{deluxetable*}{ccccclclccc}
\tablecaption{Keck/LRIS Observations \label{tab:obs_table}}
\tabletypesize{\footnotesize}
\tablehead{
\colhead{Object} & \colhead{R.A.} & \colhead{Decl.} & \colhead{$z$} & \colhead{Spatial Scale} & \colhead{PA} & \colhead{Dichroic} & \colhead{Grating} & \colhead{Cen. Wave.} & \colhead{Exp. Time} & \colhead{Obs. Date} \\
\colhead{} & \colhead{(J2000)} & \colhead{(J2000)} & \colhead{} & \colhead{(kpc pix$^{-1}$)} & \colhead{(deg)} & \colhead{(\angstrom)} & \colhead{(l mm$^{-1}$)} & \colhead{(\angstrom)} & \colhead{(s)} & \colhead{(yyyy mm dd)}
}
\colnumbers
\startdata
J000338.94+160220.6   & 00:03:38.94 & +16:02:20.65 & 0.11681 & 0.281 & 22  & 500 & 600/5000  & 6500 & 1200 & 2015 Dec 03 \\
J001340.21+152312.0   & 00:13:40.21 & +15:23:12.04 & 0.12006 & 0.288 & 113 & 500 & 600/5000  & 6500 & 1200 & 2015 Dec 03 \\
J015516.17$-$094555.9 & 01:55:16.17 & -09:45:55.94 & 0.56425 & 0.875 & 166 & 560 & 600/5000  & 6500 & 2400 & 2015 Dec 04 \\
J040210.90$-$054630.3 & 04:02:10.90 & -05:46:30.35 & 0.27065 & 0.554 & 109 & 560 & 600/5000  & 6500 & 1200 & 2015 Dec 04 \\
J073505.66+423545.6   & 07:35:05.66 & +42:35:45.68 & 0.08646 & 0.215 & 110 & 460 & 1200/7500 & 5360 & 1200 & 2015 Mar 24 \\
J092438.88+560746.8   & 09:24:38.88 & +56:07:46.84 & 0.02548 & 0.067 & 96  & 460 & 1200/7500 & 5360 & 600	 & 2015 Mar 24 \\
J093829.38+034826.6   & 09:38:29.38 & +03:48:26.69 & 0.11961 & 0.287 & 127 & 460 & 1200/7500 & 5360 & 1200 & 2015 Mar 24 \\
J095819.87+022903.5   & 09:58:19.87 & +02:29:03.51 & 0.34643 & 0.657 & 95  & 560 & 1200/7500 & 6800 & 1200 & 2015 Mar 24 \\
J100234.85+024253.1   & 10:02:34.85 & +02:42:53.17 & 0.19659 & 0.434 & 101 & 560 & 600/5000  & 7150 & 1200 & 2015 Mar 25 \\
J101527.25+625911.5   & 10:15:27.25 & +62:59:11.59 & 0.35064 & 0.663 & 94  & 560 & 1200/7500 & 6800 & 2400 & 2015 Mar 24 \\
J113657.68+411318.5   & 11:36:57.68 & +41:13:18.51 & 0.07200 & 0.182 & 28  & 500 & 1200/7500 & 5760 & 2400 & 2015 Mar 24 \\
J114851.61+514528.7   & 11:48:51.61	& +51:45:28.73 & 0.06742 & 0.171 & 66  & 500 & 1200/7500 & 5760 & 1200 & 2015 Mar 24 \\
J120814.35+641047.5   & 12:08:14.35 & +64:10:47.57 & 0.10555 & 0.257 & 95  & 560 & 900/5500  & 6640 & 1200 & 2015 Mar 25 \\
J123228.08+141558.7   & 12:32:28.08 & +14:15:58.75 & 0.42692 & 0.750 & 101 & 560 & 1200/7500 & 7200 & 2400 & 2015 Mar 24 \\
J123349.92+634957.2   & 12:33:49.92 & +63:49:57.23 & 0.13407 & 0.316 & 92  & 560 & 900/5500  & 6640 & 1200 & 2015 Mar 25 \\
J123455.90+153356.2   & 12:34:55.90 & +15:33:56.28 & 0.04637 & 0.120 & 141 & 500 & 1200/7500 & 5760 & 600  & 2015 Mar 24 \\
J132504.63+542942.3   & 13:25:04.63 & +54:29:42.38 & 0.14974 & 0.347 & 145 & 560 & 600/5000  & 7150 & 1200 & 2015 Mar 25 \\
J132943.60+315336.7   & 13:29:43.60 & +31:53:36.76 & 0.09265 & 0.229 & 131 & 560 & 900/5500  & 6640 & 1200 & 2015 Mar 25 \\
J141234.67$-$003500.0 & 14:12:34.67 & -00:35:00.06 & 0.12724 & 0.302 & 94  & 500 & 1200/7500 & 5760 & 1200 & 2015 Mar 24 \\
J142543.20+344952.9   & 14:25:43.20 & +34:49:52.91 & 0.17927 & 0.403 & 53  & 560 & 600/5000  & 7150 & 1200 & 2015 Mar 25 \\
J145640.99+524727.2   & 14:56:40.99 & +52:47:27.24 & 0.27792 & 0.565 & 43  & 560 & 600/5000  & 7150 & 2400 & 2015 Mar 25 \\
J160044.99+505213.6   & 16:00:44.99 & +50:52:13.60 & 0.10104 & 0.247 & 112 & 500 & 1200/7500 & 5760 & 2400 & 2015 Mar 24 \\
J171806.84+593313.3   & 17:18:06.84 & +59:33:13.32 & 0.27356 & 0.558 & 90  & 560 & 600/5000  & 7150 & 1200 & 2015 Mar 25 \\
\enddata
\tablecomments{Summary of Keck/LRIS observations.  Column 1: object.  Column 2: R.A. Column 3: decl.  Column 4: redshift.  Column 5: spatial scale per pixel.  Column 6: position angle of slit during observations measured E of N.  Column 7: dichroic cutoff wavelength.  Column 8: LRIS-R grating.  Column 9: central wavelength of grating.  Column 10: exposure time.  Column 10: observation date.}
\end{deluxetable*}
\end{rotatetable*}

\begin{figure*}[pht!]
\centering
\includegraphics[trim={1cm 5.0cm 2cm 2.5cm},scale=0.60,clip]{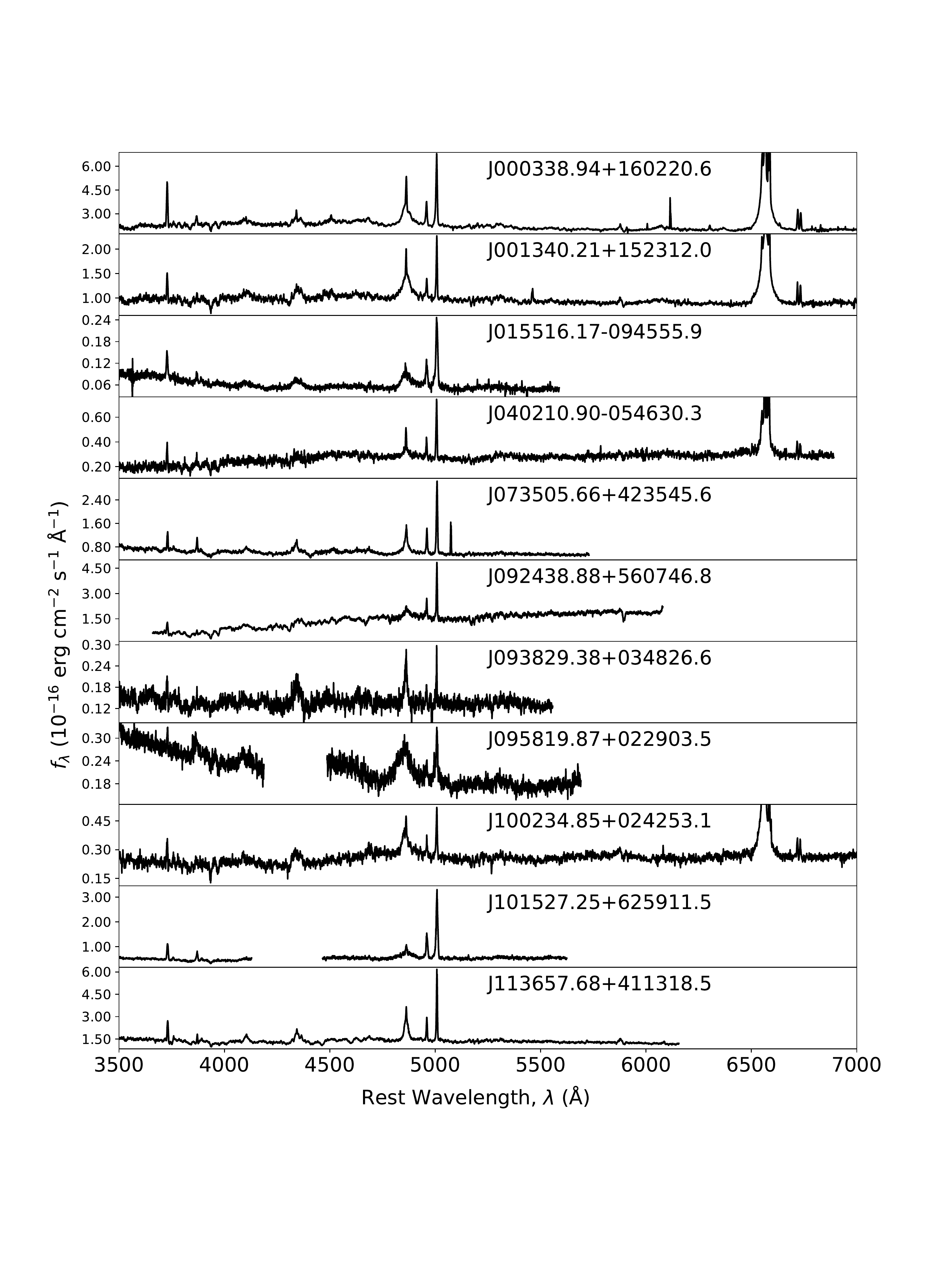}
\caption{
Spectra of our sample from Keck/LRIS observations.  Gaps in coverage correspond to the dichroic cutoff wavelength and the grating coverage, which depend on the resolution and central wavelength of the grating chosen for each object.
\label{fig:observations}}
\end{figure*}

\renewcommand{\thefigure}{\arabic{figure}}
\addtocounter{figure}{-1}
\begin{figure*}[pht!]
\centering
\includegraphics[trim={1cm 5.cm 2cm 2.5cm},scale=0.60,clip]{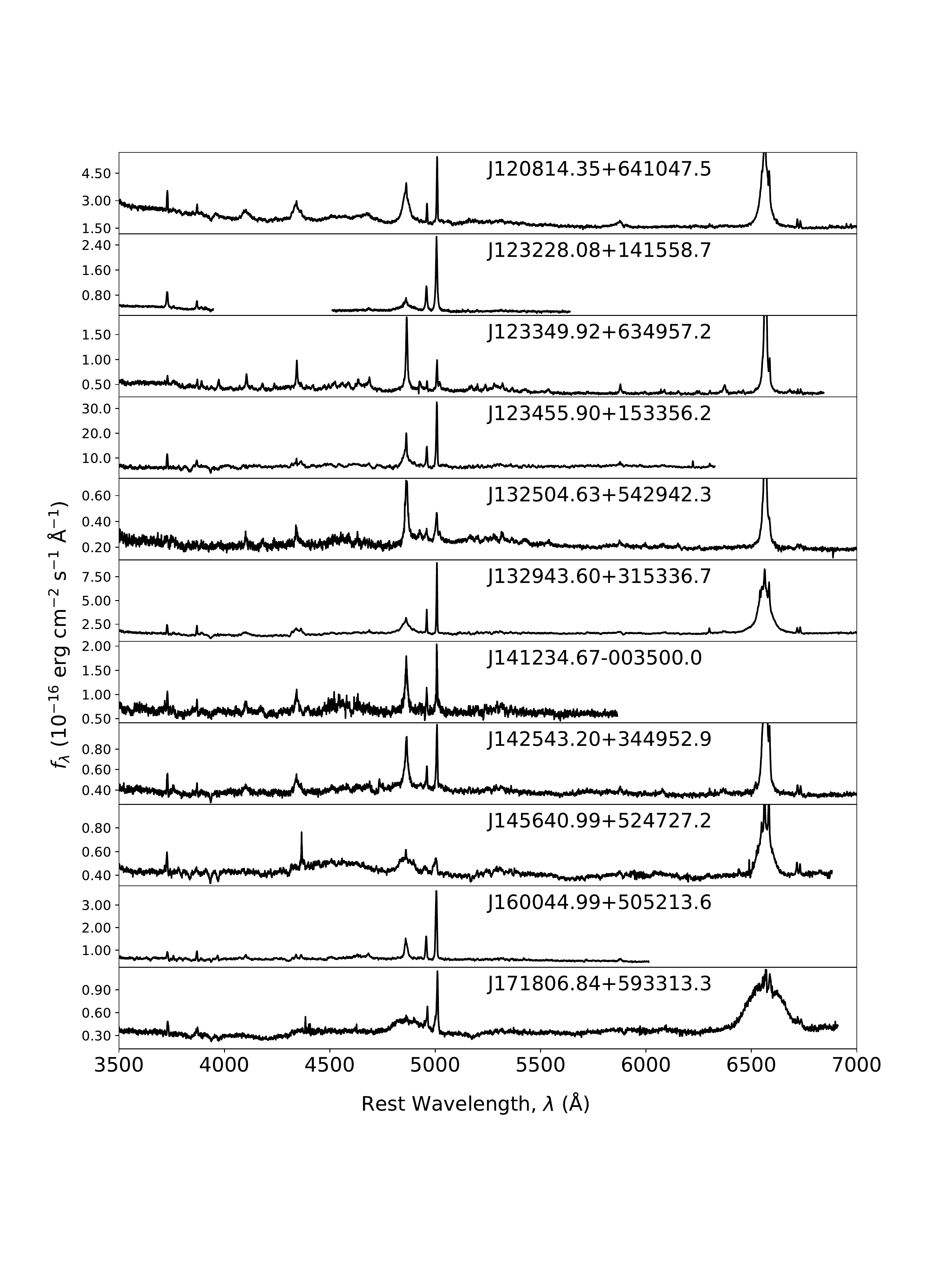}
\caption{\textit{Continued}.}
\end{figure*}
\renewcommand{\thefigure}{\arabic{figure}}

\subsection{Data Reduction} \label{sec:dataredux}

Spectroscopic data reduction was performed using standard techniques with a combination of \textsc{IRAF} \citep{Valdes1984} and {Python} scripts.  Separate reductions were performed for each of the nine LRIS observing configurations in our sample, based on the dichroic, grating, and central wavelength chosen for each object (Table \ref{tab:obs_table}).  After bias subtraction and flat-fielding, cosmic ray removal was performed using \textit{L.A.Cosmic} \citep{lacosmic}.  Any leftover cosmic ray artifacts were manually removed using the \textsc{IRAF} task \textit{imedit}.  Wavelength calibration was performed using Hg, Cd, and Zn arc lamps on the LRIS-B side, while Ne and Ar arc lamps were used on the LRIS-R side.  Sky emission lines were then used to correct for small linear shifts in the wavelength axis due to flexure.  Sky emission lines were subsequently removed by fitting the background with a low-order polynomial.  The two-dimensional spectra were then rectified in the spatial direction by tracing the signal of each object along the wavelength direction.  Flux calibration was performed using spectrophotometric standards from \citet{Massey1} and \citet{Massey2}.  Telluric correction was performed for spectra that exhibited strong contamination from atmospheric absorption.  Objects with two exposures were averaged together using the \textsc{IRAF
} task \textit{imcombine}.\\
\indent The spectra were extracted with an aperture equal to the effective radius $r_\mathrm{eff}$ of the bulge component measured from \textit{HST} imaging using \textsc{GALFIT} (see Section \ref{sec:L5100}).  The LRIS-B and LRIS-R spectra were then combined into a single spectrum. This was done for two reasons: (1) to maximize wavelength coverage to accurately model the AGN power-law continuum, and (2) to accurately model any \ion{Fe}{2} emission between 4400 and 5500 \angstrom\ that may contaminate the H$\beta$/ \ion{Mg}{1b} complex.  To do this, we convolved the higher-resolution side with a Gaussian to the same resolution as its respective lower-resolution side.  We modeled the noise from the lower-resolution spectra to artificially populate the subsequently smoothed higher-resolution side with normally distributed noise of the same standard deviation.  The wavelength axis of the higher-resolution side was then interpolated to the same dispersion as the lower-resolution side so they could be combined.  Finally, the combined spectrum was logarithmically rebinned to constant velocity scale.  For velocity dispersion measurements of the \ion{Mg}{1b} region we used the uncombined LRIS-R spectra, which in cases where the 1200/7500 or the 900/5500 grating was used, have a higher resolution than the LRIS-B grism (see Section \ref{sec:sigma}).  Figure \ref{fig:observations} shows the final extracted and combined rest-frame spectrum of each object of the final sample.  Gaps in spectral coverage between the LRIS-B and LRIS-R sides occur in three spectra, and are caused by the choice of specific LRIS-R configuration used during observations.  

\subsection{HST Archival Data} \label{sec:hst_archive}

Imaging data for each object were obtained via the Hubble Legacy Archive (HLA), which provides enhanced data products that are fully reduced, corrected for artifacts and cosmic rays, drizzled, and combined for all \textit{HST} instruments.  Because our sample includes data from a variety of \textit{HST} instruments, filters, and depths, it is crucial that the data for each object be reduced in a consistent and optimized manner for each instrument.  Furthermore, HLA data products include robust uncertainty estimates for image data, which are necessary for accurate deconvolution of the AGN PSF from the host galaxy using \textsc{GALFIT} (see \S \ref{sec:L5100}).  Details of the \textit{HST} imaging used for each object are given in Table \ref{tab:hst_table}.

\begin{deluxetable*}{ccccccc}
\tablecaption{{\textit{HST}} Archival Data\tablenotemark{a} \label{tab:hst_table}}
\tablehead{
\colhead{Object} & \colhead{Instrument} & \colhead{Camera/Channel} & \colhead{Filter} & \colhead{Spatial Scale} & \colhead{Exposure Time} & \colhead{Proposal ID} \\
\colhead{} & \colhead{} & \colhead{} & \colhead{} & \colhead{(kpc pix$^{-1}$)} & \colhead{(s)} & \colhead{}
}
\colnumbers
\startdata
J000338.94+160220.6   & ACS   & WFC  & F606W & 0.10 & 2084 & 10889 \\
J001340.21+152312.0   & WFC3  & UV   & F475W & 0.09 & 2268 & 12233 \\
J015516.17$-$094555.9 & NIC   & NIC2 & F110W & 0.32 & 5120 & 11208 \\
J040210.90$-$054630.3 & ACS   & WFC  & F606W & 0.21 & 720  & 10588 \\
J073505.66+423545.6   & WFPC2 & PC   & F814W & 0.08 & 1230 & 11130 \\
J092438.88+560746.8   & WFPC2 & PC   & F814W & 0.02 & 1230 & 11130 \\
J093829.38+034826.6   & WFPC2 & PC   & F814W & 0.11 & 1230 & 11130 \\
J095819.87+022903.5   & ACS   & WFC  & F814W & 0.24 & 2028 & 10092 \\
J100234.85+024253.1   & ACS   & WFC  & F814W & 0.16 & 2028 & 10092 \\
J101527.25+625911.5   & ACS   & WFC  & F775W & 0.25 & 2360 & 10216 \\
J113657.68+411318.5   & WFPC2 & PC   & F814W & 0.07 & 1230 & 11130 \\
J120814.35+641047.5   & WFPC2 & PC   & F814W & 0.10 & 600  & 6361  \\
J123228.08+141558.7   & WFPC2 & PC   & F606W & 0.28 & 2700 & 8805  \\
J123349.92+634957.2   & WFPC2 & PC   & F814W & 0.12 & 1230 & 11130 \\
J123455.90+153356.2   & WFPC2 & PC   & F814W & 0.04 & 600  & 6361  \\
J132504.63+542942.3   & WFPC2 & PC   & F814W & 0.13 & 1230 & 11130 \\
J132943.60+315336.7   & WFPC2 & WF   & F814W & 0.17 & 600  & 6361  \\
J141234.67$-$003500.0 & ACS   & WFC  & F814W & 0.11 & 1090 & 10596 \\
J142543.20+344952.9   & WFPC2 & PC   & F814W & 0.15 & 1230 & 11130 \\
J145640.99+524727.2   & ACS   & WFC  & F606W & 0.21 & 720  & 10588 \\
J160044.99+505213.6   & WFPC2 & PC   & F814W & 0.09 & 1230 & 11130 \\
J171806.84+593313.3   & ACS   & WFC  & F814W & 0.21 & 2040 & 9753  \\
\enddata
\tablenotetext{a}{Archival data obtained from Hubble Legacy Archive}
\tablecomments{Summary of \textit{HST} archival data obtained for our sample.  Column 1: object. Column 2: instrument.  Column 3: camera/channel.  Column 4: filter.  Column 5: spatial scale per side pixel.  Column 6: exposure time.  Column 7: proposal ID.}

\end{deluxetable*}


\section{Analysis} \label{sec:analysis}

In the following sections we discuss the necessary measurements required to analyze our sample on the $M_{\rm{BH}}-\sigma_*$ relation.  We first discuss quantities obtained from spectral analysis beginning with stellar velocity dispersion, which include the effects of host-galaxy inclination and \ion{Fe}{2} contamination in our spectra.  We then discuss in detail our multi-component fitting methods, how we measure broad H$\beta$ widths for calculation of BH masses, as well as investigate the use of [\ion{O}{3}]$\lambda 5007$ as a surrogate for $\sigma_*$.  Next we discuss measurements obtained from \textit{HST} imaging, which include surface brightness decomposition and measurements of the AGN luminosity at 5100 \angstrom.  Finally, we derive the equation used to calculate BH masses for our sample.    

\subsection{Stellar Velocity Dispersion} \label{sec:sigma}

Stellar velocity dispersions were measured using the penalized pixel-fitting (\textsc{pPXF}; \citet{ppxf1},\citet{ppxf2}) technique, which convolves a series of stellar templates with a Gauss-Hermite kernel to fit the line-of-sight velocity distribution (LOSVD) of the integrated spectrum of stellar light from galaxies.  
To minimize the possibility of template mismatch, a total of 636 stellar templates with minimal gaps in wavelength coverage were chosen from the Indo-US Library of Coud\'e Feed Stellar Spectra \citep{IndoUS}, which have a FWHM resolution of $\sim$1\angstrom\;and wavelength range between 3465 and 9469 \angstrom.  Additionally, we generated 20 narrow \ion{Fe}{2} templates of widths ranging from $50$ to $1000$ km s$^{-1}$ and 91 broad \ion{Fe}{2} templates of widths ranging from $1100$ to $10,000$ km s$^{-1}$ using the template from \citet{Veron-Cetty2004} to account for possible \ion{Fe}{2} contamination and included them with the stellar templates.  We note that the choice of \ion{Fe}{2} template used to remove \ion{Fe}{2} contamination can result in differences in the quality of the subtraction.  For example, \citet{Barth2013} notes that the \citet{Veron-Cetty2004} \ion{Fe}{2} template better accounts for \ion{Fe}{2} emission by modeling emission lines with Lorentzian profiles and includes only \ion{Fe}{2} emission features that are commonly found in Seyfert 1 galaxies, as opposed to other \ion{Fe}{2} templates which specifically model the \ion{Fe}{2} emission of I Zw 1.  We find that the inclusion of low-order additive and multiplicative polynomials has no significant effect on our fits; this can be attributed to the inclusion of broad \ion{Fe}{2} templates which can account for broad variations in the stellar continuum.  We fit the entire \ion{Mg}{1b}/\ion{Fe}{2} region from 5025 to 5800 \angstrom when possible, or as much of this region as our wavelength coverage allows.\\
\indent The algorithm utilizes a penalty function, controlled by a user-input \textit{bias} parameter, which acts to bias the fit toward a Gaussian LOSVD.  Monte Carlo simulations were performed to determine the behavior of the penalty function and determine the maximum \textit{bias} parameter at values of $\sigma_*>3\times(\mathrm{velocity~scale})$ for which the difference between output and input parameters was within the scatter of the simulation.  For all of our objects, the optimal \textit{bias} value was determined to be $\sim0.1$.\\
\indent Figure \ref{fig:ppxf} shows the best-fit \textsc{pPXF} solution for each object in our sample.  The best-fit values of the stellar velocity dispersion for each object are reported in Table \ref{tab:mbh_table}.  Uncertainties are determined using Monte Carlo methods by generating 1000 mock spectra using the noise-added best-fit model and re-fitting using \textsc{pPXF}.  \\

\subsection{Effect of Host Galaxy Inclination on Stellar Velocity Dispersion Measurements}
\label{sec:inclination_corrections}

\indent Given that 15 out of the 22 objects in our sample contain a visible disk morphology in \textit{HST} imaging (see Section \ref{sec:galfit}), we must consider the possible bias in our measurements of $\sigma_*$ due to disk contamination.  Kinematically ``cold" disk components can contaminate bulge dispersion measurements and act to increase the measured value of $\sigma_*$, especially at intermediate to edge-on inclinations. \citet{Hartmann2014} found that $\sigma_*$ can be biased by as much as 25\% for edge-on systems, and \citet{Bellovary2014} found that considerable scatter in the $M_{\rm{BH}}-\sigma_*$ relation can be explained by $\sigma_*$ measurements that do not account for disk inclination.  In our sample, we observe that objects that host disk morphologies have systematically higher values of $\sigma_*$ on the $M_{\rm{BH}}-\sigma_*$ relation than objects with no visible disk morphology.  Disk inclinations were measured using \textsc{GALFIT} surface brightness decomposition of \textit{HST} imaging (see Section \ref{sec:galfit}) and using the relation between disk axis ratio $(b/a)$ and inclination $i$ from \citet{Pizagno2007}, which takes into account a disk of finite thickness \citep{Haynes1984}.  \citet{Bellovary2014} used cosmological $N$-body simulations of disk galaxies to estimate the effect of inclination on measurements of $\sigma_*$ in bulges that grow naturally over time without making any assumptions on their kinematics, providing an equation to correct for inclination effects as a function of disk rotational velocity $v_{\rm{rot}}$ and bulge anisotropy $(v/\sigma)_{\rm{spec}}$.  Disk luminosities from \textsc{GALFIT} (see Table \ref{tab:galfit_table}) were corrected for Galactic extinction, as well as intrinsic extinction estimated from measurements of Balmer emission line ratios.  We also applied $k$-corrections and filter transformations from each \textit{HST} filter to SDSS-$r$ using \textit{pysynphot} \citep{pysynphot}.  Finally, we corrected for passive evolution using the online passive evolution calculator from \citet{vanDokkum2001} by assuming a single stellar population formed at $z>>1$.  We infer rotational velocities $v_{\rm{rot}}$ that are typical of disk luminosities in our sample, and assume an anisotropy parameter of $(v/\sigma)_{\rm{spec}}=0.6$ for a fast-rotating late-type galaxy \citep{FalconBorroso2017}.  Finally, we obtain a correction for $\sigma_*$ as a function of $i$ using the prescription from \citet{Bellovary2014} with an adopted uncertainty of 10\% for this correction.  We find that varying the parameters $v_{\rm{rot}}$ and $(v/\sigma)_{\rm{spec}}$ do not considerably change the magnitude of the correction for $\sigma_*$.  After correcting for inclination, the affected velocity dispersions decrease by 10\% on average, but do not significantly change the scatter of our sample on the $M_{\rm{BH}}-\sigma_*$ relation.  If we did not correct for inclination, the majority of our sample would reside below the relation. Stellar velocity measurements for objects with disk morphologies listed in Table \ref{tab:mbh_table} have been corrected for the effects of inclination.

\subsection{Effect of \ion{Fe}{2} Contamination on Stellar Velocity Dispersion Measurements} \label{sec:feii}

Contamination from \ion{Fe}{2} emission is present to some degree in all objects in our sample.  This is especially apparent for objects J$123349$ and J$132504$, which both exhibit strong narrow \ion{Fe}{2} emission between 4400 and 5500 \angstrom\ (see Figure \ref{fig:observations}).  In such cases, \textsc{pPXF} is prone to mistaking narrow \ion{Fe}{2} emission for variations in a stellar continuum at some different systemic velocity than real stellar absorption features and can lead to an overestimate of the stellar velocity dispersion.  We find that, while broad \ion{Fe}{2} emission can easily be subtracted off prior to stellar template fitting without affecting the fit, narrow \ion{Fe}{2} emission can make determination of its relative contribution to the host galaxy nearly impossible.  To accurately determine the relative contribution of \ion{Fe}{2} emission and its effects on our measurements of the LOSVD, we use \textsc{pPXF} to fit \ion{Fe}{2} and stellar templates simultaneously.  We find that if the total (broad + narrow) \ion{Fe}{2} fraction of the total flux within the \ion{Mg}{1b}/\ion{Fe}{2} region exceeds $\sim$5\%, the stellar velocity dispersion can be overestimated by as much as 50-90\%, due mainly to the presence of strong narrow \ion{Fe}{2}.  For our sample, the average uncertainty due to the presence of broad and narrow \ion{Fe}{2} emission is $\sim$8\%.  We further discuss the possible biases in our stellar velocity dispersion measurements due to \ion{Fe}{2} contamination in Section \ref{sec:sigma_uncertainties}.
 
\begin{figure*}
\center
\includegraphics[trim={2.cm 7.5cm 2.5cm 2.5cm},scale=0.65,clip]{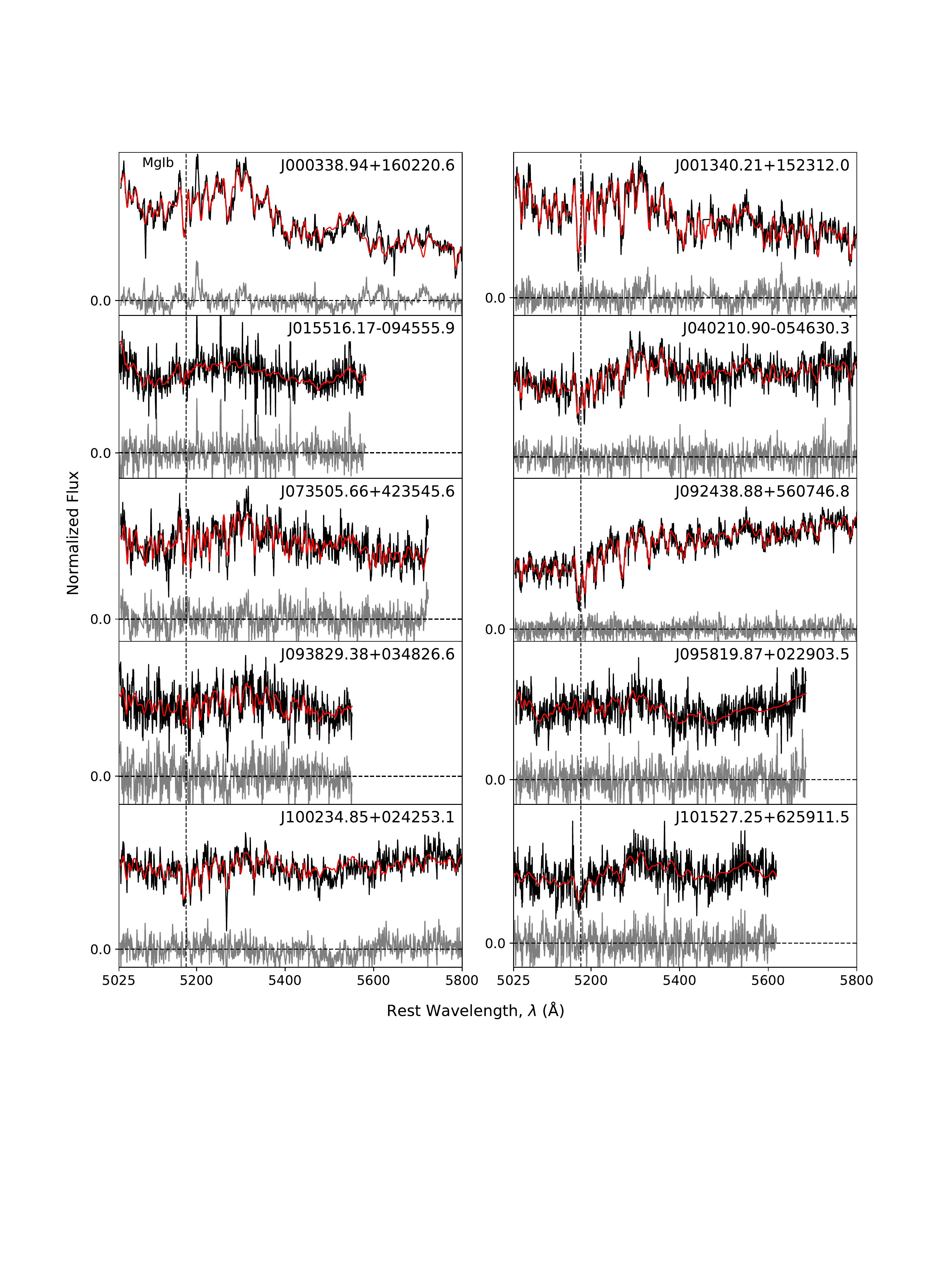}
\caption{
Velocity dispersion measurements of the \ion{Mg}{1b}/\ion{Fe}{2} region spanning from 5025 to 5800 \angstrom\ (coverage permitting).  Each spectrum (black) are median normalized to 1 for fitting purposes and overplot with the best-fit (red) to the line-of-sight velocity distribution. Residuals are shown below each fit in gray.  
\label{fig:ppxf}}
\end{figure*}


\renewcommand{\thefigure}{\arabic{figure}}
\addtocounter{figure}{-1}
\begin{figure*}
\center
\includegraphics[trim={2cm 2.5cm 2.5cm 2.5cm},scale=0.65,clip]{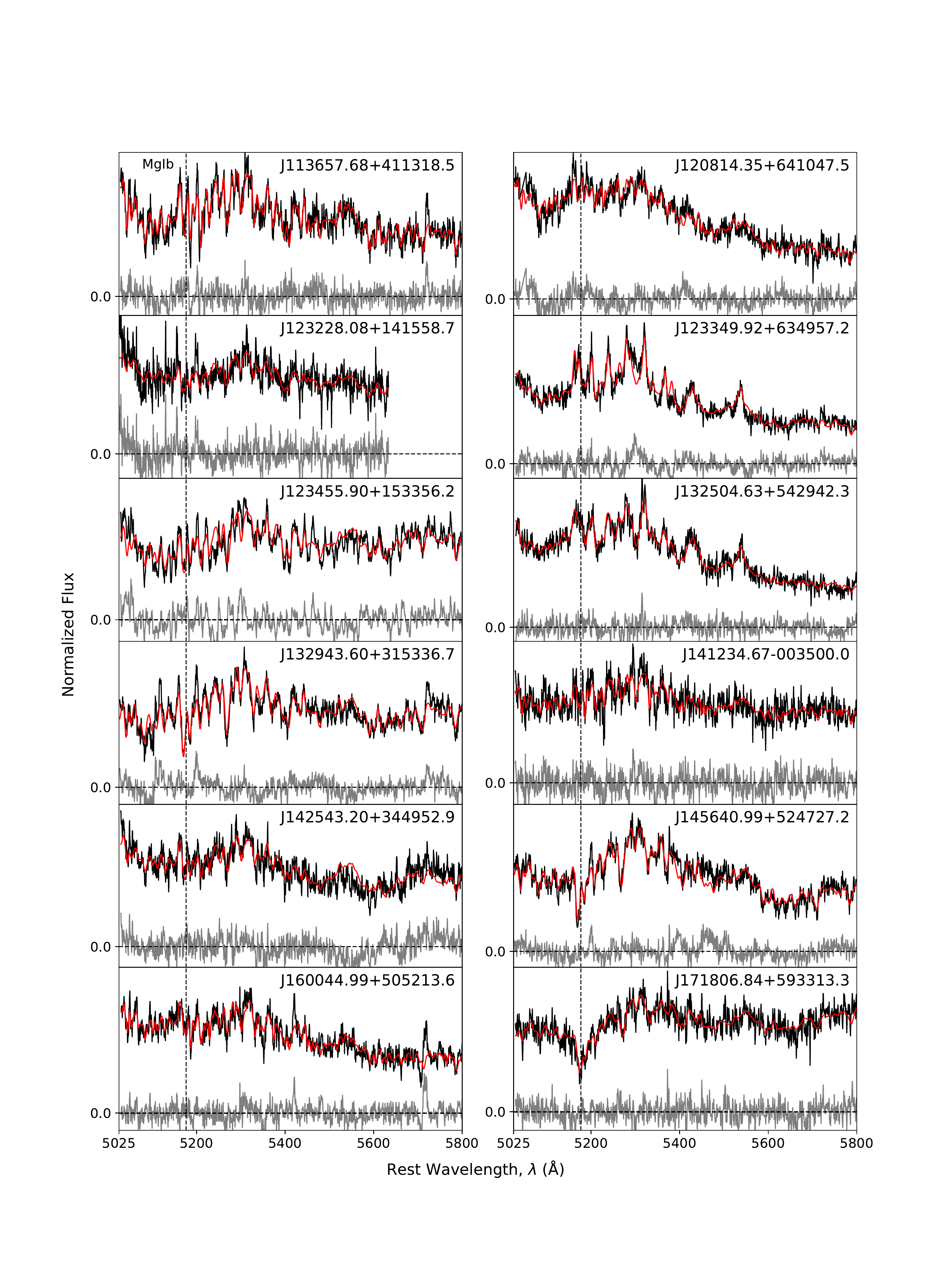}
\caption{\textit{Continued}.}
\end{figure*}
\renewcommand{\thefigure}{\arabic{figure}}
 
\subsection{Multi-Component Spectral Fitting} \label{sec:emline}

The variable and complex nature of optical AGN spectra necessitate the use of simultaneous multi-component fitting to accurately constrain the relative contributions of each of the spectral components present.  As with velocity dispersion measurements, the contribution from broad and narrow \ion{Fe}{2} emission can further affect measurements of broad H$\beta$ and [\ion{O}{3}] emission features.  Broad \ion{Fe}{2} emission between H$\beta$ and [\ion{O}{3}]$\lambda4959$ can cause H$\beta$ to appear more broad and asymmetric if unaccounted for.  Similarly, narrow \ion{Fe}{2} can be present on either side of [\ion{O}{3}]$\lambda5007$ and complicate width measurements.  In addition to \ion{Fe}{2} emission, stellar absorption from the host galaxy can cause significant asymmetries in the line profile of broad H$\beta$.  Finally, the relative strength of the AGN continuum can dilute the strength of stellar continuum, and therefore must be accounted for \citep{Greene2005}.\\
\indent To perform simultaneous fitting of all components, the fitting region is chosen to span from rest-frame $4400-5800$\;\angstrom, large enough such that the relative contribution from \ion{Fe}{2} and stellar emission can be adequately constrained from both sides of the H$\beta$/[\ion{O}{3}] region.  The stellar continuum across the fitting region is modeled using the same 636 stellar templates used to measure stellar velocity dispersion; however, we constrain the LOSVD solution to that found in the previous step (see Section \ref{sec:sigma}) and allow \textsc{pPXF} to determine the best-fit stellar templates to match the spectrum.  The \ion{Fe}{2} component is modeled using the broad and narrow template from \citet{Veron-Cetty2004}.  Each \ion{Fe}{2} template is parameterized by an amplitude, width, and velocity offset, all of which are free parameters during the fitting process.  The AGN continuum is modeled using a simple power law with an amplitude and power-law index as free parameters.  The amplitude is constrained to be positive and the power-law slope is constrained to the range $[-4,2]$.  Finally, broad and narrow H$\beta$ and [\ion{O}{3}]$\lambda\lambda4959,5007$ emission features are fit.  The amplitude ratio of the [\ion{O}{3}]$\lambda\lambda4959,5007$ lines were held at a 1:3 constant ratio as per theoretical calculations and empirical observations \citep{Dimitrijevic2007}, while the amplitude of the narrow H$\beta$ line was left as a free parameter.  The widths of narrow H$\beta$ and [\ion{O}{3}]$\lambda4959$ were tied to the width of [\ion{O}{3}]$\lambda5007$.  The velocity offsets of the [\ion{O}{3}] lines were tied, but the narrow H$\beta$ velocity offset was left as a free parameter.  Velocity offsets are measured with respect to best-fit redshift determined from the fit to stellar absorption features described in Section \ref{sec:sigma}.  Blueshifted wing components are included in the fits to narrow emission lines, and are constrained to have a width greater than their narrow core counterpart.  If the fitting algorithm cannot adequately fit a blue-wing component with the narrower core emission line, or if the resulting core component has a width less than than the intrinsic FWHM resolution of the instrument configuration, blue-wing components are removed from the model. \\
\indent All components of the model are fit simultaneously using a custom Bayesian maximum-likelihood algorithm implemented in Python, with uncertainties estimated via Markov Chain Monte Carlo (MCMC) using the affine invariant MCMC ensemble sampler \textit{emcee} \citep{emcee}.  The small size of our sample allows us to initialize parameters on an individual object basis to ensure accurate modeling of all components.  First, an initial model is constructed for the emission lines, \ion{Fe}{2} templates, and power-law continuum using reasonable starting values, and are subsequently subtracted off from the original data.  The remaining flux, which is assumed to contain a non-negligible fraction of stellar continuum, is then fit with \textsc{pPXF} \citep{ppxf1,ppxf2} to obtain the best-fit stellar templates.  Initial conditions for each parameter are determined using a least-squares numerical optimization routine which maximizes the likelihood function given by
\begin{equation}
\chi^2 =\sum_{i=1}^N\frac{(y_{{\rm{data}},i}-y_{{\rm{model}},i})^2}{\sigma_i^2},  
\end{equation}
where $\sigma_i$ is the $1\sigma$ uncertainty for each datum $y_{{\rm{data}},i}$, and $y_{{\rm{model}},i}$ is the value of the model at each datum.  Upper and lower limits on parameters, for example minimum and maximum broad-line widths, are also chosen to serve as priors to constrain fitting parameters.  Once adequate initial values and bounds have been determined, \textit{emcee} is used to sample the parameter space of each parameter to determine their posterior distributions, from which the best-fit values and uncertainties are calculated.  The number of MCMC iterations performed is ultimately determined by how well individual model components are initially fit, with the most degenerate components requiring longer runtimes.  Each object is fit with a minimum of 2500 iterations, but each parameter generally converges on a solution in less than 1000 iterations. \\
\indent We find that the use of an MCMC algorithm is advantageous over simpler least-squares methods since the high number of free parameters can lead to numerous degeneracies in parameter solutions thus requiring the algorithm to exhaustively explore each parameter space. Our MCMC implementation allows one to visualize how individual parameters approach or diverge from a solution, or if degeneracies exist.  The most common degeneracy observed during the fitting process is that of the width of broad \ion{Fe}{2}, which is due to overlapping broad \ion{Fe}{2} features; however, we have found that broad \ion{Fe}{2} does not strongly affect stellar velocity dispersion measurements because the features are too broad to mimic narrower stellar absorption features.  In general, most degeneracies resolve themselves after a sufficient number of MCMC iterations, usually after higher signal-to-noise ratio (S/N) features, such as emission lines, have converged on a stable solution, allowing less constrained features, such as \ion{Fe}{2} or stellar emission, to subsequently converge on their respective solutions.  Large degeneracies, if present, emerge in the posterior distributions of each affected parameter, and are reflected in our uncertainties. \\
\indent Figure \ref{fig:emline} shows the best-fit model, individual component models, and residuals for each object using our multi-component fitting method.  For one object, J145640, the [\ion{O}{3}] complex appears to be significantly attenuated, and we therefore mask the [\ion{O}{3}] complex during the fitting process.  See the Appendix \ref{sec:indiv_objects} for further discussion on the spectrum and fitting of object J145640. \\

\subsubsection{Measuring Broad H$\beta$ FWHM} \label{sec:hbeta}
\indent A number of objects in our sample exhibit asymmetric broad H$\beta$ emission lines.  Ordinarily, such an asymmetric profile would require multiple Gaussian components or fitting the dispersion directly from the line profile.  However, the inclusion of the stellar continuum and \ion{Fe}{2} emission, and modeling the line with a single Gaussian, fully accounts for any asymmetries or non-Gaussian shape in the line profile.  Prominent examples of this asymmetry are shown in the spectra of J$000338$, J$100234$, J$145640$, and J$171806$.  We find that the multi-component fitting technique is consistent with techniques that do not fit the stellar continuum or \ion{Fe}{2} emission simultaneously.  We find that uncertainties in H$\beta$ line widths decrease by a factor of 2.3 on average for our sample compared to line widths measured conventionally where multi-component fitting is not implemented.  This is likely due to the requirement of a more complex line models (two or more Gaussian components) needed to fully account for the asymmetric broad H$\beta$ profile.  Uncertainties in the fit for broad H$\beta$ widths in our sample range from 1\% to 6\%, and depend largely on the S/N of the spectrum and how well other components of the model are constrained.\\
\indent Variability of the line profile of H$\beta$ can also contribute to the random uncertainties of single-epoch width measurements.  \citet{Woo2007} found a 7\% rms scatter when comparing Lick rms H$\beta$ FWHM measurements to Keck single-epoch measurements, which we adopt in our random uncertainties for measured H$\beta$ FWHM.  The total random uncertainty for our H$\beta$ FWHM measurements is $\sim$8\%.


\begin{figure*}[pht!]
\center
\includegraphics[trim={1cm 1.0cm 2cm 2.5cm},scale=0.95,clip]{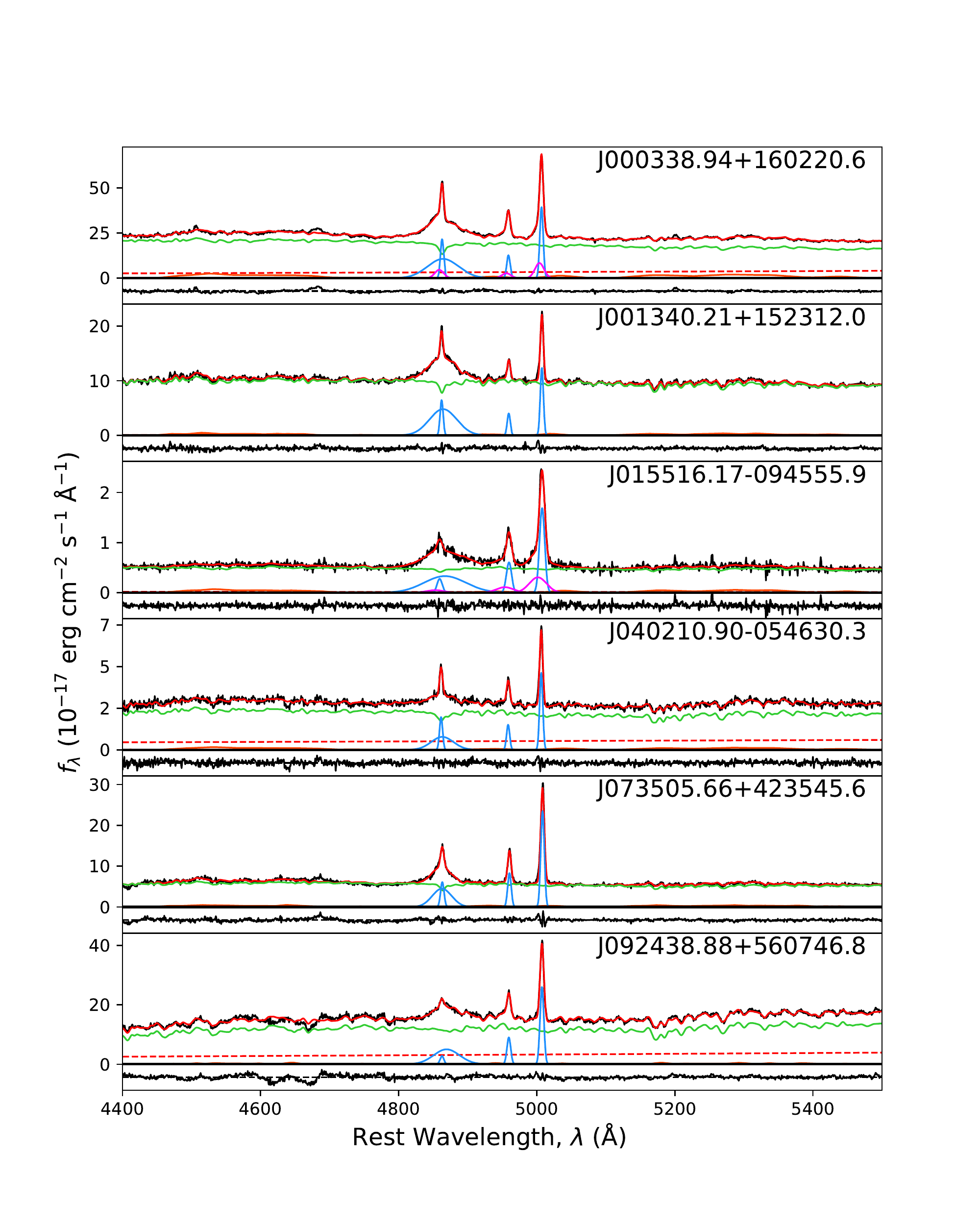}
\caption{ 
Multi-component fitting of the the H$\beta$/[\ion{O}{3}] region.  The reduced Keck/LRIS spectra (black) are overplotted with the total best-fit model (red), comprised of the stellar continuum (green), the AGN power-law continuum (dashed red), broad and narrow \ion{Fe}{2} emission (orange), broad and narrow emission lines (blue) and their corresponding blue-wing outflow components (magenta) if present.  Residuals are shown below each spectrum. 
\label{fig:emline}}
\end{figure*}


\renewcommand{\thefigure}{\arabic{figure}}
\addtocounter{figure}{-1}
\begin{figure*}[pht!]
\center
\includegraphics[trim={1cm 1.cm 2cm 2.5cm},scale=0.95,clip]{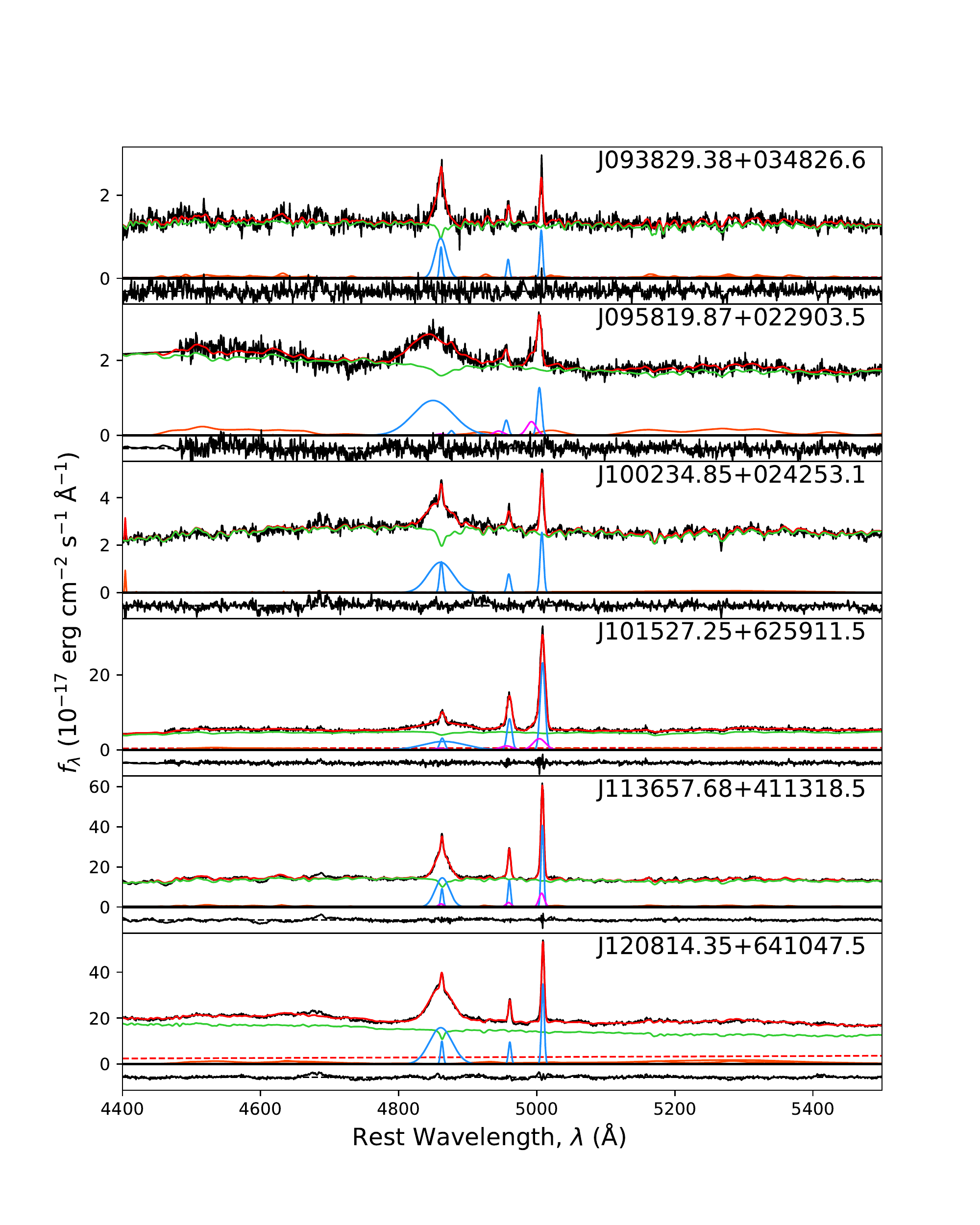}
\caption{\textit{Continued}.}
\end{figure*}
\renewcommand{\thefigure}{\arabic{figure}}

\renewcommand{\thefigure}{\arabic{figure}}
\addtocounter{figure}{-1}
\begin{figure*}[pht!]
\center
\includegraphics[trim={1cm 1.cm 2cm 2.5cm},scale=0.95,clip]{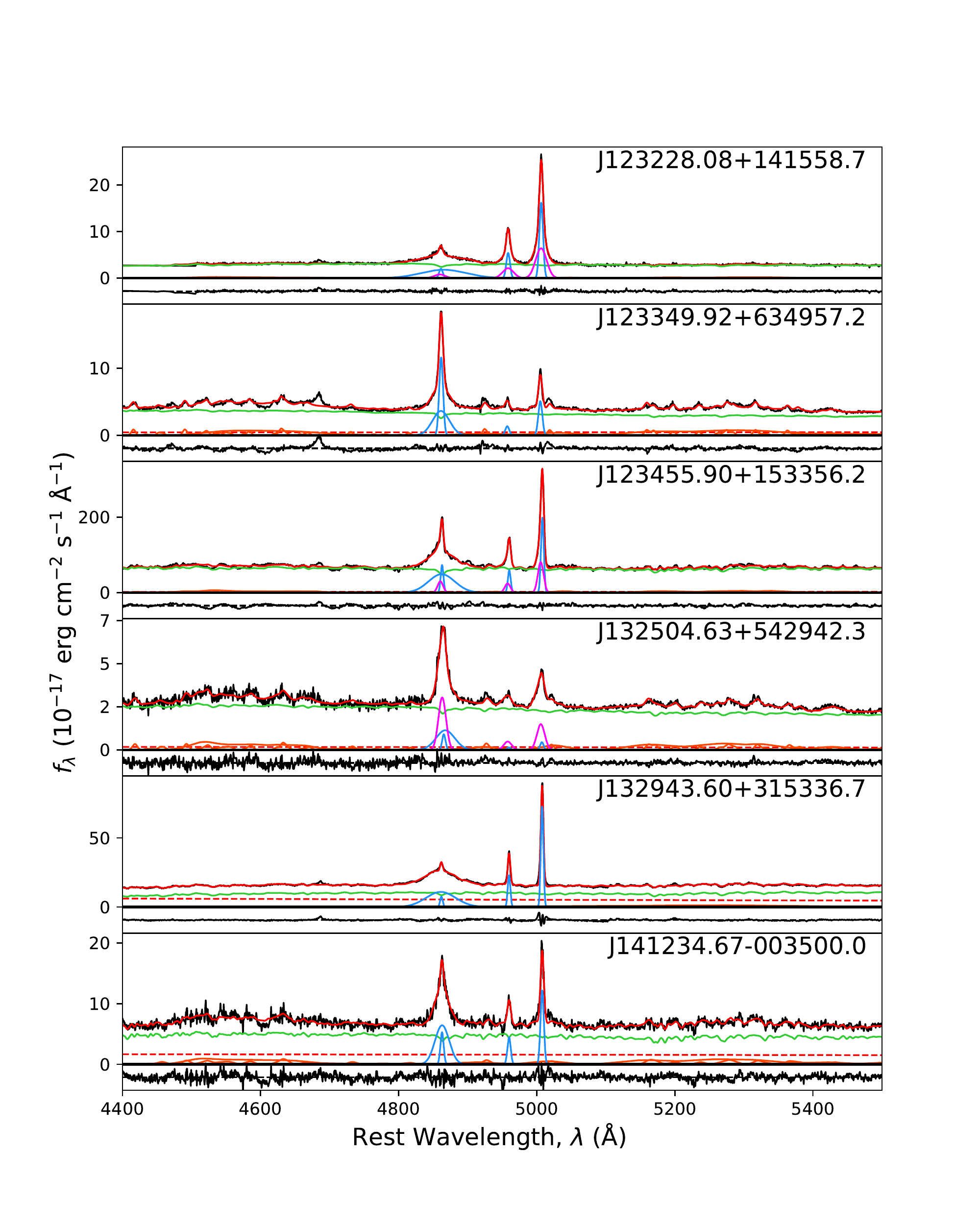}
\caption{\textit{Continued}.}
\end{figure*}
\renewcommand{\thefigure}{\arabic{figure}}

\renewcommand{\thefigure}{\arabic{figure}}
\addtocounter{figure}{-1}
\begin{figure*}[pht!]
\center
\includegraphics[trim={1cm 8.cm 2cm 2.5cm},scale=0.95,clip]{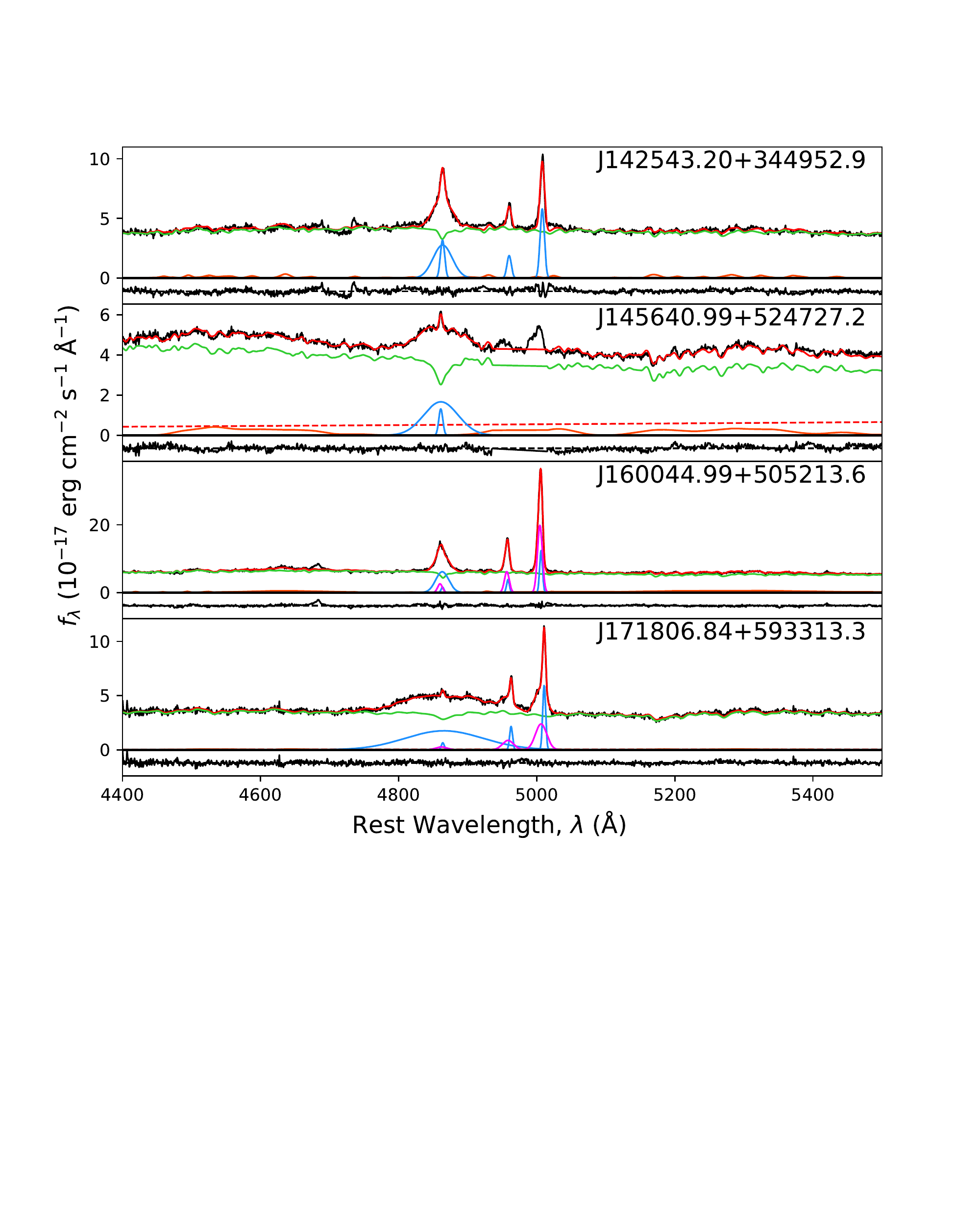}
\caption{\textit{Continued}.}
\end{figure*}
\renewcommand{\thefigure}{\arabic{figure}}



\subsubsection{[\ion{O}{3}]$\lambda5007$ as a Surrogate for Stellar Velocity Dispersion} \label{sec:oiii}

\begin{figure*}[ht!]
\center
\includegraphics[trim={0cm 0cm 0cm 0cm},scale=0.70,clip]{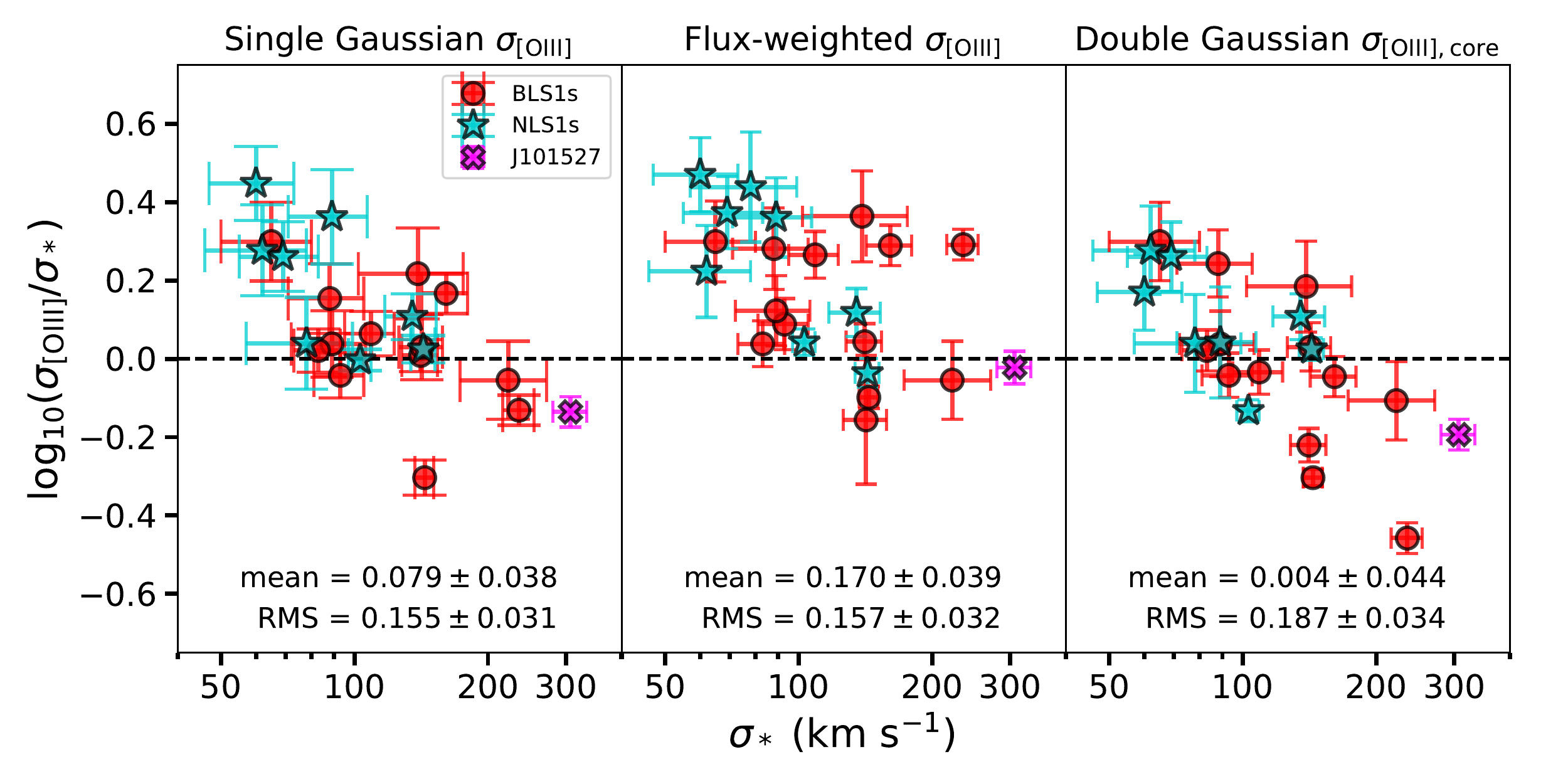}
\caption{
  Comparisons of different models for $\sigma_{\rm{[O\ III]}}$ vs. $\sigma_*$.  {Left}: single-gaussian model from which $\sigma_{\rm{[O\ III]}}$ is measured from the FWHM.  The offset mean is caused by the presence of asymmetric blue wings which act to increase the width of the line profile.  {Center}:  the flux-weighted dispersion, which is more sensitive to the presence of blue-wing profiles than the single-gaussian model.  {Right}: double-gaussian decomposition of the line profile, which results in the best agreement with $\sigma_*$ on average.
\label{fig:oiii}}
\end{figure*}

Measurements of stellar velocity dispersion for Type 1 AGNs at $z>0.4$ are often complicated by the large light fraction from the AGN coupled with surface brightness dimming of the host galaxy, resulting in stellar absorption features that are difficult or impossible to measure.  Previous studies have suggested that the widths of strong narrow-line region (NLR) emission lines, such as [\ion{O}{3}]$\lambda5007$, may be suitable surrogates for the stellar velocity dispersion if the NLR velocity field is strongly coupled with the gravitational potential of the bulge \citep{Nelson1996}.  However, non-gravitational kinematic components in ionized-gas emission can be present, manifested as a broad and blueshifted wing component indicative of possible gas outflows \citep{Heckman1981,Nelson1996}.  Non-gravitational kinematics can also manifest themselves as a blueshift of the entire [\ion{O}{3}] line profile, which comes with a dramatic line profile broadening \citep{Komossa2008a,Komossa2018}, again likely indicating strong outflows.  Studies with large surveys such as the SDSS show considerable scatter in a linear relation between $\sigma_\mathrm{[O\ III]}$ and $\sigma_*$, even after blue-wing outflow components have been removed \citep{Boroson2003,Greene2005}.  However, the scatter decreases significantly after removing sources which have their whole [\ion{O}{3}] line profile blueshifted (so-called ``blue outliers''; Figure 1 of \citet{Komossa2007}).  More recently, \citet{Woo2016} investigated [\ion{O}{3}]$\lambda5007$ kinematics in a sample of $\sim$39,000 Type 2 AGNs at $z<0.3$, accounting for outflows in $\sim$44\% of their sample.  In addition to confirming a broad correlation between $\sigma_\mathrm{[O\ III]}$ and $\sigma_*$, they found that objects with non-gravitational outflow components do not follow a linear correlation, and instead have higher $\sigma_\mathrm{[O\ III]}/\sigma_*$ ratios for higher $\sigma_*$.  In a subsequent study, \citet{Rakshit2018} found similar results to \citet{Woo2016} for $\sim5000$ Type 1 AGNs.  \citet{Bennert2018} also performed a comprehensive analysis on the use of [\ion{O}{3}] as a surrogate for $\sigma_*$ on the $M_{\rm{BH}}-\sigma_*$ relation, finding that there is good statistical agreement between relations plotted with $\sigma_{\rm{[O\ III]}}$ versus $\sigma_*$, but only after blueshifted wing components are removed.  \\
\indent Higher-resolution spectra allow us the opportunity to revisit the significance of any correlation between $\sigma_{\rm{[O\ III]}}$ and $\sigma_*$, as well as investigate the influence, and possible bias, outflow components may introduce.  In addition to fitting for outflow kinematics in [\ion{O}{3}], we attempt to fit for any broad or narrow \ion{Fe}{2} contamination within the H$\beta$ region which may bias measurements of [\ion{O}{3}] to higher widths.\\
\indent Out of the 22 objects in our sample, 10 objects exhibit line-profile asymmetry in [\ion{O}{3}] consistent with a blueshifted wing component.  Following \citet{Woo2006}, we compare the [\ion{O}{3}]$\lambda5007$ dispersion $\sigma_{\rm{[O\ III]}}$ as a function of stellar velocity dispersion $\sigma_*$ using three methods: (1) fitting a single-Gaussian model, (2) measuring the flux-weighted dispersion of the full line profile, and (3) fitting a double-Gaussian model.  The flux-weighted dispersion is calculated using the same method as \citet{Woo2016}, which calculated the second-order Gaussian moment of the sum of the full (core+blue wing) best-fit model to [\ion{O}{3}]$\lambda5007$.  The double-Gaussian model is a decomposition of the broader blue-wing component from the narrower core component, and the core component is chosen as the proxy for $\sigma_*$.  The single-Gaussian fit results in slight disagreement with $\sigma_*$, with a mean of $0.079\pm0.038$ and RMS of $0.155\pm0.031$ in $\log_{10}(\sigma_{\rm{[O\ III]}}/\sigma_*)$.  The flux-weighted measurements result in worse agreement with a mean of $0.17\pm0.04$ and comparable RMS.  Flux-weighted measurements produce, on average, higher widths than the single-Gaussian model, due to the inclusion of flux from the blue-wing component.  The best agreement with $\sigma_*$ resulted from the double-Gaussian decomposition of the [\ion{O}{3}] line profile, with a mean of $0.004\pm0.044$ and an RMS of $0.187\pm0.034$.  Despite the extra consideration in taking into account the stellar and \ion{Fe}{2} components, the RMS scatter is consistent with respect to $\sigma_*$ for all three fitting methods.  In the best case we find that a double-Gaussian decomposition of the [\ion{O}{3}] line profile results in a $\sim$30\% difference with respect to $\sigma_*$ on average for our sample.  Despite its limited size, our sample covers a wide range in $\sigma_*$, and we find good agreement with \citet{Bennert2018} that there is good statistical agreement on average when using [\ion{O}{3}] as a surrogate for $\sigma_*$, provided that blueshifted wing components are removed and \ion{Fe}{2} contamination is accounted for.  \citet{Komossa2007} traced back the remaining offsets in NLS1s to the effect of [\ion{O}{3}] blue outliers in those NLS1s. Once removed, $\sigma_{\rm{[O\ III]}}$ and $\sigma_*$ showed similar scatter.  Of the three deviating NLS1s in our sample (rightmost panel of Figure \ref{fig:oiii}), only one shows a significant kinematic shift in [\ion{O}{3}] with respect to stellar absorption features.  We did not find any other trends with blue outliers in our sample which can further reduce the scatter.  We therefore caution the use of the [\ion{O}{3}] line as a reliable  surrogate for $\sigma_*$, and agree with \citet{Bennert2018} in that it should only be used in a statistical - and not individual - proxy for $\sigma_*$ on the $M_{\rm{BH}}-\sigma_*$ relation.


\subsection{Surface Brightness Decomposition}\label{sec:galfit}

To obtain a robust measure of the AGN luminosity, archival \textit{HST} imaging was used for accurate deconvolution of the AGN PSF uncontaminated by the host galaxy.  To do this, we used the two-dimensional surface brightness fitting algorithm \textsc{GALFIT} \citep{galfit}, which convolves a given PSF with an analytical model (e.g., disk, exponential, S\'ersic, etc.) to estimate model parameters, such as flux and effective radius, of the surface brightness profile of a galaxy.\\
\indent Accurate deconvolution of galaxy surface brightness components requires a PSF that closely matches the signal response particular to each image.  Additionally, each image undergoes numerous transformations during the data reduction process or suffers from age-dependent peculiarities (such as degrading charge transfer efficiency).  We determine that an empirical PSF is the best suited to match each image.  Ideally, the empirical PSF would be obtained from a stellar PSF from the same image data as each galaxy; however, in some cases where the galaxy was imaged with the WFPC2/PC instrument, stellar PSFs were not available.  In these cases, we obtain stellar PSFs from an image of the same instrument, camera, filter, exposure time, and observation date. We use \textit{sewpy}, a {Python} wrapper for \textsc{SExtractor} \citep{sextractor} to identify stellar sources within each {HLA} image.  The brightest of these sources are examined by eye to insure each extraction is free of background contamination or saturation, and then stacked to obtain an average empirical PSF of the image.  The \text{HLA} pipeline also provides a separate image of the $1\sigma$ uncertainty for each science image, which is needed as input for \textsc{GALFIT}.  Segmentation masks are also created using \textsc{SExtractor} to mask contaminating objects (other galaxies or stars) and fed into \textsc{GALFIT}.  Segmentation maps allow us to maximize the size of the usable image for \textsc{GALFIT} to accurately fit the background.\\ 
\indent An iterative process was used to determine the number of models used to decompose each object.  Each object was initially given a PSF component to model the AGN contribution, a single S\'ersic model \citep{Sersic1963} for the host galaxy, and a background sky component.  Residuals were then examined to determine if an additional S\'ersic component was necessary, such as in the case of a disk component.  Initially, we allow the S\'ersic index for the host galaxy components to be a free parameter if it converges on a S\'ersic index of $2<n<6$ for a bulge or $0.9<n<2$ for a disk component \citep{Fisher2008,Gadotti2009}, and reinforce these using soft constraints.  If \textsc{GALFIT} does not freely converge on a reasonable S\'ersic index consistent with a bulge component, the object is refit with the S\'ersic index held constant to a value of $n=4$.  This behavior occurs when \textsc{GALFIT} cannot reconcile contaminating sky or neighboring flux with the extended profiles of high S\'ersic index models.  For the majority of cases in our sample, a free S\'ersic index reaches the upper boundary of the S\'ersic index constraint, which is resolved by holding the S\'ersic index constant and/or including additional components.  Residuals are visually inspected and additional components are added when necessary.  S\'ersic components that do not satisfy the aforementioned definitions of a bulge or disk are designated as ``other''.\\
\indent The results of the surface brightness decomposition for each object are listed in Table \ref{tab:galfit_table}.  Reported magnitudes and surface brightness values from \textsc{GALFIT} are corrected for Galactic extinction, intrinsic host galaxy extinction using the Balmer decrement, and $k$-corrected.  The uncertainties output by \textsc{GALFIT} unrealistically assume that any residual flux in the image is due purely to Poisson noise, and does not take into account deviations from the S\'ersic model which may be due to spiral arms, dust lanes, star-formation regions, or neighboring flux.  As a result, uncertainties quoted by \textsc{GALFIT} in magnitude measurements are generally low, $\sim0.02$ mag on average.  Masking was used to mediate any possible contaminating flux near our objects.  In general, we find that higher surface brightness components, such as the PSF and bulge, have lower quoted uncertainties than lower surface brightness components, such as disks.  With the exception of disturbed systems in our sample, the residuals of the surface brightness decompositions shown in Figure \ref{fig:galfit} would indicate fluctuations in the residuals are on the order of $\sim0.1$ mag, which we include in our uncertainties.  Mismatch between the empirical PSF and the intrinsic PSF of the image can be another significant source of uncertainty of our measurements. Following \citet{Canalizo2012}, we performed direct subtraction of the PSF to determine the upper and lower bounds of the residual flux and found that the average uncertainty in PSF mismatch to be $\sim$0.1 mag, in agreement with Canalizo et al.

\startlongtable
\begin{deluxetable*}{cccllccc}
\tablecaption{Results from surface brightness decomposition using \textsc{GALFIT} \label{tab:galfit_table}}
\tablehead{
\colhead{Object} & \colhead{Filter} & \colhead{Component} & \colhead{$m_\mathrm{ST}$} & \colhead{$\mu_\mathrm{eff}$} & \colhead{$r_\mathrm{eff}$} & \colhead{$r_\mathrm{eff}$} & \colhead{$n$} \\
\colhead{} & &  \colhead{} & \colhead{(mag)} & \colhead{($\frac{\mathrm{mag}}{\mathrm{arcsec}^{2}}$)} & \colhead{($''$)} & \colhead{(kpc)} & \colhead{}
}
\colnumbers
\startdata
J000338.94+160220.6 & F606W & PSF     & 19.38 & 	   & & & \\
                            & & Bulge & 17.47 & 21.60 & 1.63 $\pm$ 0.02 & 3.39 $\pm$ 0.04 & 5.93 \\
                            & & Other & 17.95 & 17.94 & 8.51 $\pm$ 0.23 & 17.69 $\pm$ 0.47 & 2.54 \\
\hline
J001340.21+152312.0 & F475W & PSF     & 19.79 & 	  & & & \\
                            & & Bulge & 17.89 & 23.45 & 2.60 $\pm$ 0.06 & 5.55 $\pm$ 0.13 & 5.31 \\
\hline
J015516.17-094555.9 & F110W & PSF     & 24.36 & 	  & & & \\
                            & & Bulge & 21.72 & 22.05 & 0.31 $\pm$ 0.01 & 1.99 $\pm$ 0.03 & 4 (fixed) \\
                            & & Disk  & 22.37 & 24.57 & 1.03 $\pm$ 0.02 & 6.69 $\pm$ 0.12 & 1 (fixed) \\
\hline
J040210.90-054630.3 & F606W & PSF     & 21.68 & 	  & & & \\
					        & & Bulge & 19.60 & 23.43 & 1.46 $\pm$ 0.05 & 6.00 $\pm$ 0.18 & 4 (fixed) \\
					        & & Other & 19.40 & 22.01 & 1.21 $\pm$ 0.01 & 4.98 $\pm$ 0.03 & 0.46 \\
\hline
J073505.66+423545.6 & F814W & PSF     & 20.55 & 	  & & & \\
					        & & Bulge & 18.70 & 19.78 & 0.35 $\pm$ 0.02 & 0.56 $\pm$ 0.04 & 4 (fixed) \\
					        & & Other & 18.97 & 22.42 & 1.51 $\pm$ 0.02 & 2.40 $\pm$ 0.03 & 0.82 \\
\hline
J092438.88+560746.8 & F814W & PSF     & 20.81 & 	  & & & \\								
	                &       & Bulge   & 14.19 & 23.32 & $18.18\pm0.44$ & $9.20\pm0.22$  & 4.86 \\
	                &       & Sp. Arm & 16.66 & 21.49 & $6.48\pm0.72$  & $3.28\pm0.36$  & 0.14 \\
	                		&         &               & -B. Mode & 1: -61.5, 2.7 & (shear) & 3: 27.4, 0.1 & (S-shape) \\
	                &       & Disk    & 14.94 & 23.74 & $41.99\pm0.94$ & $21.25\pm0.48$ & 1.02 \\
\hline
J093829.38+034826.6 & F814W & PSF     & 21.19 & 	  & & & \\
					        & & Bulge & 20.15 & 20.94 & 0.33 $\pm$ 0.04 & 0.71 $\pm$ 0.09 & 4 (fixed) \\
					        & & Other & 19.04 & 21.93 & 1.25 $\pm$ 0.01 & 2.66 $\pm$ 0.01 & 0.6 \\
\hline
J095819.87+022903.5 & F814W & PSF     & 20.93 & 	  & & & \\
					        & & Bulge & 20.94 & 21.73 & 0.50 $\pm$ 0.04 & 2.45 $\pm$ 0.21 & 4 (fixed) \\
					        & & Disk  & 18.89 & 22.16 & 1.52 $\pm$ 0.01 & 7.40 $\pm$ 0.03 & 1 (fixed) \\
\hline
J100234.85+024253.1 & F814W & PSF     & 22.06 & 	  & & & \\
					        & & Bulge & 20.59 & 22.18 & 0.78 $\pm$ 0.04 & 2.52 $\pm$ 0.14 & 4 (fixed) \\
					        & & Disk  & 19.16 & 22.68 & 1.60 $\pm$ 0.01 & 5.14 $\pm$ 0.02 & 1 (fixed) \\
\hline
J101527.25+625911.5 & F775W & PSF     & 20.45 & 	  & & & \\
					        & & Bulge & 18.55 & 21.81 & 1.19 $\pm$ 0.01 & 5.85 $\pm$ 0.06 & 4 (fixed) \\
					        & & Disk  & 19.21 & 24.18 & 4.54 $\pm$ 0.05 & 22.28 $\pm$ 0.24 & 1 (fixed) \\
\hline
J113657.68+411318.5 & F814W & PSF     & 20.35 & 	  & & & \\
					        & & Disk  & 18.44 & 20.67 & 0.98 $\pm$ 0.01 & 1.32 $\pm$ 0.01 & 1.38 \\
\hline
J120814.35+641047.5 & F814W & PSF     & 19.52 & 	  & & & \\
					        & & Bulge & 18.22 & 23.32 & 2.29 $\pm$ 0.05 & 4.37 $\pm$ 0.10 & 4 (fixed) \\
					        & & Other & 19.57 & 21.82 & 1.24 $\pm$ 0.01 & 2.37 $\pm$ 0.02 & 0.11 \\
\hline
J123228.08+141558.7 & F606W & PSF     & 20.44 & 	  & & & \\
					        & & Bulge & 18.85 & 21.69 & 0.90 $\pm$ 0.03 & 5.02 $\pm$ 0.17 & 4 (fixed) \\
\hline
J123349.92+634957.2 & F814W & PSF     & 20.45 & 	  & & & \\
					        & & Bulge & 20.80 & 21.26 & 0.34 $\pm$ 0.04 & 0.79 $\pm$ 0.09 & 4 (fixed) \\
					        & & Disk  & 19.41 & 22.41 & 1.31 $\pm$ 0.01 & 3.06 $\pm$ 0.03 & 0.99 \\
\hline
J123455.90+153356.2 & F814W & PSF     & 18.48 & 	  & & & \\
					        & & Bulge & 16.11 & 22.26 & 4.60 $\pm$ 0.14 & 4.09 $\pm$ 0.13 & 4 (fixed) \\
					        & & Other & 18.41 & 17.17 & 0.21 $\pm$ 0.00 & 0.19 $\pm$ 0.00 & 0.41 \\
					        & & Other & 15.39 & 21.73 & 6.70 $\pm$ 0.01 & 5.95 $\pm$ 0.01 & 0.43 \\
\hline
J132504.63+542942.3 & F814W & PSF     & 20.31 & 	   & & & \\
					        & & Bulge & 20.66 & 18.21 & 0.08 $\pm$ 0.01 & 0.19 $\pm$ 0.02 & 4 (fixed) \\
					        & & Other & 18.76 & 21.57 & 1.15 $\pm$ 0.02 & 2.95 $\pm$ 0.04 & 1.85 \\
\hline
J132943.60+315336.7 & F814W & PSF     & 20.28 & 	  & & & \\
					        & & Bulge & 17.63 & 18.78 & 0.43 $\pm$ 0.02 & 0.74 $\pm$ 0.03 & 4 (fixed) \\
					        & & Other & 18.64 & 21.08 & 1.55 $\pm$ 0.01 & 2.63 $\pm$ 0.01 & 0.16 \\
					        & & Other & 17.69 & 22.75 & 4.31 $\pm$ 0.01 & 7.33 $\pm$ 0.02 & 0.23 \\
\hline
J141234.67-003500.0 & F814W & PSF     & 20.63 & 	  & & & \\
					        & & Bulge & 19.18 & 23.87 & 2.61 $\pm$ 0.04 & 5.84 $\pm$ 0.09 & 4 (fixed) \\
					        & & Other & 18.55 & 22.57 & 2.86 $\pm$ 0.00 & 6.40 $\pm$ 0.01 & 0.4 \\
\hline
J142543.20+344952.9 & F814W & PSF     & 20.90 & 	  & & & \\
					        & & Bulge & 19.34 & 20.33 & 0.40 $\pm$ 0.01 & 1.20 $\pm$ 0.04 & 4 (fixed) \\
					        & & Other & 20.16 & 22.45 & 1.28 $\pm$ 0.01 & 3.82 $\pm$ 0.04 & 0.39 \\
\hline
J145640.99+524727.2 & F606W & PSF     & 20.86 & 	  & & & \\
					        & & Bulge & 19.22 & 19.99 & 0.51 $\pm$ 0.02 & 2.16 $\pm$ 0.06 & 4 (fixed) \\
					        & & Disk  & 18.16 & 21.26 & 1.22 $\pm$ 0.01 & 5.11 $\pm$ 0.02 & 1 (fixed) \\
\hline
J160044.99+505213.6 & F814W & PSF     & 20.99 & 	  & & & \\
					        & & Bulge & 19.34 & 19.80 & 0.30 $\pm$ 0.02 & 0.55 $\pm$ 0.04 & 4 (fixed) \\
					        & & Other & 19.08 & 21.08 & 0.85 $\pm$ 0.01 & 1.55 $\pm$ 0.01 & 0.77\\
\hline
J171806.84+593313.3 & F814W & PSF     & 21.87 & 	  & & & \\
					        & & Bulge & 19.05 & 22.35 & 1.15 $\pm$ 0.01 & 4.74 $\pm$ 0.03 & 3.9 \\
\enddata
\tablecomments{Results from surface brightness profile measurements Using \textsc{GALFIT}.  Column 1: object.  Column 2: \textit{HST} filter.  Column 3: morphological component type from surface brightness decomposition.  Column 4: extinction-corrected and $k$-corrected ST magnitude.  Column 5: dust extinction-corrected and $k$-corrected effective surface brightness.  Column 6: effective radius in arcseconds.  Column 7: effective radius in kiloparsecs.  Column 8: morphological component S\'ersic index.}
\end{deluxetable*}

\begin{figure*}
\center
\includegraphics[trim={0 8cm 0 0 },scale=0.60,clip]{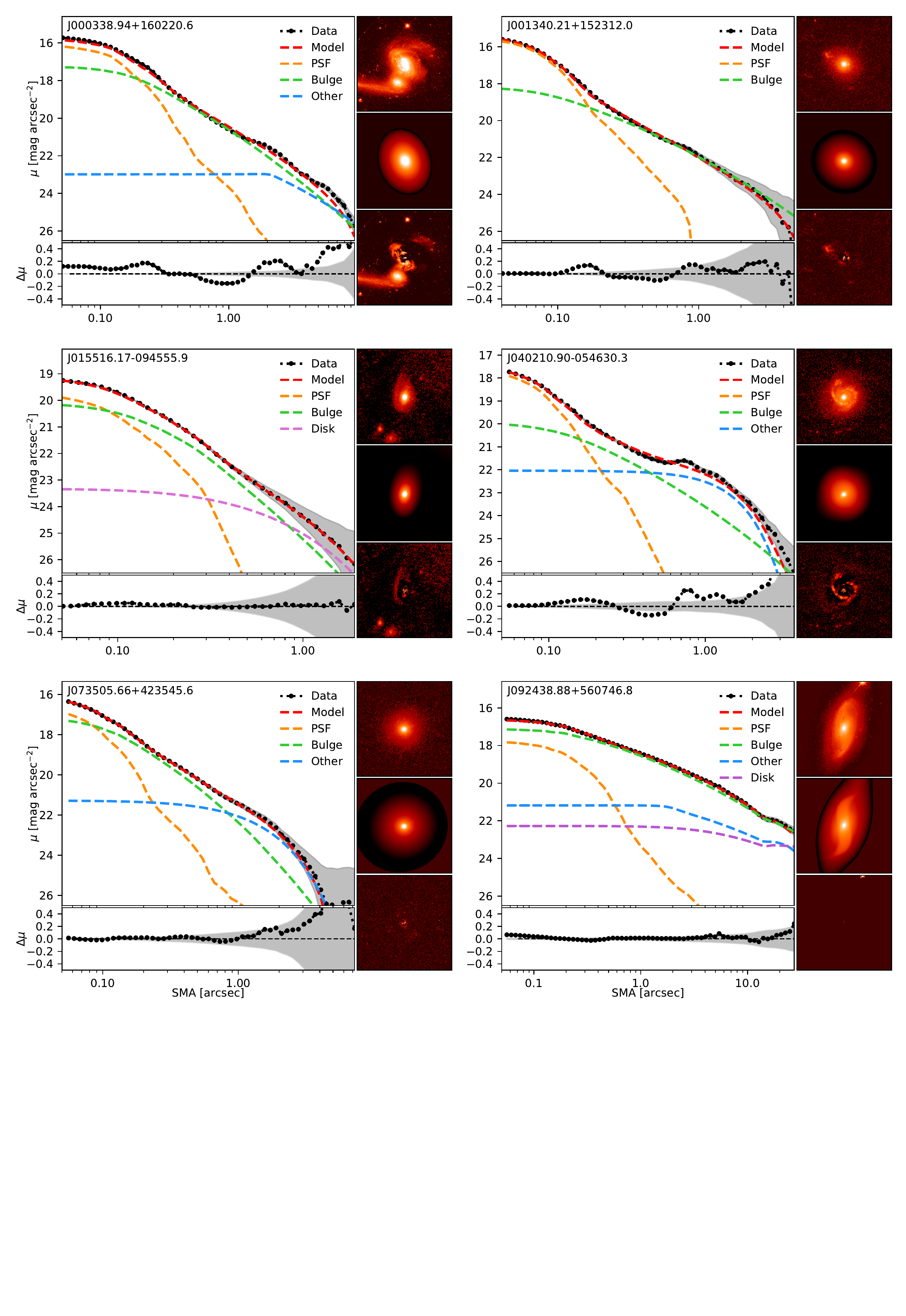}
\caption{
Surface brightness decomposition of the AGN from the host galaxy for our sample.  The large inset is the surface brightness profile, including each of the modeled components, with residuals plotted in the inset below.  The gray shaded region represents the $1\sigma$ uncertainty.  On the right, from top to bottom, the \textit{HST} image of each object, the model output from \textsc{GALFIT}, and the corresponding residuals. 
\label{fig:galfit}}
\end{figure*}

\renewcommand{\thefigure}{\arabic{figure}}
\addtocounter{figure}{-1}
\begin{figure*}
\center
\includegraphics[trim={0 0 0 0 },scale=0.60,clip]{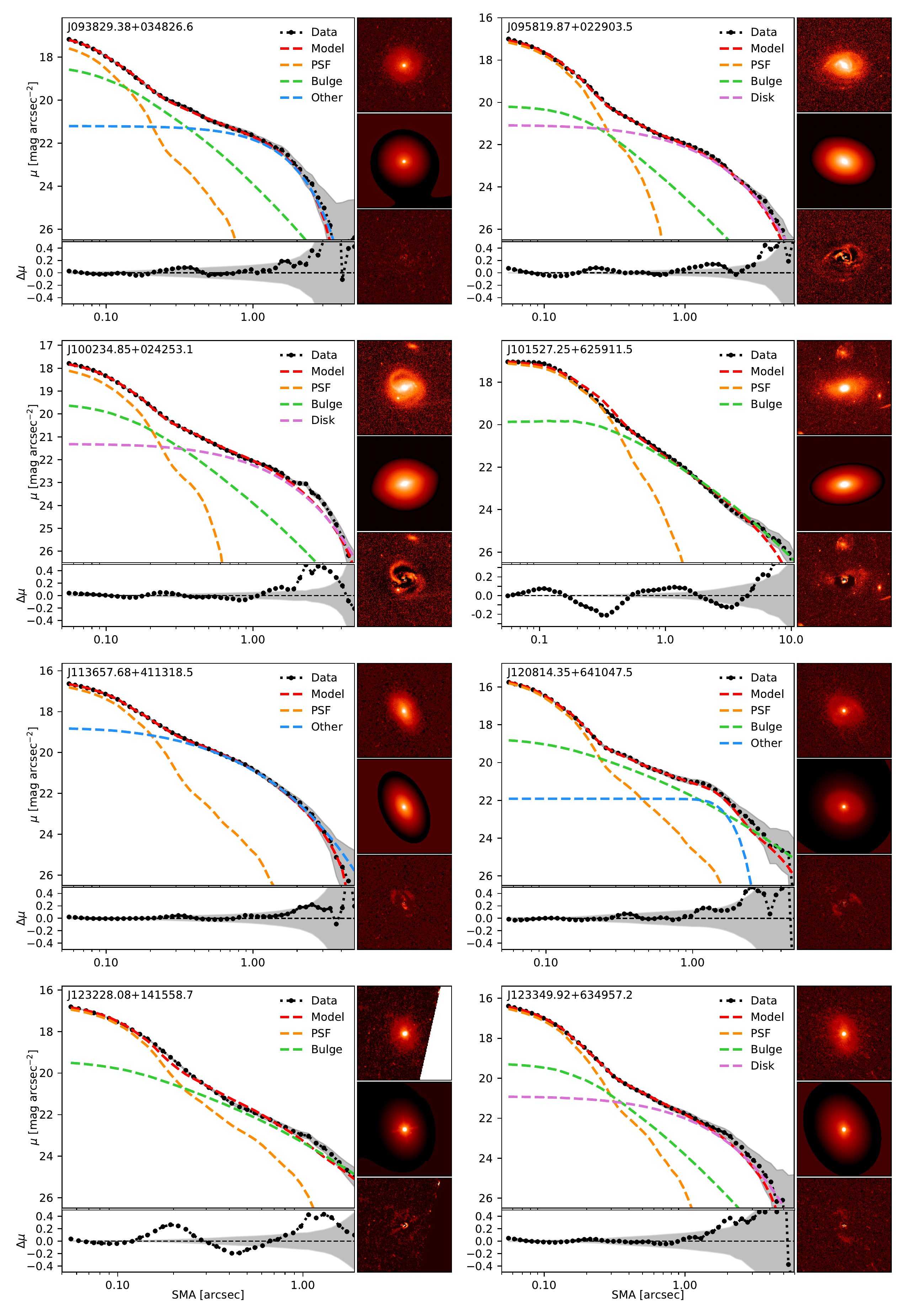}
\caption{\textit{Continued}.}
\end{figure*}
\renewcommand{\thefigure}{\arabic{figure}}

\renewcommand{\thefigure}{\arabic{figure}}
\addtocounter{figure}{-1}
\begin{figure*}
\center
\includegraphics[trim={0 0 0 0 },scale=0.60,clip]{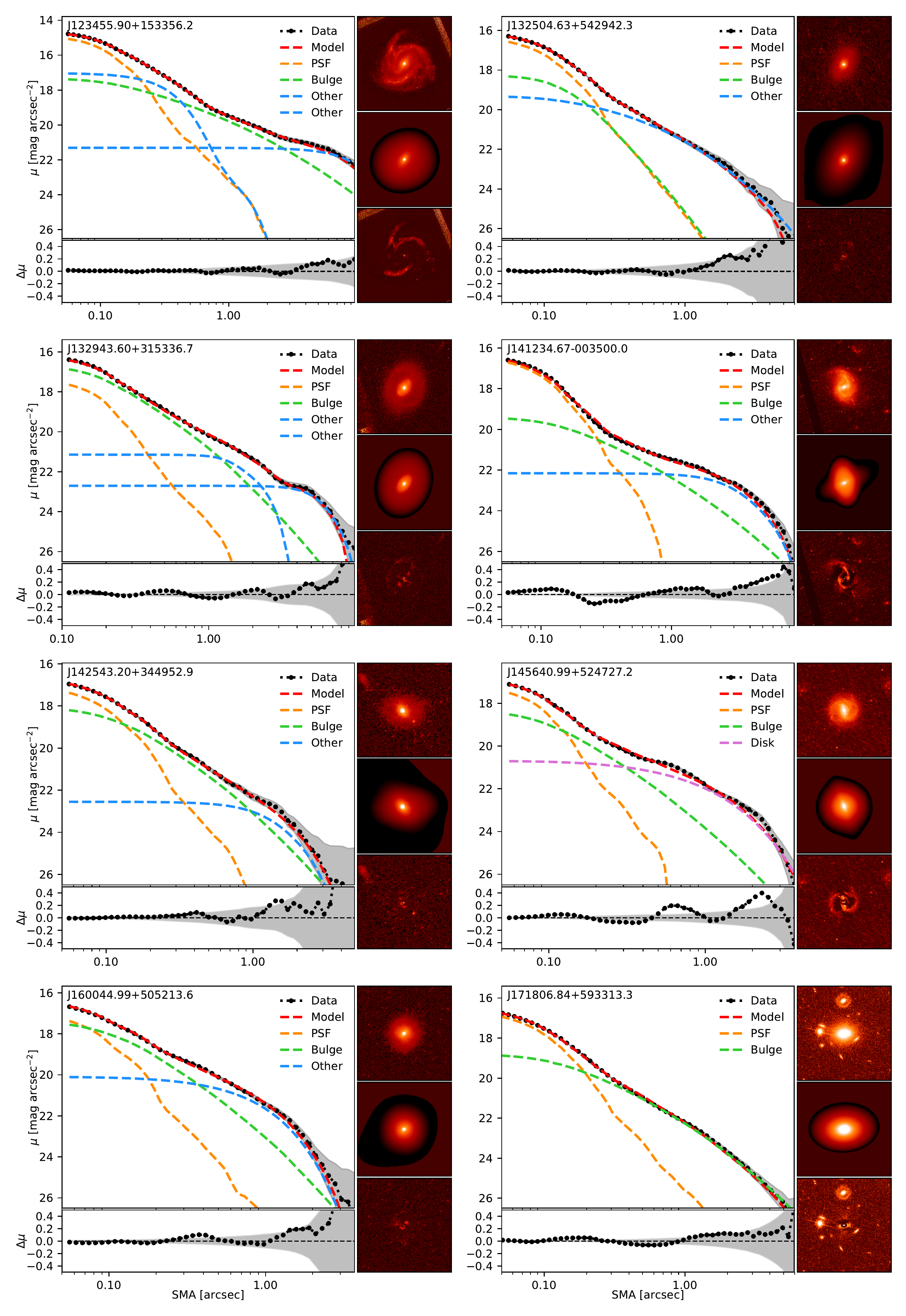}
\caption{\textit{Continued}.}
\end{figure*}
\renewcommand{\thefigure}{\arabic{figure}}

\subsection{Measuring $\lambda L_{5100}$} \label{sec:L5100}

Surface brightness decomposition of \textit{HST} imaging using \textsc{GALFIT}  was used to obtain an estimate of the optical continuum AGN luminosity at 5100 \angstrom, uncontaminated by the host galaxy.  To do this, the AGN component is modeled using a single PSF component, and other S\'ersic components are added to minimize residuals (see Section \ref{sec:galfit}).  The PSF magnitudes are then corrected for Galactic extinction, intrinsic host galaxy extinction using the Balmer decrement, and $k$-corrected.  To obtain the luminosity at 5100 \angstrom\;, we model the full (LRIS-B + LRIS-R) spectrum for each object using the IDL-based multi-component quasar spectrum fitting software \textsc{QSFit} \citep{qsfit}.  \textsc{QSFit} differs from the multi-component fitting method described in Section \ref{sec:emline} in that it only fits a single galaxy template and uses a least-squares minimization technique, providing a means to fit full spectra with a large number of free parameters in a computationally efficient way.  Using \textsc{QSFit}, we fit each object's full spectrum (shown in Figure \ref{fig:observations}) with the default settings, which include a 5 Gyr elliptical galaxy template \citep{Silva1998,Polletta2007}, \ion{Fe}{2} templates from \citep{Veron-Cetty2004}, a simple power-law model for the AGN continuum, and all known emission lines from 3500 to 7000 \angstrom\;.  Uncertainties in the power-law slope were estimated using the Monte Carlo resampling option included in \textsc{QSFit}.  The power-law model was then used to scale the AGN luminosity at the observed $HST$ filter wavelength to a luminosity at 5100 \angstrom.  Using this method, we expect uncertainties in $\lambda L_{5100}$ to be smaller if the pivot wavelength of the \textit{HST} filter is close to 5100 \angstrom, while filters with pivot wavelengths farther from 5100 \angstrom\; are dependent on how accurately the AGN continuum model can be determined (i.e., the slope of the adopted simple power-law continuum model).  The use of a single host galaxy template does not have a significant effect on our measurements since luminosities measured from \textit{HST} imaging are measured at filter pivot wavelengths close to - but typically at longer wavelengths than - 5100 \angstrom, where the effects of the power-law slope vary appreciably less than at shorter wavelengths. \\
\indent Uncertainty in measured luminosities due to variability can be appreciable and vary significantly (5-30\%) from object to object.  Detailed analysis on the flux variability of our objects would require detailed reverberation mapping which is currently unavailable.  We therefore adopt a median uncertainty of 15\% from reverberation-mapped objects from \citet{Bentz2013} as an additional uncertainty due to AGN variability.\\
\indent We estimated the total uncertainty in our measured luminosities to be $\sim$30\% on average for our sample.  Measured values of $\lambda L_{5100}$ are given in Table \ref{tab:mbh_table}. 

\subsection{BH Mass} \label{sec:bhmass}

Single-epoch BH masses are estimated using the virial relation commonly used within the context of reverberation studies (see \citet{Peterson2004}) given as
\begin{equation}
    M_\mathrm{BH}=f\frac{(\Delta V)^2R_\mathrm{BLR}}{G}
\end{equation}
where $f$ is the virial coefficient, $\Delta V$ velocity of the BLR gas at radius $R_\mathrm{BLR}$, and $G$ is the gravitational constant. The $R_\mathrm{BLR}$ is estimated empirically via proxy using the optical luminosity of the AGN \citep{Kaspi2000,Kaspi2005,Bentz2009,Bentz2013}. Following \citet{Woo2015}, we adopt the most recent measurements of the $R_\mathrm{BLR}-\lambda L_{5100}$ relation from \citet{Bentz2013} given as 
\begin{eqnarray}
    \log\left(\frac{R_\mathrm{BLR}}{1\text{lt-day}}\right)=K+\alpha\log\left(\frac{\lambda L_{5100}}{10^{44}\text{ L}_\odot}\right)
\end{eqnarray}
where $K=1.527^{+0.031}_{-0.031}$ is the zero point, and $\alpha=0.533^{+0.035}_{-0.033}$ is the slope of the log-linear relation.  The velocity of the BLR is typically measured via the line dispersion $\sigma_{\mathrm{H}\beta}$; however, it is often easier to measure FWHM$_{\mathrm{H}\beta}$ in low-S/N spectra and convert to $\sigma_{\mathrm{H}\beta}$ using a constant factor.  It is well known that the FWHM$/\sigma$ ratio is velocity dependent \citep{Peterson2004,Collin2006,Kollatschny2011}.  To account for any systematic uncertainties in choice of velocity proxy, \citet{Woo2015} derived separate virial coefficients for $\sigma_{\mathrm{H}\beta}$ and FWHM$_{\mathrm{H}\beta}$. Since our sample consists of spectra with variable S/N, we measure line widths using a Gaussian FWHM and adopt the appropriate virial coefficient of $\log f = 0.05\pm 0.12$ from \citet{Woo2015}.  By adopting the aforementioned relations, the BH mass equation becomes 
\begin{align}
    M_\mathrm{BH}=&10^{6.867^{+0.155}_{-0.153}}\left(\frac{\text{FWHM}_{\mathrm{H}\beta}}{10^3\text{ km s}^{-1}}\right)^2\nonumber\\
    &\times\left(\frac{\lambda L_{5100}}{10^{44}\text{ erg s}^{-1}}\right)^{0.533^{+0.035}_{-0.033}}\ M_\odot
\end{align} \label{eqn:bh_mass}

Values for calculated BH masses for our sample can be found in Table \ref{tab:mbh_table}.  The uncertainties quoted for BH mass in Table \ref{tab:mbh_table} include uncertainties from measurements of FWHM$_{\rm{H}\beta}$ and $\lambda L_{5100}$, as well as the uncertainties derived from the virial coefficient $f$ and the $R_{\rm{BLR}}-\lambda L_{5100}$ relation.  The most significant contribution to the uncertainties in BH mass is derived from the calibration of the virial coefficient $f$. 

\section{Results} \label{sec:results}

\subsection{The $M_{\rm{BH}}-\sigma_*$ Relation} \label{sec:m_sigma_relation}

\begin{figure*}[ht!]
\center
\includegraphics[trim={0 0 0 0},scale=0.75,clip]{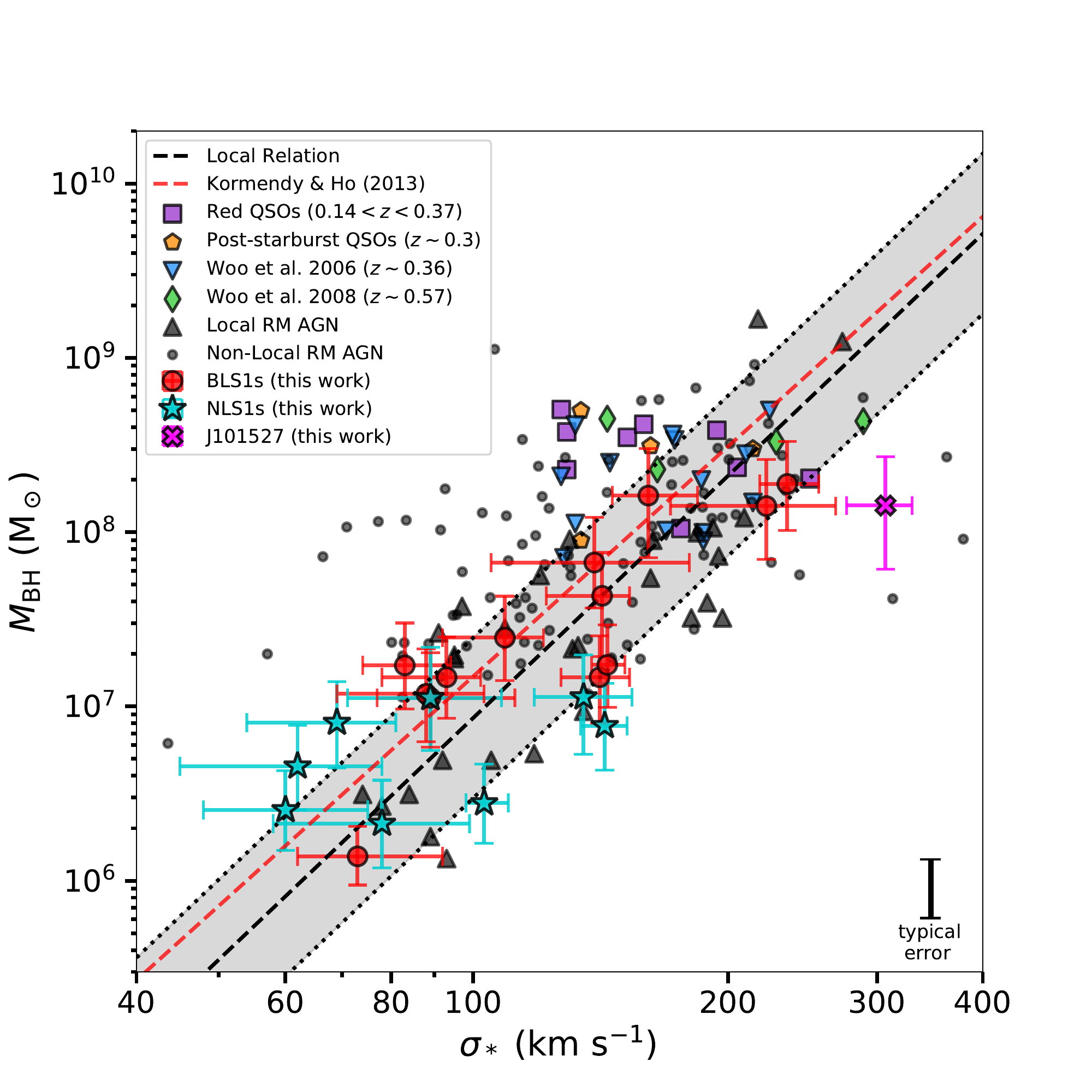}
\caption{
The $M_{\rm{BH}}-\sigma_*$ relation for our sample, including objects from selected non-local studies.  The black dashed line indicates the local relation we derived using BH masses of local AGNs from \citet{Bennert2011a}, local inactive galaxies from \citet{McConnell2013}, and local reverberation-mapped AGNs from \citet{Woo2015}, using the most recent BH mass calibration from \citet{Woo2015} (Equation \ref{eqn:local_msigma}). The black dotted lines and shaded area represent the local scatter.  The red dashed line represents the relation from \citet{Kormendy2013}, based on BH masses of local inactive galaxies measured using stellar and gas kinematics.  Additionally, we include local reverberation-mapped objects from \citet{Woo2015} and non-local reverberation-mapped objects from \citet{Shen2015} for comparison.
\label{fig:m-sigma}}
\end{figure*}

We plot the results of our measurements for the $M_{\rm{BH}}-\sigma_*$ relation in Figure \ref{fig:m-sigma}.  We include other non-local objects from previous studies of red 2MASS quasars at $0.14<z<0.37$ \citep{Canalizo2012}, post-starburst quasars at $z\sim0.3$ \citep{Hiner2012}, Seyfert 1 galaxies at $z=0.36$ and $z=0.57$ \citep{Woo2006,Woo2008}, as well as local and non-local reverberation-mapped AGN samples from \citet{Woo2015} and \citet{Shen2015}, respectively, for comparison. \\
\indent To compare our measurements to the local relation, we recalculate BH masses for all objects with $z<0.1$ for the combined sample of AGNs from \citet{Bennert2011a}, local inactive galaxies from \citet{McConnell2013}, and local reverberation-mapped BH masses from \citet{Woo2015} using the most recent BH mass calibration from \citet{Woo2015}, which adopts a virial coefficient of $\log f=0.05\pm0.12$ for H$\beta$ line widths measured using a Gaussian FWHM.  The local comparison sample consists of a total of 124 objects ranging in mass from $6.1$ to $10.3$ in $\log_{10}(M_{\rm{BH}})$.  We perform linear regression using a maximum-likelihood approach and estimate uncertainties using MCMC.  The linear fit to the local $M_{\rm{BH}}-\sigma_*$ relation is given by
\begin{align}\label{eqn:local_msigma}
\log_{10}&\left(\frac{M_{\rm{BH}}}{M_{\odot}}\right) = (8.323^{+0.046}_{-0.046})\nonumber\\
& + (4.613^{+0.230}_{-0.231})\log_{10}\left(\frac{\sigma_*}{200\text{ km s}^{-1}}\right)
\end{align} 
with an intrinsic scatter of $\epsilon_{0}=0.427^{+0.033}_{-0.032}$.  The best-fit local relation is plotted as a black dashed line, and the $68\%$ confidence interval is given by the dotted lines and shaded region in Figure \ref{fig:m-sigma}.  Our local relation has a shallower slope than that of \citet{McConnell2013} ($\beta=5.64\pm0.32$) and is nearly consistent with that of the \citet{Woo2015} updated reverberation-mapped sample ($\beta=4.97\pm0.28$).  Additionally, we plot the relation from \citet{Kormendy2013} (red dashed line), which measured local BH masses in inactive galaxies using stellar and gas kinematics.  Since single-epoch BH masses are calibrated using local inactive galaxies, the good agreement between the \citet{Kormendy2013} relation and our local relation indicates that BH masses for AGNs are well-calibrated.\\
\indent The 22 objects in our sample span a mass range of two orders of magnitude from $6.1$ to $8.3$ in $\log_{10}(M_{\rm{BH}})$. The mean offset of our sample from the local relation is $-0.018^{+0.111}_{-0.108}$ dex, with a scatter of $0.385^{+0.099}_{-0.112}$ dex.  The scatter in our sample is comparable to that of the $0.43\pm0.03$ dex found for local reverberation-mapped objects \citep{Woo2015} as well as the $0.38$ dex for objects from stellar dynamical measurements \citep{McConnell2013}. Overall, the distribution of objects in our sample does not preferentially lie above or below the local relation. NLS1s in our sample span a mass range from $6.3$ to $7.1$ in $\log_{10}(M_{\rm{BH}})$ and, on average, fall on the local relation with comparable scatter to the overall sample.  Overall, our sample expands on the non-local relation by occupying the lower to intermediate SMBH mass range with a scatter comparable to the local relation.

\subsection{Evolution in the $M_{\rm{BH}}-\sigma_*$ Relation} \label{sec:m_sigma_evolution}

\begin{figure*}[ht!]
\center
\includegraphics[trim={0 0 0 0},scale=0.6,clip]{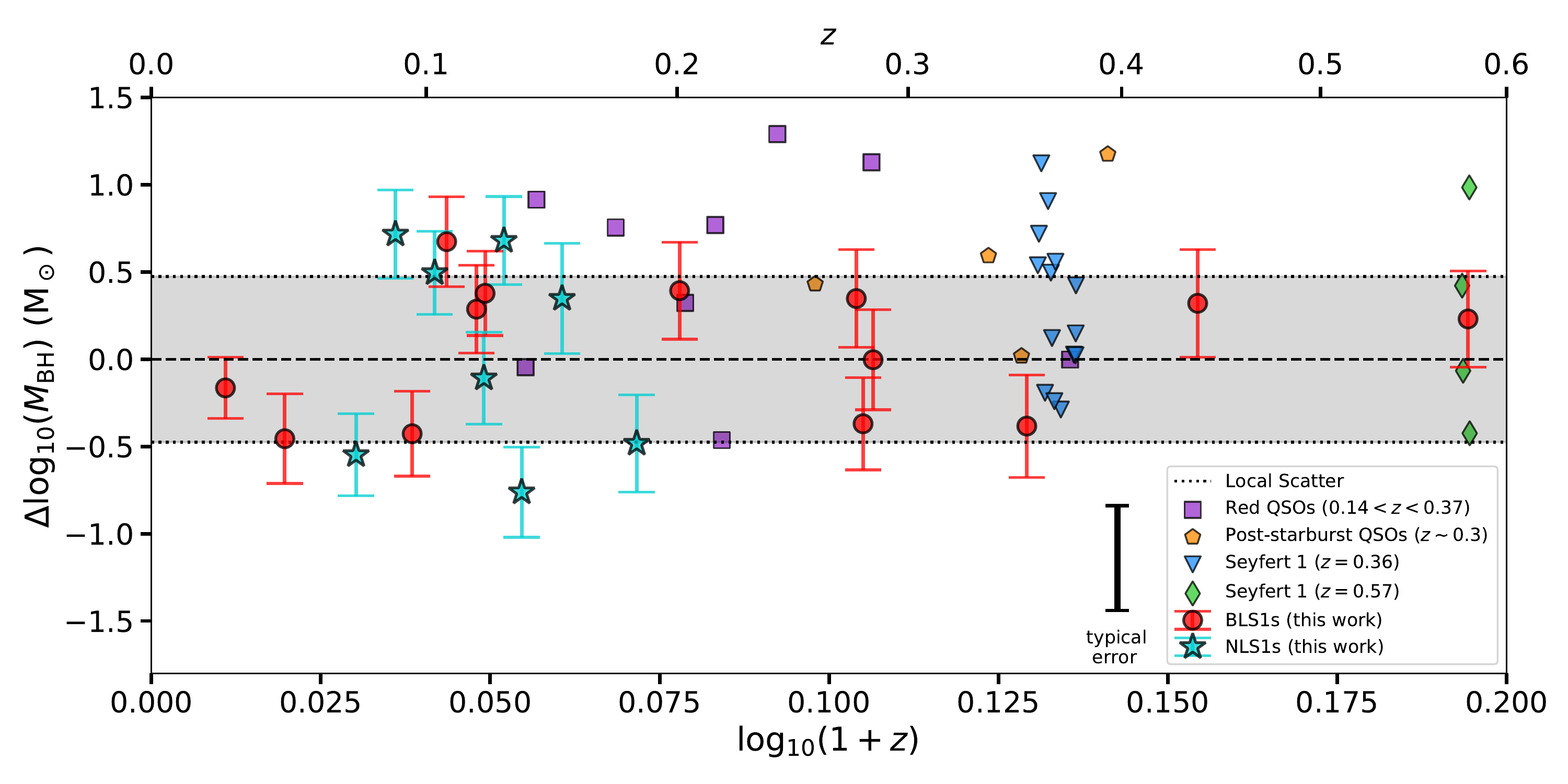}
\caption{
$\Delta\log_{10}(M_{\rm{BH}})$ as a function of redshift with respect to the local $M_{\rm{BH}}-\sigma*$ relation to investigate any offset.  The gray shaded area corresponds to the scatter in the local relation.  The majority of non-local objects reside well within the local scatter.  While the slope of $\gamma=2.16\pm0.62$ can be fit, implying a significant offset, the correlation is weak due to the consistent scatter across all sampled redshifts, indicating that the offset is driven by only a small number of objects at the high-redshift regime.  Alternatively, the average offset of non-local objects, found by fitting a constant, is $C=0.19\pm0.08$, and results in an identical residual scatter as a linear fit.  While the offset remains non-zero, this does not necessarily indicate evolution, as we see greater offset in lower-redshift objects on average than higher-redshift objects, indicating that we may still be sampling the higher-mass BH regime due to selection effects.
\label{fig:evolution}}
\end{figure*}

\indent In Figure \ref{fig:evolution} we plot $\Delta\log_{10}(M_{\rm{BH}})$ as a function of redshift.  Following \citet{Woo2006,Woo2008}, we investigate the possibility of evolution in the $M_{\rm{BH}}-\sigma_*$ relation, by performing linear regression of $\Delta\log_{10}(M_{\rm{BH}})$ with respect to to the local $M_{\rm{BH}}-\sigma_*$ relation as a function of redshift following the linear model used by \citet{Park2015}, given by 
\begin{equation}
\Delta\log_{10}(M_{\rm{BH}}) = \gamma\log_{10}(z+1).    
\end{equation}
Since we have defined $\Delta\log_{10}(M_{\rm{BH}})$ with respect to the local relation, we exclude an intercept as free parameter.  We also avoid binning BH masses by redshift to avoid introducing any biases due to the fact that our objects are not at discrete redshift intervals, unlike the samples of \citet{Woo2006,Woo2008}, which were - by design - selected at discrete intervals of $z=0.36$ and $z=0.57$, respectively.  We perform maximum-likelihood regression and estimate uncertainties using MCMC finding the value in the best-fit slope to be $\gamma=2.16\pm0.62$ with a scatter of $\epsilon_0=0.43\pm0.03$, which implies a $3.5\sigma$ confidence for a non-zero positive slope. \\
\indent Previous analysis by \citet{Woo2008} compared $z=0.36$ and $z=0.57$ Seyfert 1 objects \citep{Woo2006,Woo2008} found a slope of $\gamma=3.1\pm1.5$, however, they compared their sample to the local inactive relation fit available at the time by \citet{Tremaine2002}, which is a shallower local relation (a slope of $\beta=4.02\pm0.44$), and enhances the apparent offset in $M_{\rm{BH}}$ by $0.43$ dex and $0.63$ dex at $z=0.36$ and $z=0.57$, respectively.  If we perform the same analysis of the non-local objects from \citet{Woo2008} with our revised local relation we find a slope of $\gamma=1.93\pm0.73$ or $2.6\sigma$ confidence for a non-zero slope and an offset in $M_{\rm{BH}}$ of only $0.26$ dex and $0.38$ dex at $z=0.36$ and $z=0.57$, respectively, which is less than the $\sim0.4$ dex scatter of these data at these redshifts.  Including dust-reddened 2MASS QSOs from \citet{Canalizo2012}, and post-starburst QSOs from \citet{Hiner2012} enhances the slope further to $\gamma=2.43\pm0.68$ ($3.6\sigma$ confidence) due to these objects being preferentially above the relation by $\sim$0.5 dex in  $\Delta\log_{10}(M_{\rm{BH}})$ at $z\sim0.3$. With the inclusion of our objects, the significance of a non-zero slope decreases slightly to $3.5\sigma$ confidence.  If we were to omit higher-luminosity QSOs and consider only Seyfert 1 objects the slope decreases to $\gamma=1.69\pm0.65$ ($2.6\sigma$ confidence).\\
\indent From Figure \ref{fig:evolution}, there is ample reason to be skeptical of any underlying trend in $\Delta\log_{10}(M_{\rm{BH}})$ as a function of $z$, as it is clear that there remains considerable scatter in $\Delta\log_{10}(M_{\rm{BH}})$.  We can quantify the strength of a linear correlation for our data in the context of the scatter by computing the nonparametric Spearman's correlation coefficient, assuming there exists some monotonically increasing relationship in $\Delta\log_{10}(M_{\rm{BH}})$ as a function of $z$.  We calculate the Spearman's coefficient and its uncertainty using Monte Carlo methods.  Spearman's correlation coefficient of the non-local sample, including our objects, is $r_s = 0.23\pm0.04$, indicating a very weak to weak positive correlation.  The weakness of the correlation is due primarily to the consistent scatter of $\sim$0.4 dex across the entire sampled redshift range.  In other words, the scatter we observe locally and at low redshifts is comparable to the scatter we observe at the highest redshifts, which implies there is - at best - a weak dependence of  $\Delta\log_{10}(M_{\rm{BH}})$ on redshift.  If we instead fit a constant model to $\Delta\log_{10}(M_{\rm{BH}})$ to all non-local objects, we find that the constant offset from the local relation is $C=0.19\pm0.08$ with a scatter of $\epsilon_0=0.40\pm0.07$.  Most importantly, the residual scatter is nearly identical regardless of the model chosen, due solely to the large amount of scatter at all redshifts.  Additionally, because the intercept of the linear fit to $\Delta\log_{10}(M_{\rm{BH}})$ is held constant to zero (because we are comparing it to the local relation at $z=0$), any datum at high redshift can have considerable influence on the slope of the linear fit, especially for our small sample.  
Considering the level of scatter across the sampled redshift range, the weak correlation of  $\Delta\log_{10}(M_{\rm{BH}})$ with respect to $z$, and the fact that the majority of these data reside well within the local scatter (see Figure \ref{fig:evolution}), we conclude that any evolution in the $M_{\rm{BH}}-\sigma*$ relation in the past 6 Gyr is very weak at best.

\begin{deluxetable*}{ccclclc}
\tablecaption{$M_{\rm{BH}}$ and $\sigma_*$ Measurements \label{tab:mbh_table}}
\tablehead{
\colhead{Object} & \colhead{$z$} & \colhead{$E(B-V)$} & \colhead{FWHM$_{\mathrm{H}\beta}$} & \colhead{$\log_{10}(\lambda L_{5100})$} & \colhead{$\sigma_*$} & \colhead{$\log_{10}(M_\mathrm{BH})$} \\
\colhead{} & \colhead{} & \colhead{} &\colhead{(km s$^{-1}$)} & \colhead{(erg s$^{-1}$)} & \colhead{(km s$^{-1}$)} & \colhead{($M_\odot$)}
}
\colnumbers
\startdata
J000338.94+160220.6   & 0.11681 & 0.010     & $3253_{-233}^{+233}$   & $43.07_{-0.10}^{+0.11}$ & $109_{-17}^{+12}$  & $7.39_{-0.19}^{+0.32}$ \\
J001340.21+152312.0   & 0.12006 & \nodata   & $2854_{-206}^{+205}$   & $42.85_{-0.10}^{+0.11}$ & $93 _{-15}^{+9 }$  & $7.17_{-0.18}^{+0.30}$ \\
J015516.17$-$094555.9 & 0.56425 & 0.019     & $4350_{-333}^{+366}$   & $43.40_{-0.10}^{+0.11}$ & $139_{-34}^{+41}$  & $7.83_{-0.20}^{+0.36}$ \\
J040210.90$-$054630.3 & 0.27065 & 0.051     & $2308_{-204}^{+232}$   & $42.98_{-0.14}^{+0.11}$ & $89 _{-12}^{+23}$  & $7.05_{-0.21}^{+0.35}$ \\
J073505.66+423545.6   & 0.08646 & \nodata   & $2019_{-155}^{+148}$   & $42.93_{-0.13}^{+0.11}$ & $69 _{-15}^{+12}$  & $6.91_{-0.20}^{+0.31}$ \\
J092438.88+560746.8   & 0.02548 & \nodata   & $2650_{-199}^{+209}$   & $41.05_{-0.08}^{+0.11}$ & $65 _{-11}^{+19}$  & $6.14_{-0.14}^{+0.21}$ \\
J093829.38+034826.6   & 0.11961 & \nodata   & $1186_{-102}^{+118}$   & $42.71_{-0.11}^{+0.11}$ & $78 _{-20}^{+21}$  & $6.33_{-0.19}^{+0.33}$ \\
J095819.87+022903.5   & 0.34643 & \nodata   & $4845_{-360}^{+361}$   & $43.83_{-0.15}^{+0.11}$ & $222_{-51}^{+46}$  & $8.15_{-0.22}^{+0.37}$ \\
J100234.85+024253.1   & 0.19659 & 0.115     & $2607_{-186}^{+187}$   & $42.82_{-0.15}^{+0.18}$ & $88 _{-19}^{+15}$  & $7.07_{-0.20}^{+0.35}$ \\
J101527.25+625911.5   & 0.35064 & \nodata   & $4379_{-324}^{+369}$   & $44.01_{-0.21}^{+0.11}$ & $307_{-31}^{+23}$  & $8.15_{-0.25}^{+0.39}$ \\
J113657.68+411318.5   & 0.07200 & \nodata   & $1476_{-104}^{+109}$   & $42.57_{-0.11}^{+0.11}$ & $103_{-5 }^{+7 }$  & $6.45_{-0.18}^{+0.29}$ \\
J120814.35+641047.5   & 0.10555 & 0.140     & $2400_{-170}^{+169}$   & $43.26_{-0.10}^{+0.11}$ & $83 _{-9 }^{+11}$  & $7.24_{-0.19}^{+0.33}$ \\
J123228.08+141558.7   & 0.42692 & \nodata   & $4704_{-334}^{+333}$   & $44.00_{-0.20}^{+0.11}$ & $161_{-15}^{+23}$  & $8.21_{-0.24}^{+0.37}$ \\
J123349.92+634957.2   & 0.13407 & \nodata   & $1767_{-143}^{+133}$   & $43.11_{-0.10}^{+0.11}$ & $143_{-9 }^{+9 }$  & $6.89_{-0.19}^{+0.32}$ \\
J123455.90+153356.2   & 0.04637 & \nodata   & $2742_{-206}^{+211}$   & $42.92_{-0.13}^{+0.11}$ & $141_{-14}^{+12}$  & $7.17_{-0.20}^{+0.32}$ \\
J132504.63+542942.3   & 0.14974 & 0.039     & $1929_{-198}^{+248}$   & $43.27_{-0.13}^{+0.11}$ & $89 _{-18}^{+19}$  & $7.05_{-0.22}^{+0.41}$ \\
J132943.60+315336.7   & 0.09265 & 0.021     & $3122_{-224}^{+221}$   & $42.84_{-0.11}^{+0.11}$ & $144_{-6 }^{+7 }$  & $7.24_{-0.19}^{+0.30}$ \\
J141234.67$-$003500.0 & 0.12724 & \nodata   & $1475_{-125}^{+109}$   & $42.97_{-0.10}^{+0.11}$ & $62 _{-17}^{+16}$  & $6.66_{-0.19}^{+0.31}$ \\
J142543.20+344952.9   & 0.17927 & 0.144     & $2004_{-150}^{+145}$   & $43.21_{-0.20}^{+0.11}$ & $135_{-17}^{+19}$  & $7.05_{-0.23}^{+0.33}$ \\
J145640.99+524727.2   & 0.27792 & 0.166     & $3608_{-256}^{+263}$   & $43.35_{-0.22}^{+0.11}$ & $142_{-20}^{+11}$  & $7.63_{-0.24}^{+0.34}$ \\
J160044.99+505213.6   & 0.10104 & \nodata   & $1363_{-101}^{+102}$   & $42.63_{-0.10}^{+0.11}$ & $60 _{-12}^{+15}$  & $6.41_{-0.18}^{+0.30}$ \\
J171806.84+593313.3   & 0.27356 & 0.031     & $8179_{-598}^{+600}$   & $43.22_{-0.12}^{+0.11}$ & $235_{-17}^{+21}$  & $8.28_{-0.20}^{+0.33}$ \\
\enddata
\tablecomments{
Measurements of $M_{\rm{BH}}$ and $\sigma_*$.  Column 1: object.  Column 2: redshift as measured from stellar absorption features, repeated here for reference.  Column 3: intrinsic extinction as measured from the Balmer decrement.  Column 4: H$\beta$ FWHM.  Column 5: base 10 logarithm of the AGN luminosity at 5100 \angstrom, as measured from \textsc{GALFIT} surface brightness decomposition.  Column 6: inclination-corrected stellar velocity dispersion.  Column 7: base 10 logarithm of calculated BH mass from Equation \ref{eqn:bh_mass}. 
}
\end{deluxetable*}

\section{Systematics} \label{sec:systematics}

The following sections outline possible systematic uncertainties and selection effects that may affect our measurements.

\subsection{H$\beta$ Width Measurements}

Previous studies \citep{Woo2006,Woo2008} use the second moment of the H$\beta$ emission line, showing that line measurements from single-epoch spectra are consistent with those of reverberation studies; however, it is often easier to measure FWHM in lower-S/N spectra.  One caveat of adopting a FWHM parameterization for the H$\beta$ width is the fact that the relationship between FWHM and $\sigma$ is not necessarily FWHM/$\sigma=2.355$, and previous studies have attempted to account for the discrepancy \citep{Park2012}.  \citet{Woo2015} derived a virial factor that takes into account the systematic uncertainty added to mass estimates derived from calibrations from reverberation studies, given by $f=0.05\pm0.12$, which we adopt here.  We found that asymmetries in the broad H$\beta$ line profile are due to underlying stellar absorption, and that when broad H$\beta$ and the stellar continuum are fit simultaneously, a single Gaussian component fully accounts for any line asymmetries.  We find that our single-component Gaussian measurements are consistent with measurements using multiple Gaussian components to account for line asymmetries.  We also find that the uncertainties in our estimates of the FWHM decrease by a factor of 2.3 when fit simultaneously with the stellar continuum and \ion{Fe}{2} emission.  Uncertainty due to variability of the FWHM of H$\beta$ with respect to rms line widths from reverberation studies are estimated to be $7$\% \citep{Woo2007}, which we add to our random uncertainties in quadrature.  On average, the total uncertainty in our measurements for broad H$\beta$ FWHM is $\sim$8\%, corresponding to a $0.06$ dex uncertainty in $M_{\rm{BH}}$.  One object in our sample, J015516, was observed independently by \citet{Woo2008} to have $\sigma_{\rm{H}\beta}=2103$ km s$^{-1}$, which is consistent with our measurement of FWHM$_{\rm{H}\beta}=4350$ km s$^{-1}$ if we assume FWHM/$\sigma\sim2$.  We conclude that our estimates for H$\beta$ width measured from the FWHM of the line profile are not a significant source of systematic uncertainty, and do not significantly affect estimates of $M_{\rm{BH}}$. 

\subsection{$\lambda L_{5100}$ Measurements}

Residuals of surface brightness photometry performed on \textit{HST} imaging show there is very good agreement of the empirically constructed PSF and the central surface brightness of the AGN for each object.  Large residuals in surface brightness profiles are at most $\Delta\mu=0.3$ mag arcsec$^{-2}$ and appear to result from intrinsic properties of each object, such as the presence of dust lanes and spiral arms.  On average, the uncertainty due to PSF mismatch is $\sim$0.1 mag.  We do not suspect PSF mismatch to be a significant source of error in our measurements for the AGN luminosity.  For comparison, \citet{Park2015} independently fit  J073505 from \textit{HST}/NICMOS/F110W imaging and obtained a $\lambda L_{5100}=1.02$ ($10^{44}$ erg s$^{-1}$), while we obtained $\lambda L_{5100}=1.01$ ($10^{44}$ erg s$^{-1}$) with \textit{HST}/ACS-WFC/F775W imaging. \\
\indent The simple power-law parameterization used to model the AGN continuum from the full spectrum (LRIS-B + LRIS-R) also contributes an uncertainty of $\sim$0.1 mag.  Uncertainties in various corrections, e.g. extinction, AGN fraction, $k$-correction, and passive evolution, we conservatively estimate at $\sim$0.1 mag.\\
\indent To account for uncertainty due to variability in our measured luminosities, we adopt an additional 15\% uncertainty based on the median uncertainty from reverberation-mapped luminosities from \citet{Bentz2013}.\\
\indent The overall uncertainty in our estimates for $\lambda L_{5100}$ is $\sim$30\% on average, corresponding to a $0.08$ dex uncertainty in $M_{\rm{BH}}$, consistent with the uncertainties estimated by \citet{Treu2007}.  Given that $M_{\rm{BH}}\propto \lambda L_{5100}^{0.533}$, we do not expect our measurements for $\lambda L_{5100}$ to contribute a significant offset in our estimates for $M_{\rm{BH}}$.\\
\indent  Extinction, if left unaccounted for, can also lead to an underestimate of $\lambda L_{5100}$, and therefore an underestimate of BH mass.  We correct for Galactic extinction, as well as intrinsic extinction estimated from measurements of narrow Balmer line ratios.  We do not use broad-line emission ratios to correct for extinction within the BLR.  However, given the low dependence of $\lambda L_{5100}$ on BH mass, we do not suspect extinction from the BLR to significantly affect our results except in extreme cases.  For instance, not accounting for a reddening value of $E(B-V)=0.1$ corresponds to a 0.06 dex underestimation of BH mass.
 
 \subsection{BH Mass Calibration} \label{sec:calibration}
 
 The derivation of Equation \ref{eqn:bh_mass} used to calculate single-epoch BH mass is empirically calibrated using local ($z<0.3$) reverberation-mapped AGNs to obtain the $R_{\rm{BLR}}-\lambda L_{5100}$ relation.  The behavior of the $R_{\rm{BLR}}-\lambda L_{5100}$ relation at $z>0.3$ however is still unknown due to a lack of reverberation-mapping studies at higher redshifts, which may be problematic for the high-$z$ objects in our sample.  Furthermore, it is possible that the the behavior of the $R_{\rm{BLR}}-\lambda L_{5100}$ relation may be dependent on accretion rate.  Recent reverberation-mapping measurements performed by \citet{Du2016} of super-Eddington accreting massive BHs in AGNs found that $R_{\rm{BLR}}$ scales inversely with accretion rate, i.e. higher accretion rates result in smaller $R_{\rm{BLR}}$.  If not taken into consideration, this dependence could systematically cause us to overestimate the BH mass of NLS1 objects in our sample per given $\lambda L_{5100}$, which have higher accretion rates ($12\%$ on average) than the BLS1s in our sample (4\% on average).  However, since the NLS1s in our sample have generally lower accretion rates than those studied by \citet{Du2016}, we expect the contribution of accretion rate on the calculation of BH mass for objects in our sample to be negligible.
 
\subsection{$\sigma_*$ Measurements} \label{sec:sigma_uncertainties}

\subsubsection{Template Mismatch}  

Template fitting performed to measure the LOSVD of the host galaxy is typically performed using a set of template stars observed on the same night as the science targets; however, if the stellar population of the host galaxy is not known, it can result in template mismatch which can bias measurements of $\sigma_*$.  To minimize the effects of template mismatch, we instead use a large number ($N=636$) of template stars of various types from the Indo-US Library of Coud\'e Feed Stellar Spectra \citep{IndoUS}.  The random uncertainty is estimated via Monte Carlo methods, which sample all possible templates until a stable LOSVD solution is met.  Given the large number of stellar templates used in the fit, it is unlikely template mismatch contributes to significant uncertainties in our measurements in $\sigma_*$. \\

\subsubsection{Fitting Region}

The choice of fitting region used to measure $\sigma_*$ can also potentially contribute to significant bias.  \citet{Greene2006b} investigated the viability and systematics of measuring $\sigma_*$ in the \ion{Ca}{0}H+K, \ion{Mg}{1b}, and \ion{Ca}{0}T regions and found that while the \ion{Ca}{0}T region is the least susceptible to template mismatch and \ion{Fe}{2} contamination, it is the region most affected by AGN continuum dilution, which acts to bias measurements of $\sigma_*$ to higher values (decrease line EW).  On the other hand, the \ion{Ca}{0}H+K region is the least affected by continuum dilution, but the most susceptible to template mismatch.  Additionally, both the \ion{Ca}{0}T and \ion{Ca}{0}H+K regions can be biased by their stellar populations, most notably by the presence of A stars which significantly broaden hydrogen lines.  \citet{Greene2006b} concluded that \ion{Mg}{1b} is the most practical region to measure $\sigma_*$ at redshifts $0.05<z<0.76$ under the conditions  that the amount of AGN continuum dilution is $\leq85\%$ and Eddington ratios are $\leq0.5$.  The average AGN dilution in our sample is 41\% and does not exceed 82\%, as measured by taking the AGN-to-total flux ratio from surface brightness decomposition of \textit{HST} imaging.  Figure \ref{fig:edd_feii}(a) shows that our objects have Eddington ratios well below the 50\% threshold for accurate measurements of $\sigma_*$, therefore we do not suspect continuum dilution to contribute significant bias.  While measurements of $\sigma_*$ in the \ion{Mg}{1b} region can be significantly biased by the presence of \ion{Fe}{2} emission, this effect can be mitigated by including \ion{Fe}{2} templates in our fitting process, as discussed below.  \\

\subsubsection{\ion{Fe}{2} Contamination}

\begin{figure*}
\gridline{
\fig{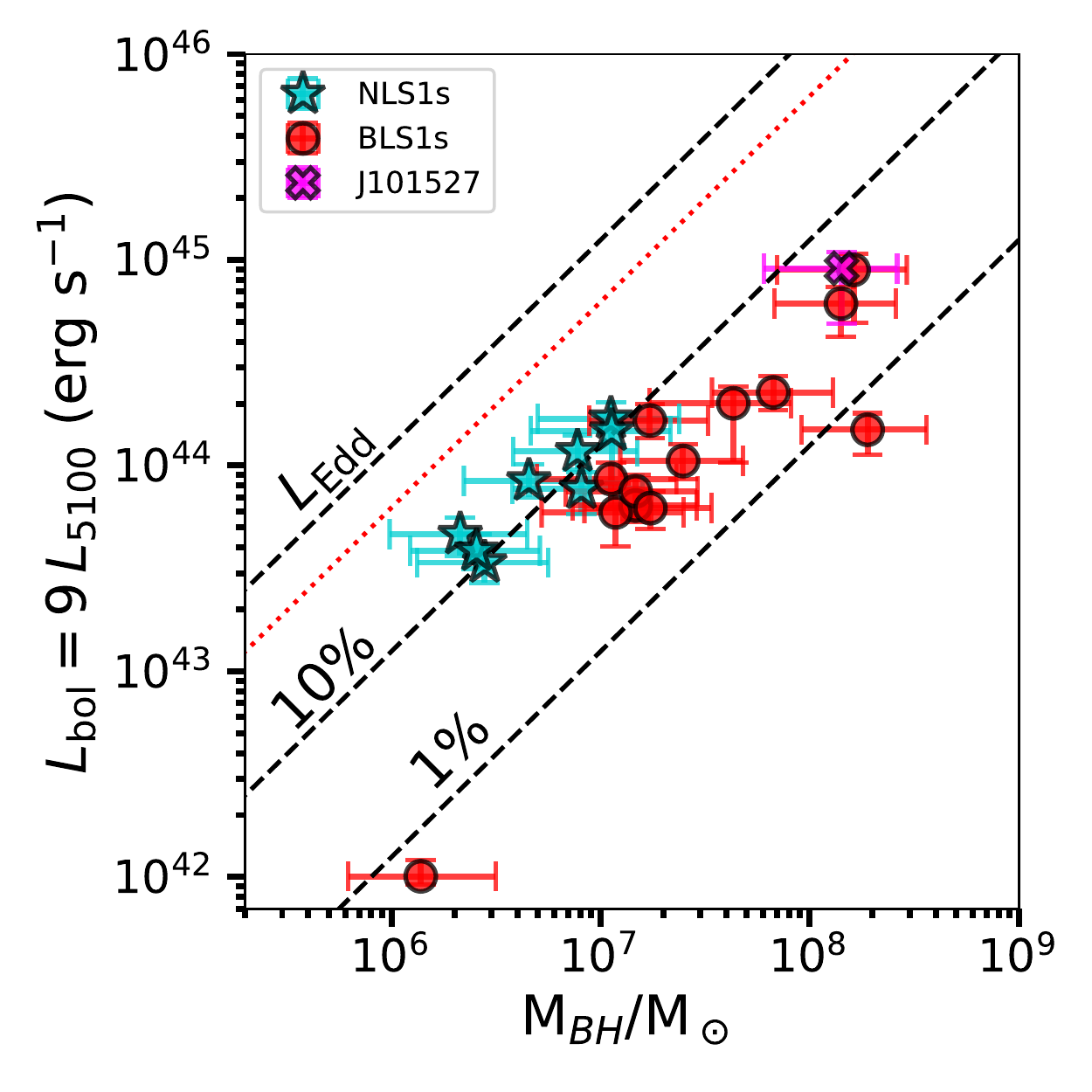}{0.5\textwidth}{(a)}
\fig{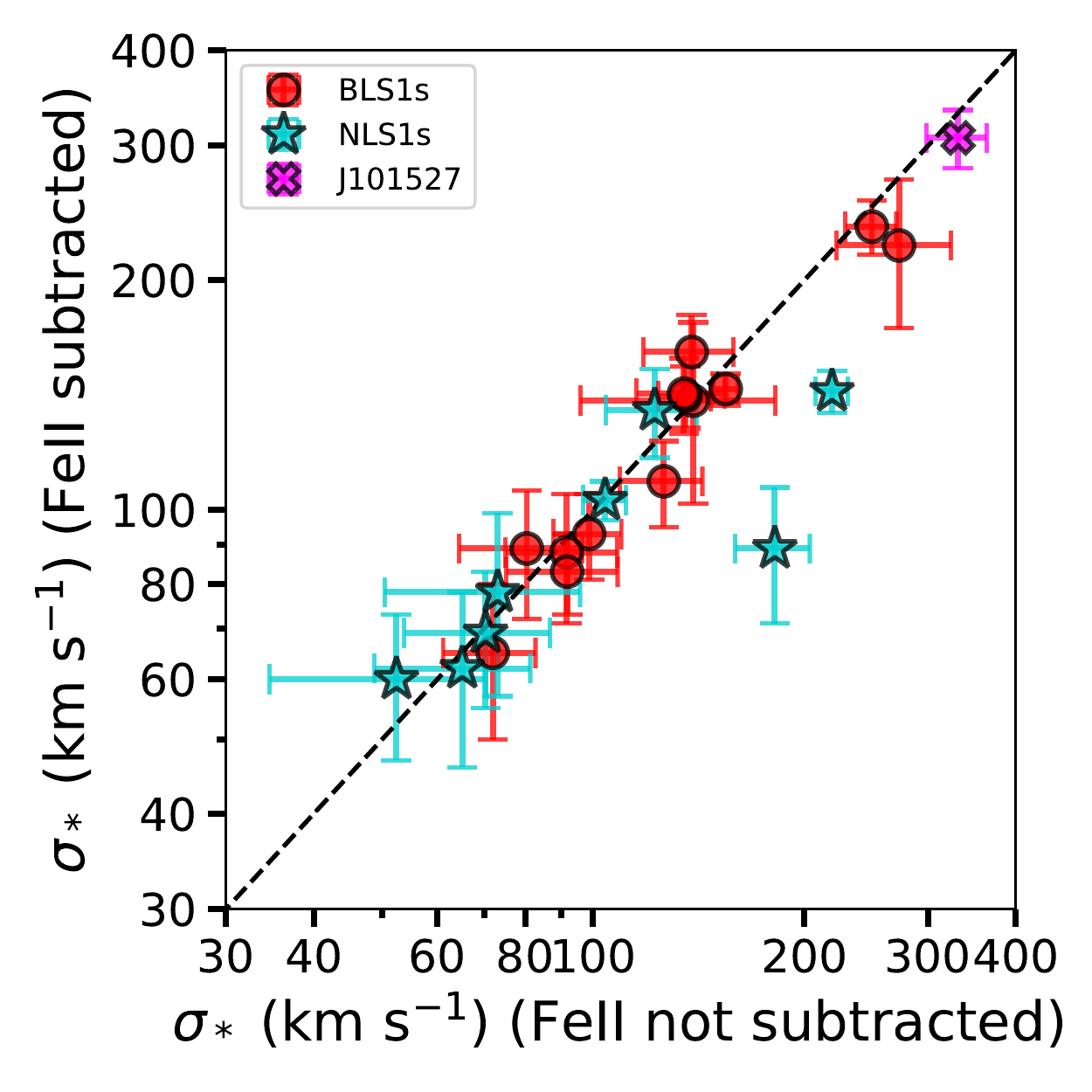}{0.5\textwidth}{(b)}
}\textbf{}
\caption{
(a) Bolometric luminosity vs. BH mass.  Dashed lines correspond to Eddington ratios.  The dotted red line indicates 50\% of the Eddington limit, beyond which it is ill-advised to measure $\sigma_*$ due to significant AGN continuum dilution.  All NLS1 objects in our sample have the highest Eddington ratios, consistent with previous studies which indicate that NLS1s have high accretion rates.  (b)  Effect of \ion{Fe}{2} emission on measurements of $\sigma_*$.  For the majority of our sample, the effect is negligible.  However, \ion{Fe}{2} contamination can significantly affect $\sigma_*$ measurements if strong narrow \ion{Fe}{2} is present, as this emission can mimic stellar absorption features.  Narrow \ion{Fe}{2} emission seems to be strongest in NLS1 objects, possibly due to their high accretion rates. 
\label{fig:edd_feii}}
\end{figure*}

Broad and narrow \ion{Fe}{2} emission is present in all objects in our sample to some extent and can have significant effects.  To account for this, we include 20 narrow and 91 broad \ion{Fe}{2} templates to be fit simultaneously with stellar templates.  We avoid subtracting off \ion{Fe}{2} emission prior to stellar template fitting due to the presence of strong narrow emission in some objects, which can mimic variations in the stellar continuum and make determination of the relative contribution of narrow \ion{Fe}{2} impossible.  Narrow emission, if unaccounted for, can bias measurements of $\sigma_*$ to larger values by as much 90\%, corresponding to a $0.2-0.3$ dex offset in $\log_{10}(\sigma_*)$ on the $M_{\rm{BH}}-\sigma_*$ relation.  We show the offset of measured values of $\sigma_*$ caused by the presence of \ion{Fe}{2} in the fitting region in Figure \ref{fig:edd_feii}(b).  We also show that NLS1s in our sample are the most affected by \ion{Fe}{2} contamination, particularly due to the presence of strong narrow \ion{Fe}{2} contamination in these objects.  There is a well known anti-correlation between the strength of \ion{Fe}{2} and other properties of NLS1 galaxies like BH mass and Eddington ratio (e.g., \citet{Grupe2004}, \citet{Komossa2008b}, \citet{Xu2012}), and we observe the same trend in our sample.  One such object, J123349, remains offset in $\log_{10}(\sigma_*)$ by $+0.16$ dex, which could be due to \ion{Fe}{2} template mismatch.  This highlights the importance of correcting for \ion{Fe}{2} emission, especially in samples of high luminosity and NLS1 (high Eddington ratio) where narrow \ion{Fe}{2} contamination is most common, as they can significantly bias $\sigma_*$ measurements.

\subsubsection{Morphology} 

\begin{figure*}
\gridline{
\fig{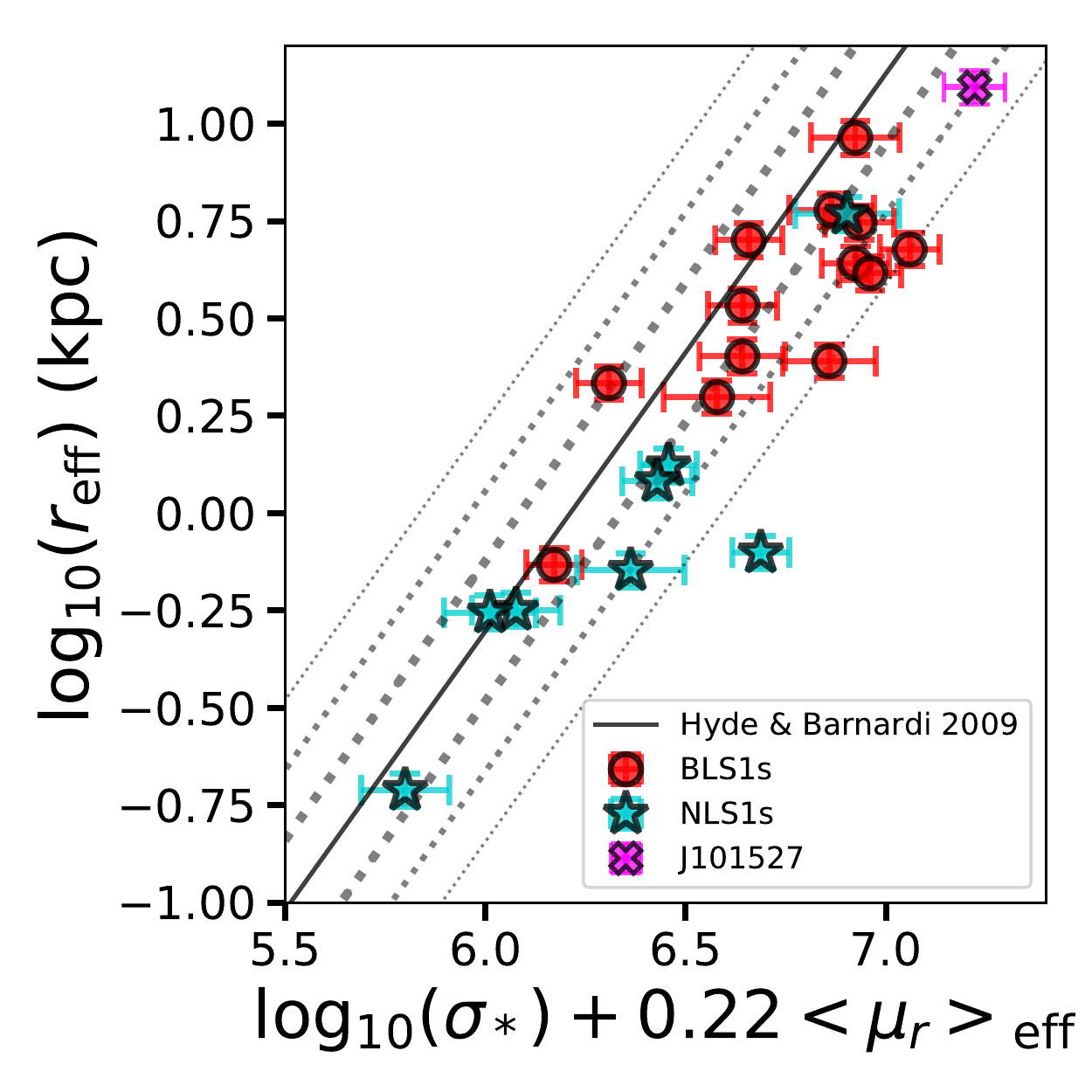}{0.5\textwidth}{(a)}
\fig{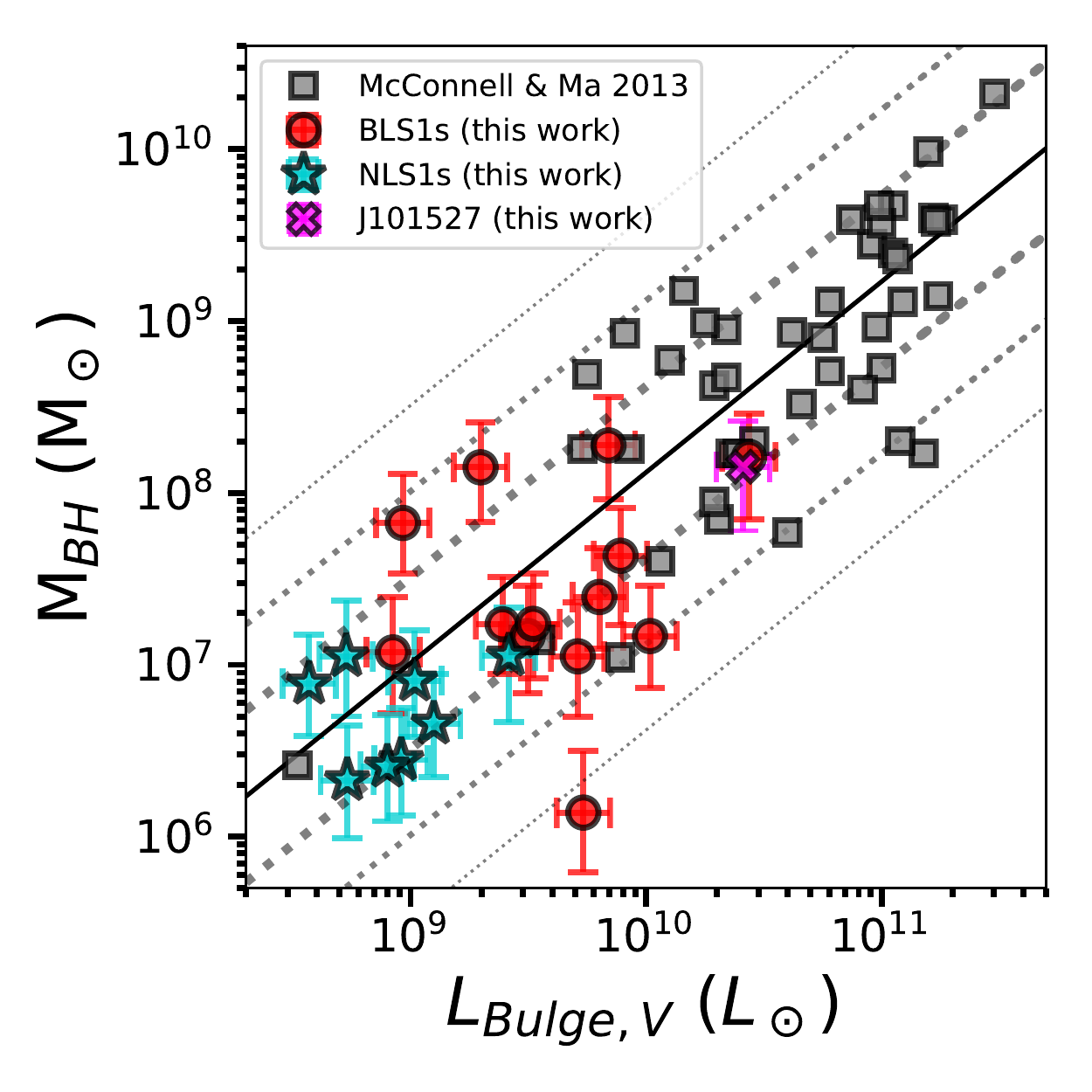}{0.5\textwidth}{(b)}
}\textbf{}
\caption{
(a) Fundamental plane relation for objects in our sample with magnitudes converted to SDSS-$r$ band.   Dashed lines enclose 68\%, 95\%, and 99\% of SDSS early-type galaxies from the orthogonal fit to $\sim$50,000 SDSS DR6 of early-type galaxies in SDSS-$r$ band at $0.0<z<0.35$ from \citet{Hyde2009}.  (b) The $M_{\rm{BH}}-L_{\rm{bulge}}$ relation for objects in our sample compared to local inactive objects from \citet{McConnell2013}.  Dashed lines represent the the 68\%, 95\% and 99\% confidence intervals of the scatter. 
\label{fig:fund_plane}}
\end{figure*}

The observed scatter in our sample could be attributed to properties such as host galaxy morphology, which can have a significant influence on the measurement of $\sigma_*$.  Morphological biases in $\sigma_*$ may arise if hosts are not elliptical or do not exhibit ``classical'' bulges.  For instance, \citet{Graham2011} showed that barred hosts tend to fall $\sim$0.5 dex below the $M_{\rm{BH}}-\sigma_*$ relation compared to non-barred hosts.  From our sample, five objects (J095819, J100234, J132943, J141234, and J145640) show clear bar morphologies within their disks; however, we see no such offset of barred hosts compared to non-barred hosts on the $M_{\rm{BH}}-\sigma_*$ relation within our sample.\\
\indent Another more obvious source of potential offset in $\sigma_*$ could be the result of a bulge that is no longer in dynamical equilibrium, such as in the case of a merger event.  Previous studies have shown that mergers in progress have been found to have increased scatter on the $M_{\rm{BH}}-\sigma_*$ relation and tend to have undermassive BHs relative to their hosts, corresponding to a larger velocity dispersion than inferred from the local relation \citep{Kormendy2011,Kormendy-Bender2013}.  More recently, high spatial resolution near-IR integral field spectroscopy performed by \citet{Medling2015} of nuclear disks of late-stage, gas-rich mergers have shown that their BHs are overmassive by a significant amount, suggesting that they grow more quickly than their hosts.  One object in our sample, J000338, appears to be in the early stages of a merger in \textit{HST} imaging and falls above the local relation by a factor of $0.31$ dex in BH mass but well within $1\sigma$ scatter of the local relation.  This is most consistent with time-resolved $N$-body simulations used to investigate the evolution of $\sigma_*$ during mergers performed by \citet{Stickley2014}, which found that $\sigma_*$ in the bulge component in the early stages of the interaction does not significantly deviate from the value of $\sigma_*$ measured before the interaction.  They also found that, while the value of $\sigma_*$ oscillates during the merger process, it is unlikely that the deviation from the equilibrium value will be large.  Considering the large separation distance between the two progenitors ($\sim11$ kpc, not considering any projection effects), and that the measured $\sigma_*$ is within 15\% of the value of $\sigma_*$ implied by the local relation, we conclude that the $\sigma_*$ measured for J000338 is consistent for a dynamically relaxed bulge and do not omit it from our analyses. \\
\indent Another object, J101527, appears to show evidence of interaction from \textit{HST} imaging, and surface brightness decomposition of J101527 also reveals a double nucleus, consisting of an AGN and another low-surface-brightness object.  \citet{Kim2017} classified J101527 as a candidate recoiling SMBH resulting from a merger, and the host galaxy is likely a bulge-dominated elliptical in the late stages of a merger (see \citet{Kim2017} for a detailed analysis of J101527).  Kim et al. estimated the stellar velocity dispersion from Keck/LRIS spectra using the [\ion{S}{2}]$\lambda6716$ width following \citet{Komossa2007}, obtaining a value of $\sigma_{[\rm{S\ II}]}=190\pm20$ km s$^{-1}$, which places J101527 very close to our local $M_{\rm{BH}}-\sigma_*$ relation.  We measure a nearly identical value using the [\ion{O}{3}]$\lambda5007$ width of $\sigma_{[\rm{O\ III}]}=197\pm3$ km s$^{-1}$ from our Keck/LRIS spectra.  Measuring $\sigma_*$ directly from the stellar continuum, we find $\sigma_*=307\pm27$ km s$^{-1}$, a 56\% difference from what is measured from the [\ion{O}{3}] width.  The large offset in $\sigma_*$ results in a BH mass that is undermassive by $\sim$1.0 dex, making it the largest outlier in our sample. However, this offset may indicate that the stellar component is not yet dynamically relaxed.  Numerical simulations indicate that measurements of $\sigma_*$ are enhanced for merging nuclei as separation distance decreases \citep{Stickley2014}.  The clear morphological peculiarities of this object, as well as the large uncertainty in measured $\sigma_*$ values, warrant the omission of J101527 from analyses when considering evolution in the $M_{\rm{BH}}-\sigma_*$ relation.  We however include its measurements in Table \ref{tab:mbh_table} as well as flag this object as a merger in our diagrams.\\
\indent To further investigate possible biases due to morphology, we consider the location of bulges of our sample on the fundamental plane relation (FP; \citet{Djorgovski1987}).  Surface brightness measurements are obtained using the \texttt{sersic2} option in \textsc{GALFIT} and appropriate corrections for extinction, $k$-correction, surface brightness dimming, filter transformations, and passive evolution are applied.  The FP relation for our sample is shown in Figure \ref{fig:fund_plane}(a).  Following \citet{Canalizo2012}, we compare our objects to the SDSS-$r$ orthogonal fit to $\sim$50,000 SDSS DR6 of early-type galaxies at $0.0<z<0.35$ from \citet{Hyde2009}, given by the solid line in Figure \ref{fig:fund_plane}(a).  We find that the majority of our sample is in good agreement with the FP relation. 
One object, J123349, falls completely outside the SDSS scatter.  This object is classified as an NLS1 and exhibits a high fraction of \ion{Fe}{2} contamination in its spectra, which is likely biasing the measurement of $\sigma_*$ to higher values despite our best efforts to account for it using \ion{Fe}{2} templates.  The measured stellar velocity dispersion of J123349 also places this object high on the Faber-Jackson relation \citep{Faber1976} relative to its bulge luminosity, indicating that \ion{Fe}{2} contamination is likely contributing to its offset on the FP relation.\\
\indent It is worth noting that the NLS1s in our sample have consistently smaller bulges than the BLS1s.  Some previous studies have suggested that NLS1s fall below the $M_{\rm{BH}}-\sigma_*$ relation \citep{Mathur2001,Grupe2004}. These were based on $\sigma_{\rm{[O\ III]}}$ as a proxy for $\sigma_*$; however, they did not remove blue outliers, which are completely dominated by outflows \citep{Komossa2007}.  After removing blue outliers, the remaining NLS1 galaxies scatter around the relation like BLS1 galaxies.  In a study of 93 local SDSS NLS1 galaxies, \citet{Woo2015} found similar agreement with the local relation when stellar velocity dispersions are measured directly from stellar absorption features.  \\
\indent The use of \textit{HST} imaging of NLS1s in our sample allows us to further investigate the morphologies of these objects in greater detail.  The location of NLS1s on the FP relation in Figure \ref{fig:fund_plane}(a) would imply that they do indeed have smaller bulges, as we would expect from their location on the $M_{\rm{BH}}-\sigma_*$ relation.  This also presents a strong case for narrow H$\beta$ emission being indicative of a lower-mass BH, and not due to any peculiar geometry of the BLR \citep{Decarli2008,Decarli2011}.  However, one caveat is that the NLS1 sample used by \citet{Decarli2008} (originating from \citet{Grupe1999}) was specific in selecting NLS1s with strong \ion{Fe}{2}, whereas our sample contains only two (of eight) NLS1s with strong \ion{Fe}{2}.  Despite this, it is worth noting that aside from measurements of $\sigma_*$ (which are directly influenced by \ion{Fe}{2} contamination), the NLS1s in our sample appear to show similar physical characteristics (lower $M_{\rm{BH}}$, higher $L/L_{\rm{Edd}}$, and $L_{\rm{Bulge},V}$) that do not appear to be a function of \ion{Fe}{2} strength.  However, a larger comparison sample of NLS1 objects of varying \ion{Fe}{2} strength is needed to make any conclusive statements on how \ion{Fe}{2} strength affects other properties of NLS1s.  \\
\indent As a final test for any morphological bias, we investigate where our sample falls on the $M_{\rm{BH}}-L_{\rm{bulge}}$ relation.  We transform bulge luminosities from each respective \textit{HST} filter to Johnson-$V$ luminosities, making all the necessary corrections, to compare our sample to the $M_{\rm{BH}}-L_{\rm{bulge}}$ relation from \citet{McConnell2013}.  We plot the $M_{\rm{BH}}-L_{\rm{bulge}}$ relation of our sample in Figure \ref{fig:fund_plane}(b) alongside points and best-fit from McConnell \& Ma, with the 68\%, 95\% and 99\% confidence intervals of the scatter.  We find good agreement with the relation from McConnell \& Ma, with our sample having a comparable scatter.  One of our objects on the $M_{\rm{BH}}-L_{\rm{bulge}}$ relation, J092438, is a clear outlier and we discuss possible reasons for its apparent undermassive BH relative to its host bulge luminosity in the Appendix.

\subsubsection{Selection Effects} \label{sec:selection_effects}

\begin{figure*}[ht!]
\center
\includegraphics[trim={0 0 0 0},scale=0.6,clip]{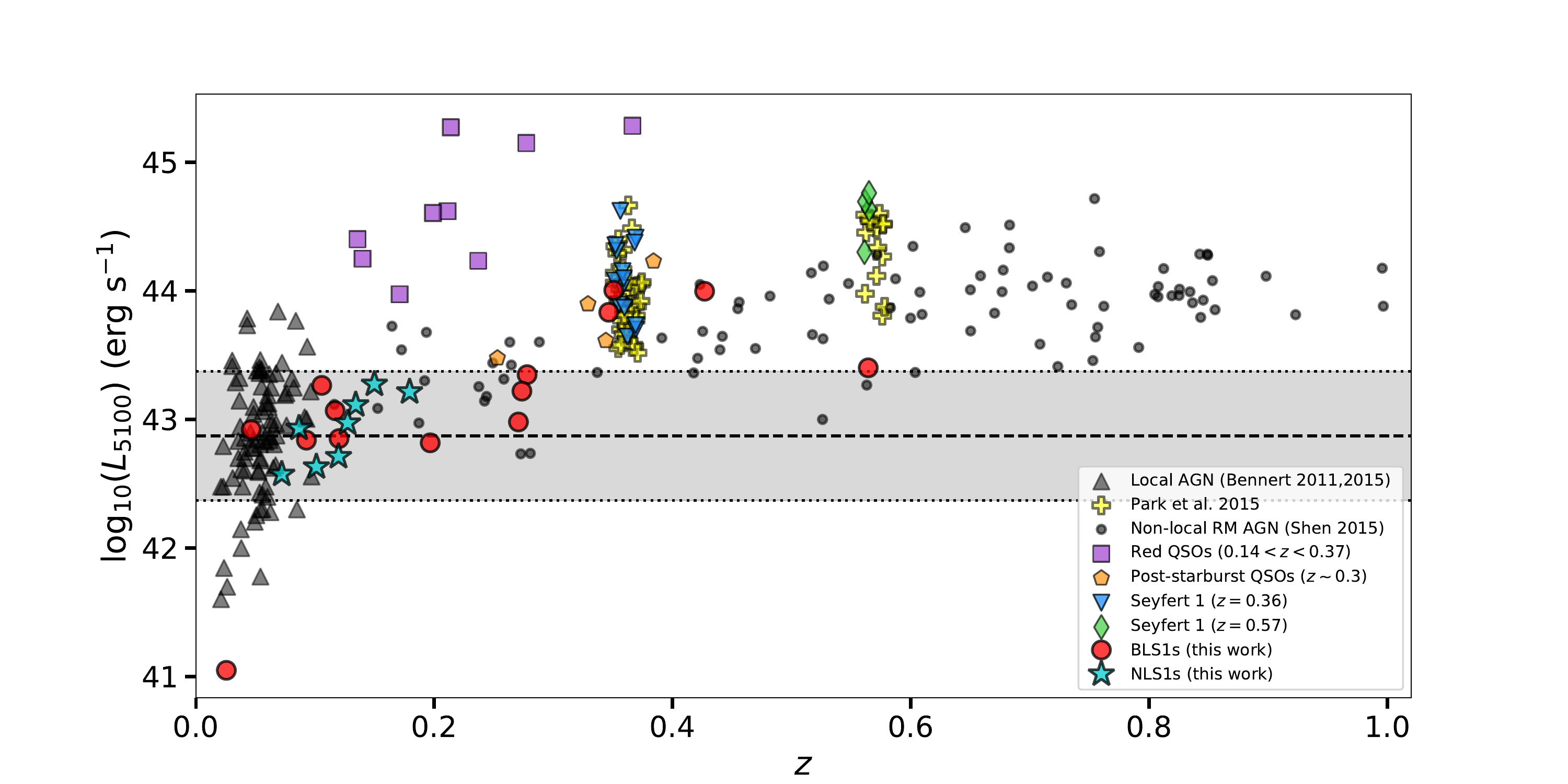}
\caption{
Comparison of AGN luminosities for non-local samples as a function of redshift.  The dashed line indicates the local average and the gray shaded area represents the scatter of local AGNs from the combined samples of \citet{Bennert2011a,Bennert2015}.  Non-local objects which have considerable offset from the local $M_{\rm{BH}}-\sigma_*$ have AGN luminosities higher than the local average, indicating that these samples may be sampling the upper envelope of BH masses at these redshifts due to the luminosity-threshold selection effects.  Our sample was able to probe lower-mass BHs out to $z\sim0.3$, however, it likely suffers from selection effects at higher redshifts. 
\label{fig:agn_luminosity_vs_redshift}}
\end{figure*}

In addition to measurement uncertainties, we investigate any possible uncertainties and biases that may result from selection criteria.  The selection criteria used for our sample required a broad H$\beta$ FWHM within the range $500\text{~km s}^{-1}\leq\mathrm{FWHM}_{\mathrm{H}\beta}\leq2000 \text{~km s}^{-1}$ to select candidate NLS1 objects, and visible stellar absorption features (typically \ion{Ca}{0}H+K equivalent width $\mathrm{EW}_{\mathrm{Ca H+K}}>0.5$ \angstrom~) to ensure that $\sigma_*$ could be accurately measured.  \\
\indent In general, the presence of broad lines may select against host galaxies with higher (edge-on) inclinations, which have systematically higher $\sigma_*$ values relative to host galaxies at lower inclinations due to the presence of contaminating disk stars along the line of sight \citep{Bellovary2014}.  We correct $\sigma_*$ for objects in our sample which contain disks to face-on $\sigma_*$ values to account for any $\sigma_*$ values that may be inflated due to the presence of a disk and thus give the appearance of an undermassive BH on the $M_{\rm{BH}}-\sigma_*$ relation (see Section \ref{sec:sigma}).\\
\indent The broad H$\beta$ line width requirement, by design, biases our sample toward lower BH masses.  Our original intention was to detect lower-mass BHs at high redshifts using NLS1 galaxies.  Unfortunately, our selection in FWHM$_{\rm{H}\beta}$ was not rigorously met due to the nature of the SDSS DR7 line-fitting algorithm, and no NLS1 objects were found at $z>0.2$.  On the other hand, our sample covers a broad range in mass: two orders of magnitude in the range $6.3<\log_{10}(M_{\rm{BH}})<8.3$.  Finally, the requirement for visible absorption lines to accurately measure $\sigma_*$ in the \ion{Mg}{1b} region also preferentially selects objects with AGN luminosities lower than or comparable to the host galaxy, therefore requiring lower average AGN luminosities and lower BH masses, assuming the AGN luminosity is due entirely to BH accretion and a function of BH mass.  Indeed, the average AGN-to-total light fraction of our sample (as measured from \textit{HST} imaging) is $41\pm 12\%$ with no objects exceeding $82\%$, which ensures we should be able to accurately measure $\sigma_*$ and should not be significantly biased toward higher dispersions due to AGN dilution of stellar absorption features.\\
\indent However, by selecting objects with visible stellar absorption features with a minimum EW (line strength), we may implicitly introduce some maximum AGN luminosity threshold relative to the host galaxy per given redshift.  A requirement for visible stellar absorption features implies that stellar light from the host galaxy is not significantly diluted by the light from the AGN, which implies there is some maximum AGN-to-total light ratio beyond which $\sigma_*$ cannot be measured, and therefore a maximum luminosity threshold for the AGN.  Additionally, the S/N requirements for visible absorption features in higher-redshift objects would require more massive, and therefore more luminous host galaxies, of which there are fewer due to the steep drop in the luminosity function of galaxies.  Therefore selection by absorption line visibility at higher redshifts could still be biased to overmassive BHs. We do see that the highest-redshift BHs in our sample tend to be overmassive, which may be the result of our implicitly imposed AGN luminosity threshold. \\ 
\indent \citet{Lauer2007} explained that intrinsic scatter in the $M_{\rm{BH}}-$bulge scaling relations implies that, for a given $L_{\rm{bulge}}$ or $\sigma_*$, there exists a range of BH masses.  However, samples selected by some given AGN luminosity threshold (as is typically done for high-redshift samples) results in a distribution of $L_{\rm{host}}$ and $\sigma_*$ per $M_{\rm{BH}}$, assuming that AGN luminosity is determined solely by BH mass.  This, combined with the steep drop in the luminosity function of galaxies, preferentially selects overmassive BHs relative to their hosts, which can thus produce a false signature of evolution in high-redshift studies.  \citet{Woo2006,Woo2008} selected objects with a minimum H$\beta$ EW of 5~\angstrom, but using Monte Carlo simulations to model a sharp selection in luminosity, \citet{Treu2007} found that this bias is negligible for the $M_{\rm{BH}}-\sigma_*$ relation unless the scatter at high redshift increases considerably. \\
\indent In addition to the \citet{Lauer2007} bias, \citet{Shen2010} suggest that luminosity-threshold samples tend to be biased toward high SE virial BH masses by as much as $0.2$ to $0.3$ dex, due to the uncorrelated variations between continuum luminosity and line widths in reverberation-mapping studies, from which we obtain BH virial mass estimates.  Together, it is possible the \citet{Lauer2007} and \citet{Shen2010} biases can account for the offset in BH mass at high redshift, with biases becoming worse as a function of $z$ for higher-luminosity thresholds.  \\
\indent We plot $\log_{10}(\lambda L_{5100})$ as a function of $z$ in Figure \ref{fig:agn_luminosity_vs_redshift}, comparing our and other non-local samples with the local AGN samples from \citet{Bennert2011a,Bennert2015}.  We also include objects from the \citet{Park2015} sample, which expanded the number of $M_{\rm{BH}}-L_{\rm{bulge}}$ measurements at $z=0.36$ and $z=0.57$.  There is clear offset in AGN luminosity as a function of redshift for all non-local samples with respect to local AGN luminosities.  As \citet{Canalizo2012} pointed out, the AGN luminosities of non-local objects tend to reside at $\log_{10}(L_{5100}/{\rm{erg~s}^{-1}})>43.6$.  Objects in our sample have AGN luminosities consistent with the local AGN sample up until to $z\sim0.3$, at which point $\log_{10}(L_{5100}/{\rm{erg~s}^{-1}})>43.6$.  At $z>0.3$, three out of the four high-luminosity objects in our sample also lie above the $M_{\rm{BH}}-\sigma_*$ relation.  The same trend was observed by \citet{Shen2015} who, by using SDSS virial BH masses to probe the $M_{\rm{BH}}-\sigma_*$ relation out to $z=1.0$, found a similar trend with BH mass and AGN luminosity as a function of $z$.  They pointed out that the limited dynamic range and higher average AGN luminosities in \citet{Woo2006,Woo2008} are responsible for the observed offset compared to local objects with considerably lower average luminosities.  This trend, which is observed here and in other studies of the non-local $M_{\rm{BH}}-\sigma_*$ relation \citep{Shen2008,Canalizo2012,Hiner2012,Park2015,Shen2015}, makes it clear that significant statistical biases are at work in luminosity threshold samples.  However, as we have shown here, these biases can be overcome by probing lower luminosities and lower BH masses at higher redshifts.  Our selection criteria, which selected objects only by broad-line width and the presence of absorption features, was able to probe lower AGN luminosities than previously observed in SE studies, out to $z\sim0.3$.  The eight NLS1s in our sample also have similar AGN luminosities to the local AGN sample up to $z\sim0.2$.  A larger sample of objects selected with similar criteria is necessary to show if we can further probe similar local luminosities at higher redshifts.  We note, however, that our ability to probe similar BH mass regimes at higher redshifts assumes that the shapes of the luminosity and BH mass functions do not significantly change with redshift.  Likewise, if the scatter in the $M_{\rm{BH}}-\sigma_*$ relation changes significantly with redshift, it may indicate the presence of additional  statistical biases which must be accounted for.

\section{Discussion} \label{sec:discussion}

We consider the effects of an evolution of the $M_{\rm{BH}}-\sigma_*$ relation as a function of cosmic time and its physical interpretation, if real.  \citet{Woo2008} measured a significant offset in $M_{\rm{BH}}$ in non-local BHs, implying a significant evolution in bulge mass over relatively short timescales.  Using the scaling relation for central velocity dispersion with stellar mass of SDSS quiescent galaxies from \citet{Zahid2016}, we estimate that the offset in $M_{\rm{BH}}$ from \citet{Woo2008} implies that bulges must grow by a factor of $\sim6$ within $5.5$ Gyr ($z=0.57$) to be consistent with the local relation at $z=0$.  Objects from \citet{Canalizo2012} and \citet{Hiner2012}, which also fall significantly above the local relation, further increase the evolution slope and required amount stellar mass assembly.  With the inclusion of our objects, we measure an evolution slope of $\gamma=2.16\pm0.62$, which relaxes the amount of bulge growth required to a factor of $\sim4$ within $5.5$ Gyr.  Despite the improved constraints in the local relation and increased non-local sample size, the required mass assembly for non-local bulges necessary to be consistent with the local $M_{\rm{BH}}-\sigma_*$ relation remains high.  The means by which bulges achieve such considerable secular mass assembly are the subject of ongoing debate.  It is possible that disruption of stellar disks and/or minor mergers can cause significant bulge growth without significantly growing the SMBH \citep{Croton2006,Jahnke2009,Bennert2011b,Cisternas2011}.  \\
\indent It is certainly possible that the evolutionary trend we measure in the $M_{\rm{BH}}-\sigma_*$ relation is not of physical origin, but the confluence of selection effects and large scatter.  In Section \ref{sec:selection_effects}, we discussed possible biases due to our selection criteria, which may lead to the overmassive BHs we observe at $z>0.4$ and consequent significant evolution we measure in the $M_{\rm{BH}}-\sigma_*$ relation. Figure \ref{fig:agn_luminosity_vs_redshift} clearly shows that we are likely still sampling the upper envelope of BH masses at higher redshifts, which could be driving the evolution slope upwards at the high-redshift end. These biases present the greatest challenge in acquiring an unbiased sample at high redshift, however, we have shown that AGN selection by H$\beta$ line width (and not luminosity) allows us to probe lower SE BH masses at higher redshifts than previously measured, assuming strong stellar absorption features are present to accurately measure $\sigma_*$.  Furthermore, the $0.44$ dex scatter we observe as a function of redshift (see Figure \ref{fig:evolution}) is significant and consistent across the entire redshift range of our sample, indicating that any linear trend - and therefore evolution - in the $M_{\rm{BH}}-\sigma_*$ relation as a function of redshift is weak at best ($r_s = 0.23\pm0.04$).  The fact that we can fit the points in Figure \ref{fig:evolution} with a constant model (as opposed to a linear model) and achieve nearly identical residual scatter indicates that we lack sufficient evidence for significant evolution in the $M_{\rm{BH}}-\sigma_*$ relation up to $z\sim0.6$, and that any measured offset is being driven by a sampling of higher-luminosity AGNs, and therefore higher-mass BHs, at higher redshifts.   \\
\indent We stress that measurements of $\sigma_*$ are the largest single source of systematic uncertainty for objects on the $M_{\rm{BH}}-\sigma_*$ relation due mainly to low S/N at high redshift.   Even if $\sigma_*$ can be directly measured from stellar absorption features, biases due to the choice of fitting region, the presence of \ion{Fe}{2} contamination, possible AGN dilution, and host galaxy morphology can introduce significant uncertainties which may contribute the scatter in the $M_{\rm{BH}}-\sigma_*$ relation.  The obvious alternative to measuring faint stellar absorption features at high redshift is to use brighter gas emission features, such as the [\ion{O}{3}]$\lambda5007$ width, as a proxy for $\sigma_*$, assuming that the NLR gas traces the stellar LOSVD of bulges.  We found that obtaining accurate measurements of the [\ion{O}{3}] width is not trivial, as one must correct for both \ion{Fe}{2} contamination and the presence of blue-wing features which may indicate the presence of gas outflows (also see, e.g., \citet{Komossa2008a}, \citet{Woo2016} and \citet{Bennert2018}).  Still, there remains considerable scatter in the relationship between $\sigma_*$ and [\ion{O}{3}] width, and therefore studies of the $M_{\rm{BH}}-\sigma_*$ relation using [\ion{O}{3}] as a proxy should only be used in statistical samples. \\
\indent We have shown that, aside from their high Eddington ratios, the host galaxies of NLS1 objects contain BHs that reside on the local $M_{\rm{BH}}-\sigma_*$ relation.  Furthermore, the NLS1s in our sample are in good agreement with lower-mass bulges on the FP relation, indicating that their observed narrow broad-lines are likely due to having lower BH masses (as the $M_{\rm{BH}}-\sigma_*$ relation would imply) and not an observational or geometrical peculiarity of the BLR as previous studies have suggested.
However, a larger sample of NLS1s is still required to determine their behavior on the $M_{\rm{BH}}-\sigma_*$ relation at higher redshifts. \\
\indent We conclude that there is insufficient evidence for evolution in the $M_{\rm{BH}}-\sigma_*$ relation up to $z\sim0.6$ due to comparable scatter of objects on the local $M_{\rm{BH}}-\sigma_*$ relation and for objects at lower redshifts.  The case for no evolution within our sampled redshift range is consistent with the results found by \citet{Schramm2013}, who found good agreement with the local $M_{\rm{BH}}-M_{\rm{bulge}}$ relation using a sample of 18 X-ray-selected AGNs at $0.5<z<1.2$.  It is still possible that there is significant offset at higher redshifts than those sampled here. \rm For instance, using gravitationally lensed quasar hosts, \citet{Peng2006} found that BHs grew significantly faster than their hosts at $z>1.7$.  \citet{Bennert2011b} also found significant offset in the $M_{\rm{BH}}-M_{\rm{bulge}}$ relation at $z\sim2$, implying the BH mass growth pre-dates bulge formation.  However, current studies of the non-local $M_{\rm{BH}}-\sigma_*$ relation still lack statistically representative samples of BHs at the currently sampled redshifts, as we are likely still sampling the upper envelope BH masses due to selection effects.  These are the hurdles that must be overcome in order to conclusively determine any evolution in the $M_{\rm{BH}}-\sigma_*$ relation, and which we will address in future studies. 

\acknowledgments
\indent We thank the anonymous referee for their thoughtful and constructive comments on this work.  We also thank Giorgio Calderone and Michele Cappellari for their email correspondence regarding spectral fitting using \textsc{QSFit} and \textsc{pPXF}, respectively.\\
\indent Partial support for this project was provided by the National Science Foundation, under grant No. AST 1817233.  \\
\indent Some of the data presented herein were obtained at the W. M. Keck Observatory, which is operated as a scientific partnership among the California Institute of Technology, the University of California and the National Aeronautics and Space Administration. The Observatory was made possible by the generous financial support of the W. M. Keck Foundation.\\
\indent The authors wish to recognize and acknowledge the very significant cultural role and reverence that the summit of Maunakea has always had within the indigenous Hawaiian community.  We are most fortunate to have the opportunity to conduct observations from this mountain.\\
\indent Some of the data presented herein were obtained using the UCI Remote Observing Facility, made possible by a generous gift from John and Ruth Ann Evans\\
\indent Based on observations made with the NASA/ESA \textit{Hubble Space Telescope}, and obtained from the Hubble Legacy Archive, which is a collaboration between the Space Telescope Science Institute (STScI/NASA), the Space Telescope European Coordinating Facility (ST-ECF/ESA) and the Canadian Astronomy Data Centre (CADC/NRC/CSA)\\
\indent Funding for the SDSS and SDSS-II has been provided by the Alfred P. Sloan Foundation, the Participating Institutions, the National Science Foundation, the U.S. Department of Energy, the National Aeronautics and Space Administration, the Japanese Monbukagakusho, the Max Planck Society, and the Higher Education Funding Council for England. The SDSS Web Site is \url{http://www.sdss.org/}.  The SDSS is managed by the Astrophysical Research Consortium for the Participating Institutions. The Participating Institutions are the American Museum of Natural History, Astrophysical Institute Potsdam, University of Basel, University of Cambridge, Case Western Reserve University, University of Chicago, Drexel University, Fermilab, the Institute for Advanced Study, the Japan Participation Group, Johns Hopkins University, the Joint Institute for Nuclear Astrophysics, the Kavli Institute for Particle Astrophysics and Cosmology, the Korean Scientist Group, the Chinese Academy of Sciences (LAMOST), Los Alamos National Laboratory, the Max Planck Institute for Astronomy (MPIA), the Max Planck Institute for Astrophysics (MPA), New Mexico State University, Ohio State University, University of Pittsburgh, University of Portsmouth, Princeton University, the United States Naval Observatory, and the University of Washington.\\
\indent While contributing to this research K. Hiner was supported by the UCR Dissertation Year Fellowship and FONDECYT grant 3140154.  J.H.W. acknowledges support by the Basic Science Research Program through the National Research Foundation of Korea government (No.2017R1A5A1070354).  E.T. acknowledges support from FONDECYT Regular 1160999, CONICYT PIA ACT172033 and Basal-CATA PFB-06/2007 and AFB170002 grants.

%

\vspace{5mm}
\facilities{
HST (ACS,NIC,WFPC2,WFC3), HLA, Keck:I (LRIS), Sloan
}


\software{
{Astropy } \citep{astropy}, \textit{emcee} \citep{emcee}, \textsc{GALFIT } \citep{galfit},  \textsc{pPXF } (\citet{ppxf1}, \citet{ppxf2}), {PyRAF } ({PyRAF} is a product of the Space Telescope Science Institute, which is operated by AURA for NASA), \textsc{SExtractor 
}}



\appendix
\section{Notes on Individual Objects}\label{sec:indiv_objects}

\subsection{J092438.88+560746.8}

This object is a clear outlier on the $M_{\rm{BH}}-L_{\rm{bulge}}$ relation shown in Figure \ref{fig:fund_plane}(b), characterized by an apparent undermassive BH relative to its host-bulge luminosity.  The \textit{HST} imaging of this object clearly shows it is host to a disk and spiral arm component, which we fit using \textsc{GALFIT} to ensure accurate deconvolution of the bulge component (see Figure \ref{fig:galfit}).  It is also worth noting that the host-disk component appears to have a considerably low axis ratio ($b/a$), which implies that the disk is being viewed at high inclination assuming that the disk is circular. As a result, we correct the measured stellar velocity dispersion for inclination in Section \ref{sec:inclination_corrections}.  Curiously, this object is not an outlier on either the $M_{\rm{BH}}-\sigma_*$ or FP relations.  The offset from the $M_{\rm{BH}}-L_{\rm{bulge}}$ relation appears to be due not from an intrinsically high bulge luminosity, but an undermassive BH stemming from a faint AGN continuum. The luminosity at 5100 \angstrom\; is the lowest measured in our sample, resulting in an Eddington ratio of $<1\%$ (the lowest point in Figure \ref{fig:edd_feii}(a)).  \\
\indent We also observe a significant difference in broad H$\beta$ line width between the Keck and SDSS spectra.  Here we classify this object as a BLS1 galaxy, since we measure $\rm{FWHM}_{{\rm{H}}\beta}=2650\pm205$ km s$^{-1}$ with the Keck spectrum; however, a fit to the SDSS spectrum of this object indicates that it is closer to an NLS1 galaxy with $\rm{FWHM}_{{\rm{H}}\beta}=2036\pm75$ km s$^{-1}$.  One possibility for the difference in line width could be due to the aperture size between Keck and SDSS observations.  Since this object has $z=0.025$, the $3''$ diameter fiber of the SDSS contains significantly more host-galaxy flux, and therefore absorption, than the $1"$ slit we used for our Keck observations, leading to a more asymmetric -- and possibly narrower -- line profile.  \\
\indent The second possibility is that J092438 could be a ``changing-look AGN''. These are AGNs showing strong variability in their broad Balmer lines, and their continuum emission, and in the most extreme cases they change completely from Seyfert 2 (no BLR at all detected) to a Seyfert 1 (very strong BLR), the cause of which remains debated.  This could be caused by cases of extreme extinction, but perhaps more likely by strong intrinsic continuum variability, to which the BLR responds.  In most cases, the Seyfert type does not change completely, but the broad-line widths do vary significantly. \citet{Runco2016} found that, in a sample of 102 local SDSS AGNs, followed up using Keck, $\sim38\%$ showed appreciable BLR line variability.\\
\indent Both Keck and SDSS observations are consistent with J092438 being in a low state, indicated by its very faint AGN continuum and Balmer lines.  If one assumes that the AGN continuum is intrinsically stronger, and its BLR lines are stronger and broader, then J092438 would be more consistent with the $M_{\rm{BH}}-L_{\rm{bulge}}$ relation and fall within the upper limits of the scatter of the local $M_{\rm{BH}}-\sigma_*$ relation.

\subsection{J145640.99+524727.2}

\begin{figure*}[ht!]
\center
\includegraphics[trim={1cm 0.0cm 1cm 1cm},scale=0.75,clip]{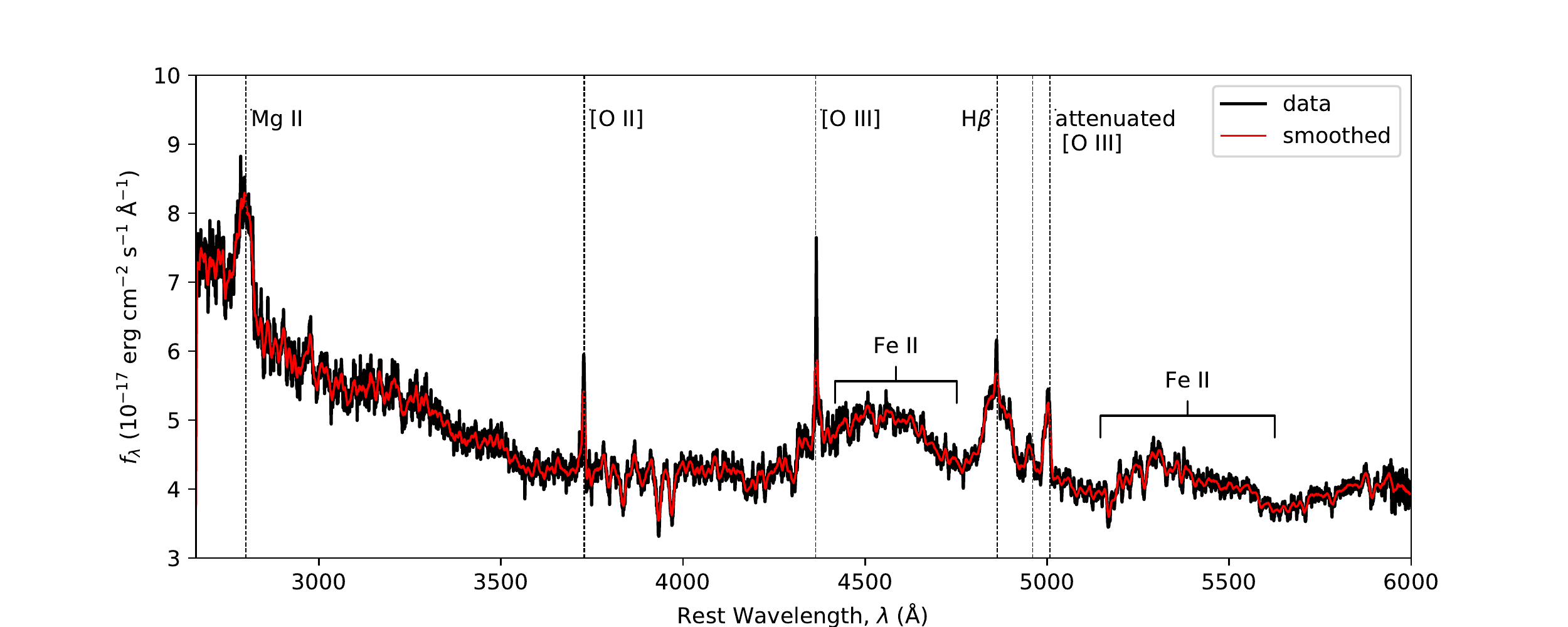}
\caption{
Full Keck/LRIS spectrum of J145640, the features of which are consistent with that of a low-ionization broad absorption line quasi-stellar object.  We have labeled strong emission features, most notably the characteristic strong broad \ion{Fe}{2} emission and attenuated [\ion{O}{3}]$\lambda\lambda4959,5007$ emission.  Absorption in \ion{Mg}{2} can be seen blueward of the emission line.
\label{fig:J145640}}
\end{figure*}

This BLS1 object is characterized by a strong attenuation in the [\ion{O}{3}] complex, requiring the region to be masked during the multi-component fitting process, as shown in Figure \ref{fig:emline}.  Narrow H$\beta$ does not appear attenuated and could be fit without constraints.  The [\ion{O}{3}]$\lambda\lambda4959,5007$ lines appear to have a considerably larger width than the narrow component of H$\beta$, and appear to be asymmetric with respect to the expected [\ion{O}{3}] line centers, implying that this object exhibits the asymmetric blue-wing components discussed in Section \ref{sec:oiii}.  The decomposition of the [\ion{O}{3}]$\lambda\lambda4959,5007$ lines could not be performed, however, due to the severe attenuation.  The strong attenuation does not appear to be the result of poor skyline subtraction during the spectral reduction process, as the SDSS spectrum appears to show the same attenuation.  \citet{Cales2013} also studied this object as part of a sample of 38 post-starburst quasars with spectra obtained from the Kitt Peak National Observatory Mayall 4 m telescope, but could not fit the [\ion{O}{3}]$\lambda\lambda4959,5007$ lines in their spectrum either.  The spectrum also exhibits very strong broad \ion{Fe}{2} emission, which can be seen on either side of the H$\beta$/[\ion{O}{3}]$\lambda\lambda4959,5007$ complex in Figure \ref{fig:observations}.  More notably, [\ion{O}{3}]$\lambda4363$, which requires higher temperatures and densities than [\ion{O}{3}]$\lambda\lambda4959,5007$ \citep{Osterbrock1989}, is very strong.  The shape of [\ion{O}{3}]$\lambda4363$ is also much more symmetric than the more attenuated and more asymmetric [\ion{O}{3}]$\lambda\lambda4959,5007$ doublet.  \citet{Zakamska2016} found similar line profiles in high-redshift red quasars, which may indicate that [\ion{O}{3}]$\lambda\lambda4959,5007$ may originate from a region of lower-density and more easily accelerated gas than [\ion{O}{3}]$\lambda4363$.  Finally, our Keck observations also reveal absorption blueward of broad \ion{Mg}{2}. These features are common characteristics in low-ionization broad absorption line (LoBAL) QSOs.  \citet{Zhang2010} studied a sample of 68 SDSS LoBAL QSOs at $0.4<z\leq 0.8$, some of which show similar \ion{Mg}{2} absorption as J145640.  No previous studies classify J145640 as a LoBAL, likely due to the fact that at $z=0.278$, \ion{Mg}{2} remains ($\sim220$ \angstrom) below the wavelength coverage of the SDSS. We plot the full Keck spectrum of J145640 in Figure \ref{fig:J145640}.  LoBALs are thought to be associated with high accretion rates and/or early stages of AGN/galaxy evolution \citep{Canalizo2001,Lazarova2012,Hamann2019}.  We find no significant differences between the properties of J145640 and those of the rest of the sample, with the exception of its Eddington ratio, which is the second lowest in the sample.

\bibliography{mybib.bib}
\bibliographystyle{aasjournal}

\end{document}